\documentclass[journal]{IEEEtran}
\usepackage{url}
\usepackage{graphicx}
\usepackage{color}
\usepackage{cite}
\usepackage{subfigure}
\usepackage{verbatim}
\usepackage{amsmath}
\usepackage{pifont}
\usepackage{amssymb}
\usepackage{amsthm}
\usepackage{multirow}
\usepackage[table,xcdraw]{xcolor}
\usepackage{adjustbox}
\usepackage[flushleft]{threeparttable}
\usepackage{algorithmic}
\usepackage[lined,boxed,ruled,linesnumbered]{algorithm2e}
\usepackage{cuted}
\usepackage[nocomma]{optidef}
\newcommand{\xmark}{\ding{55}}%
\usepackage{subfiles}
\usepackage{acro}
\DeclareAcronym{NIST}{short =NIST,long=National Institute of Standards and Technology}
\DeclareAcronym{IENs}{short = IENs,long=Intelligent Energy Networks}
\DeclareAcronym{SG2}{short = SG2,long=Smart Grid 2.0}
\DeclareAcronym{SG1}{short = SG1,long=Smart Grid 1.0}
\DeclareAcronym{ITU}{short = ITU,long=International Telecommunication Union}
\DeclareAcronym{AI}{short = AI,long=Artificial Intelligence }
\DeclareAcronym{SMS}{short =SMS,long=Short Message Service}
\DeclareAcronym{3GPP}{short =3GPP,long=3rd Generation Partnership Project}
\DeclareAcronym{V2X}{short =V2X,long=Vehicle-to-everything}
\DeclareAcronym{V2V}{short =V2V,long=Vehicle-to-Vehicle}
\DeclareAcronym{V2I}{short =V2I,long=Vehicle-to-Infrastructure}
\DeclareAcronym{DDoS}{short =DDoS,long=Distributed Denial-of-Service}
\DeclareAcronym{DoS}{short =DoS,long=Denial-of-Service}
\DeclareAcronym{DNS}{short =DNS,long=Domain Name System}
\DeclareAcronym{CCTV}{short =CCTV,long=Closed Circuit Television}
\DeclareAcronym{IoT}{short =IoT,long=Internet of Things}
\DeclareAcronym{MEC}{short =MEC,long=Multi-access Edge Computing}
\DeclareAcronym{5G}{short =5G,long= fifth-generation}
\DeclareAcronym{SDN}{short =SDN,long=Software defined Networking}
\DeclareAcronym{NFV}{short =NFV,long=Network Function Virtualization}
\DeclareAcronym{SDA}{short =SDA,long=Service-defined Architecture}
\DeclareAcronym{RAN}{short =RAN,long=Radio Access Network}
\DeclareAcronym{RSU}{short =RSU,long=Road-side Unit}
\DeclareAcronym{eMBB}{short =eMBB,long=enhanced Mobile Broad Band}
\DeclareAcronym{uRLLC}{short =uRLLC,long=Ultra-Reliable Low-Latency Communication}
\DeclareAcronym{mMTC}{short =mMTC,long=massive Machine Type Communications}
\DeclareAcronym{LPW}{short =LPW,long=Low-power Wireless}
\DeclareAcronym{LPAN}{short =LPAN,long=Low-power Personal Network}
\DeclareAcronym{LPWAN}{short =LPWAN,long=Low-power Wide Area Network}
\DeclareAcronym{EDA}{short =EDA,long=Energy depletion attacks}
\DeclareAcronym{PCF}{short =PCF,long=Policy Control Function}
\DeclareAcronym{NRF}{short =NRF,long=Network Repository Function}
\DeclareAcronym{NSSF}{short =NSSF,long=Network Slice Selection Function}
\DeclareAcronym{UDM}{short =UDM,long=Unified Data Management}
\DeclareAcronym{UPF}{short =UPF,long=User Plane Function}
\DeclareAcronym{AMF}{short =AMF,long=Access and Mobility Function}
\DeclareAcronym{SMF}{short =SMF,long=Session Management Function}
\DeclareAcronym{LADN}{short =LADN,long=Local Area Data Network }
\DeclareAcronym{MIMO}{short =MIMO,long=Multiple-input and Multiple-output}
\DeclareAcronym{SFC}{short =SFC,long=Service function chaining}
\DeclareAcronym{SECaaS}{short =SECaaS,long=Security as a service}
\DeclareAcronym{PISA}{short =PISA,long=Protocol-Independent Switch Architecture}
\DeclareAcronym{ONOS}{short =ONOS,long=Open Network Operating System}
\DeclareAcronym{IDS}{short =IDS,long=Intrusion Detection System}
\DeclareAcronym{IPS}{short =IPS,long=Intrusion Prevention System}
\DeclareAcronym{HSM}{short =HSM,long=hardware security module}
\DeclareAcronym{PKI}{short =PKI,long=Public-key Infrastructure}
\DeclareAcronym{UE}{short =UE,long=User Equippment}
\DeclareAcronym{NTP}{short =NTP,long=Network Transfer Protocol}
\DeclareAcronym{UDP}{short =UDP,long=User Diagram Protocol}
\DeclareAcronym{ICMP}{short =ICMP,long=Internet Control Message Protocol}
\DeclareAcronym{DMZ}{short =DMZ,long=Demilitarized Zone}
\DeclareAcronym{DPI}{short =DPI,long=Industrial Deep Packet Inspection}
\DeclareAcronym{LTE}{short =LTE,long=Long-term Evolution}
\DeclareAcronym{TEID}{short =TEID,long=Tunnel Endpoint Identifier}
\DeclareAcronym{GTP}{short =GTP,long=General Packet Radio Service tunneling protocol}
\DeclareAcronym{GTP-U}{short =GTP-U,long=GPRS Tunnelling Protocol – User}
\DeclareAcronym{GTP-C}{short =GTP-C,long=GPRS Tunnelling Protocol – Control}
\DeclareAcronym{SGW}{short =SGW,long=Serving Gateway}
\DeclareAcronym{PGW}{short =PGW,long=Packet Gateway}
\DeclareAcronym{PDP}{short =PDP,long=Packet Data Protocol}
\DeclareAcronym{MSISDN}{short =MSISDN,long=Mobile Station International Subscriber Directory Number}
\DeclareAcronym{IMSI}{short =IMSI,long=International Mobile Subscriber Identity}
\DeclareAcronym{MME}{short =MME,long=Mobility Management Entity}
\DeclareAcronym{WSMP}{short =WSMP,long=Wave Short Message Protocol}
\DeclareAcronym{ADAS}{short =ADAS,long=Advanced Driver-Assistance Systems}
\DeclareAcronym{CACC}{short =CACC,long=Cooperative Adaptive Cruise Control}
\DeclareAcronym{VANET}{short =VANET,long=Vehicular Ad-hoc Networks}
\DeclareAcronym{RSSI}{short =RSSI,long=Radio Signal Strength Indicator}
\DeclareAcronym{AoA}{short =AoA,long=Angle of radio Arrival}
\DeclareAcronym{TDOA}{short =TDOA,long=Time Difference of Arrival}
\DeclareAcronym{LOS}{short =LOS,long=Light-of-Sight}
\DeclareAcronym{non-LOS}{short =non-LOS,long=non-Light-of-Sight}
\DeclareAcronym{NLOS}{short =NLOS,long=non-Light-of-Sight}
\DeclareAcronym{C-ITS}{short =C-ITS,long=Cooperative Intelligent Transportation Systems}
\DeclareAcronym{LIDAR}{short =LIDAR,long=LIght Detection and Ranging}
\DeclareAcronym{VLC}{short =VLC,long=visible light communication}
\DeclareAcronym{BSM}{short =BSM,long=Basic Safety Message}
\DeclareAcronym{CAM}{short =CAM,long=Cooperative Awareness Message}
\DeclareAcronym{OFDMA}{short =OFDMA,long=Orthogonal Frequency Division Multiple Access}
\DeclareAcronym{ULA}{short =ULA,long=Uniform Linear Array}
\DeclareAcronym{PEB}{short =PEB,long=Position Error Bound}
\DeclareAcronym{UKF}{short =UKF,long=Unscented Kalman Filter}
\DeclareAcronym{HD}{short =HD,long=High-resolution Dynamic}
\DeclareAcronym{CV}{short =CV,long=Constant Velocity}
\DeclareAcronym{CA}{short =CA,long=Constant Acceleration}
\DeclareAcronym{CTRA}{short =CTRA,long=Constant Turn Rate and Acceleration}
\DeclareAcronym{CSAV}{short =CSAV,long=Constant Steering Angle and Velocity}
\DeclareAcronym{CCA}{short =CCA,long=Constant Curvature and Acceleration}
\DeclareAcronym{IMM}{short =IMM,long=Interacting Multiple Model}
\DeclareAcronym{NIES}{short =NIES,long=Normalised Innovation Error Squared}
\DeclareAcronym{LIS}{short =LIS,long=Large Intelligent Surface}
\DeclareAcronym{FOV}{short =FOV,long=Field of View}
\DeclareAcronym{6G}{short =6G,long=sixth-generation}
\DeclareAcronym{4G}{short =4G,long=fourth-generation}
\DeclareAcronym{AIML}{short =AI/ML,long=Artificial Intelligence and Machine Learning}
\DeclareAcronym{GSM}{short =GSM,long=Global System for Mobile}
\DeclareAcronym{UMTS}{short =UMTS,long=Universal Mobile Telecommunications System}
\DeclareAcronym{SIM}{short =SIM,long=Subscriber Identity Module}
\DeclareAcronym{USIM}{short =USIM,long=Universal Subscriber Identity Module}
\DeclareAcronym{TMSI}{short =TMSI,long=Temporary Mobile Subscriber Identity}
\DeclareAcronym{HSPA}{short =HSPA,long=High-Speed Packet Access}
\DeclareAcronym{AKA}{short =AKA,long=Authentication and Key Agreement}
\DeclareAcronym{EPS}{short=EPS,long=Evolved Packet System}
\DeclareAcronym{E-UTRAN}{short=E-UTRAN,long=Evolved-Universal Terrestrial Radio Access Network}
\DeclareAcronym{ESN}{short =ESN,long=Electronic Serial Numbers}
\DeclareAcronym{MDN}{short =MDN,long=Mobile Directory Numbers}
\DeclareAcronym{MSC}{short =MSC,long=Mobile Switching Center}
\DeclareAcronym{AuC}{short =AuC,long=Authentication Center}
\DeclareAcronym{HLR}{short =HLR,long=Home Location Register}
\DeclareAcronym{VLR}{short =VLR,long=Visitor Location Register}
\DeclareAcronym{M2M}{short =M2M,long=Machine-to-Machine}
\DeclareAcronym{OFDM}{short =OFDM,long=Orthogonal Frequency Division Multiplexing}
\DeclareAcronym{eMBMS}{short =eMBMS,long=enhanced Multimedia Broadcast Multicast Service}
\DeclareAcronym{UMTS-FDD}{short =UMTS-FDD,long= UMTS–frequency-division duplexing}
\DeclareAcronym{AN}{short =AN,long= Access Network}
\DeclareAcronym{SN}{short =SN,long= Serving Network}
\DeclareAcronym{HN}{short =HN,long= Home Network}
\DeclareAcronym{EPC}{short =EPC,long= Evolved Packet Core}
\DeclareAcronym{ePDG}{short =ePDG,long= Evolved Packet Data Gateway}
\DeclareAcronym{HSS}{short =HSS,long=Home Subscriber Server}
\DeclareAcronym{SEAF}{short =SEAF,long=Security Anchor Function}
\DeclareAcronym{AUSF}{short =AUSF,long=Authentication Server Function}
\DeclareAcronym{ARPF}{short =ARPF,long=Authentication Credential Repository and Processing Function}
\DeclareAcronym{SUCI}{short =SUCI,long=Subscription Concealed Identifier}
\DeclareAcronym{SUPI}{short =SUPI,long=Subscription Permanent Identifier}
\DeclareAcronym{LI}{short =LI,long=Lawful Interception}
\DeclareAcronym{SIDF}{short =SIDF,long=Subscription Identifier De-concealing Function}
\DeclareAcronym{GDPR}{short =GDPR,long=General Data Protection Regulation}
\DeclareAcronym{BTS}{short =BTS,long=Base Stations}
\DeclareAcronym{CU}{short =CU,long=Centralized Unit}
\DeclareAcronym{DU}{short =DU,long=Distributed Unit}
\DeclareAcronym{BSC}{short =BSC,long=Base Station Controller}
\DeclareAcronym{SGSN}{short =SGSN,long=Serving GPRS Support Node}
\DeclareAcronym{GMSC}{short =GMSC,long=Gateway Mobile Switching Centre}
\DeclareAcronym{GGSN}{short =GGSN,long=Gateway GPRS Support Node}
\DeclareAcronym{RNC}{short =RNC,long=Radio Network Controller}
\DeclareAcronym{PCRF}{short =PCRF,long=Policy and Charging Rules Function}

\DeclareAcronym{NaaS}{short =NaaS,long=Network as a service}
\DeclareAcronym{CP}{short =CP,long=Control Plane}
\DeclareAcronym{vCU}{short =vCU,long=virtualized Central Unit}
\DeclareAcronym{LLF}{short =LLF,long=Lower-Level Function}
\DeclareAcronym{QKD}{short =QKD,long=Quantum Key Distribution}
\DeclareAcronym{GUTI}{short =GUTI,long=Global Unique Temporary Identifier}
\DeclareAcronym{NOMA}{short =NOMA,long=Non-Orthogonal Multiple Access}
\DeclareAcronym{IRS}{short =IRS,long=Intelligent Reflecting Surfaces}
\DeclareAcronym{MTD}{short =MTD,long=Moving Target Defense}
\DeclareAcronym{LDPC}{short =LDPC,long=Low-Density Parity-check Code}
\DeclareAcronym{PET}{short =PET,long=Privacy Enhancing Technologies}
\DeclareAcronym{SDWAN}{short =SDWAN,long=Software-Defined Wide Area Network}
\DeclareAcronym{WAN}{short =WAN,long=Wide Area Network}
\DeclareAcronym{SD-LAN}{short =SD-LAN,long=Software-Defined Local Area Network}
\DeclareAcronym{LAN}{short =LAN,long=Local Area Network}
\DeclareAcronym{HAN}{short =HAN,long=Home Area Network}
\DeclareAcronym{NAN}{short =NAN,long=Neighborhood Area Network} 
\DeclareAcronym{FAN}{short =FAN,long=Field Area Network}
\DeclareAcronym{BAN}{short =BAN,long=Building Area Network}
\DeclareAcronym{AMI}{short =AMI,long=Advanced Metering Infrastructure}
\DeclareAcronym{DERs}{short =DERs,long=Distributed Energy Resources}
\DeclareAcronym{CSI}{short =CSI,long=Channel State Information}
\DeclareAcronym{SA}{short =SA,long=Standalone}
\DeclareAcronym{NSA}{short =NSA,long=Non-Standalone}
\DeclareAcronym{VNF}{short =VNF,long=Virtualized Network Function}
\DeclareAcronym{VM}{short =VM,long=Virtual Machine}
\DeclareAcronym{vRAN}{short =vRAN,long=Virtualized Radio Access Network}
\DeclareAcronym{C-RAN}{short =C-RAN,long=Cloud RAN}
\DeclareAcronym{TLS}{short =TLS,long=Transport Layer Security}
\DeclareAcronym{EAP}{short =EAP,long=Extensible Authentication Protocol}
\DeclareAcronym{CEDS}{short =CEDS,long=Cybersecurity for Energy Delivery Systems}
\DeclareAcronym{NERC}{short =NERC,long=North American Electric Reliability Corporation}
\DeclareAcronym{ENISA}{short =ENISA,long=European Union Agency for Network and Information Security} 
\DeclareAcronym{IEC}{short =IEC,long=International Electrotechnical Commission}
\DeclareAcronym{SCADA}{short =SCADA,long=Supervisory Control And Data Acquisition}
\DeclareAcronym{ICS}{short =ICS,long=Industrial Control Systems}
\DeclareAcronym{AGI}{short =AGI,long=Artificial General Intelligence}
\DeclareAcronym{ESG}{short =ESG,long=Environmental- Social and Governance}
\DeclareAcronym{EV}{short =EV,long=Electric Vehicle}

\theoremstyle{definition}

\DeclareUnicodeCharacter{2061}{}

\begin{document}


 \title{Towards Secured Smart Grid 2.0: Exploring Security Threats, Protection Models, and Challenges }


    \author{\IEEEauthorblockN{Lan-Huong Nguyen$\dagger$, Van-Linh Nguyen$\dagger$, \textit{Senior Member, IEEE}, Ren-Hung Hwang, \textit{Senior Member, IEEE}, Jian-Jhih Kuo, \textit{Member, IEEE}, Yu-Wen Chen, \textit{Member, IEEE}, Chien-Chung Huang, Ping-I Pan
    \thanks{L. H. Nguyen and R. H. Hwang are with College of Artificial Intelligence, National Yang Ming Chiao Tung University, Tainan, Taiwan.} \thanks{V. L. Nguyen and J. J. Kuo are with the Department of Computer Science and Information Engineering, National Chung Cheng University (CCU), and also with the Advanced Institute of Manufacturing with High-Tech Innovations (AIM-HI), CCU, Taiwan.} \thanks{Y. W. Chen is with New York City College of Technology, New York, United States.} \thanks{C. C. Huang and P. I Pan are with Green Energy and Environment Research Laboratories, Industrial Technology Research Institute, Tainan, Taiwan} \thanks{$\dagger$ L. H. Nguyen and V.L. Nguyen contributed equally to this work}
    \thanks{Corresponding authors: R. H. Hwang} }}

	\markboth{IEEE Communications Surveys \& Tutorials}%
    {Nguyen \MakeLowercase{\textit{et al.}}: Towards Secured Smart Grids 2.0: Exploring Security Threats, Protection Models, and Challenges}
	
	\maketitle

	\begin{abstract}

     Many nations are promoting the green transition in the energy sector to attain neutral carbon emissions by 2050. \ac{SG2} is expected to explore data-driven analytics and enhance communication technologies to improve the efficiency and sustainability of distributed renewable energy systems. These features are beyond smart metering and electric surplus distribution in conventional smart grids. Given the high dependence on communication networks to connect distributed microgrids in \ac{SG2}, potential cascading failures of connectivity can cause disruption to data synchronization to the remote control systems. This paper reviews security threats and defense tactics for three stakeholders: power grid operators, communication network providers, and consumers. Through the survey, we found that \ac{SG2}'s stakeholders are particularly vulnerable to substation attacks/vandalism, malware/ransomware threats, blockchain vulnerabilities and supply chain breakdowns. Furthermore, incorporating artificial intelligence (AI) into autonomous energy management in distributed energy resources of \ac{SG2} creates new challenges. Accordingly, adversarial samples and false data injection on electricity reading and measurement sensors at power plants can fool AI-powered control functions and cause messy error-checking operations in energy storage, wrong energy estimation in electric vehicle charging, and even fraudulent transactions in peer-to-peer energy trading models. Scalable blockchain-based models, physical unclonable function, interoperable security protocols, and trustworthy AI models designed for managing distributed microgrids in \ac{SG2} are typical promising protection models for future research.

	\end{abstract}
	

\section{Introduction}
\label{sec:introduction}

\begin{figure}[t]
	\begin{center}
		\includegraphics[width=1\linewidth]{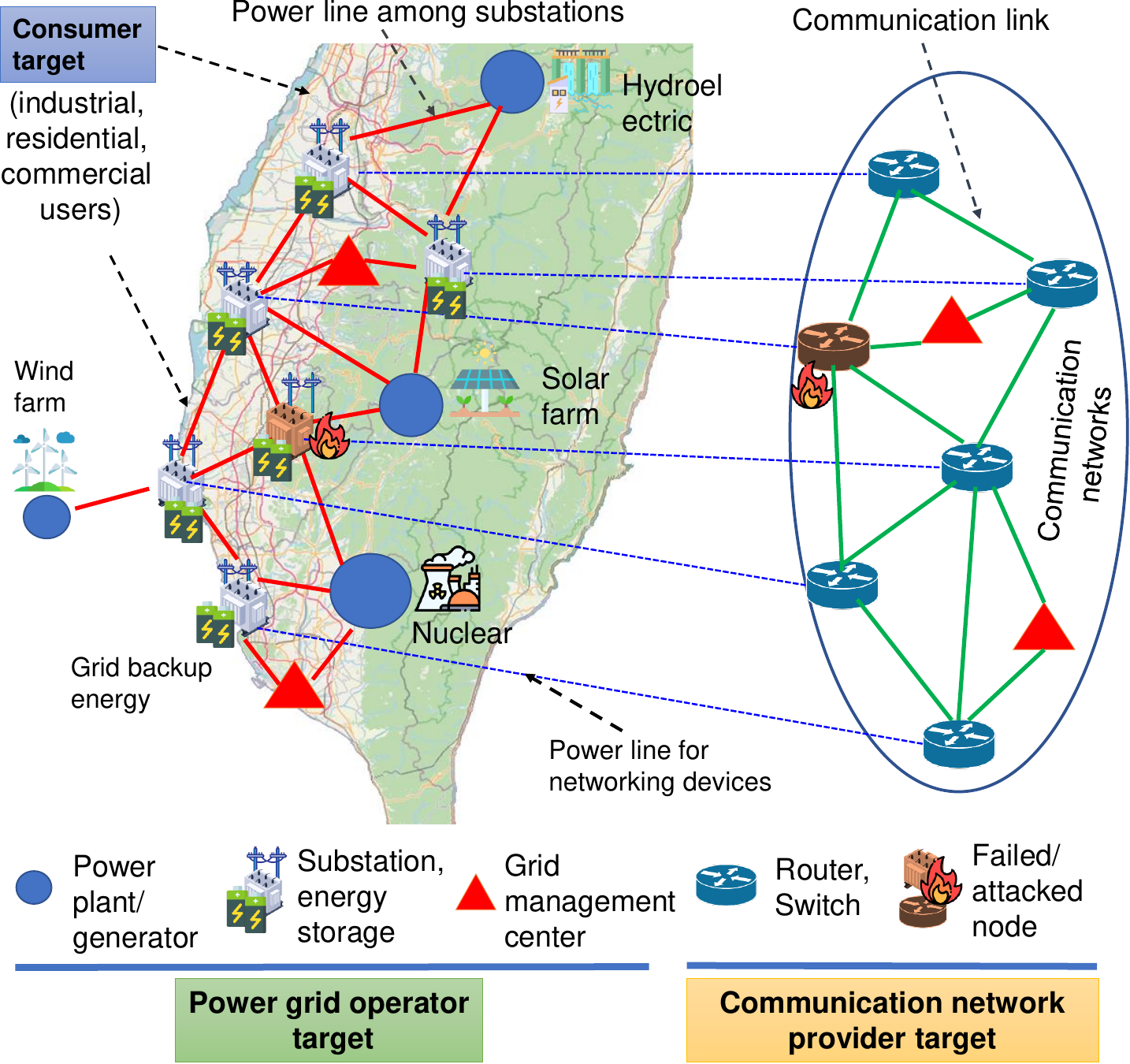}
	\end{center}
	\centering
	\caption{The illustration shows the crucial role of communication technologies in synchronizing measurement data from substations, enabling remote control capabilities for efficient power distribution. However, the dependency of power grids on communication technology creates fresh threats of security attacks to energy security, e.g., ransomware to disable control systems and denial of services against transmission lines to stop data exchange.}
	\label{fig:power-communication-relationship}
\end{figure}

Integrating distributed energy systems is a major topic in the green and renewable energy era with sustainable goals of usage efficiency, autonomous intelligence, and resilience capability against sudden failures. These new energy integration capabilities expects to be the core of Smart Grid 2.0 (SG2) \cite{GRID2.0,PorambageSG2}. \ac{SG2} aims to enhance energy distribution and usage efficiency with the help of communication technologies \cite{ORECCHINI2011,Sun16,Ma17}. These interdependent power grid and communication networks can help connect distributed electricity, gas, and cooling systems, offering unprecedented opportunities for remote control capability and flexibility \cite{ALI2021}. However, with reliance on digital communication technologies, as shown in Fig.~\ref{fig:power-communication-relationship}, SG2 faces significant threats, targeting connectivity among power grid providers, two-way communication network systems, and consumer entities (industrial, residential, and commercial users). For example, ransomware attacks have recently been recorded to cause prominent blackouts in many countries \cite{CDW}.

Recent drone attacks and supply chain risks also threaten critical facilities (economic loss) and even endanger public safety (e.g., power outages in the cooling systems of nuclear plants). Also, providing sophisticated communication networks to millions of charging stations and diverse renewable energy sources in SG2, while not overburdening the distribution network or destabilizing the grid, is also a challenge. Understanding the threats and challenges is the critical step toward developing robust defense approaches for guaranteeing energy security, and further national safety. This article aims to explore the various security threats in \ac{SG2}, particularly its communication infrastructure and \ac{SG2} enabling technologies, such as peer-to-peer energy trading and AI-powered grid network functions. In addition, the study discusses emerging strategies for safeguarding \ac{SG2} from cascading failures and for developing effective distribution grid restoration plans in disaster scenarios and severe security attacks in the future.

\subsection{State-of-the-art literature review}

\begin{figure}[t]
	\begin{center}
		\includegraphics[width=1\linewidth]{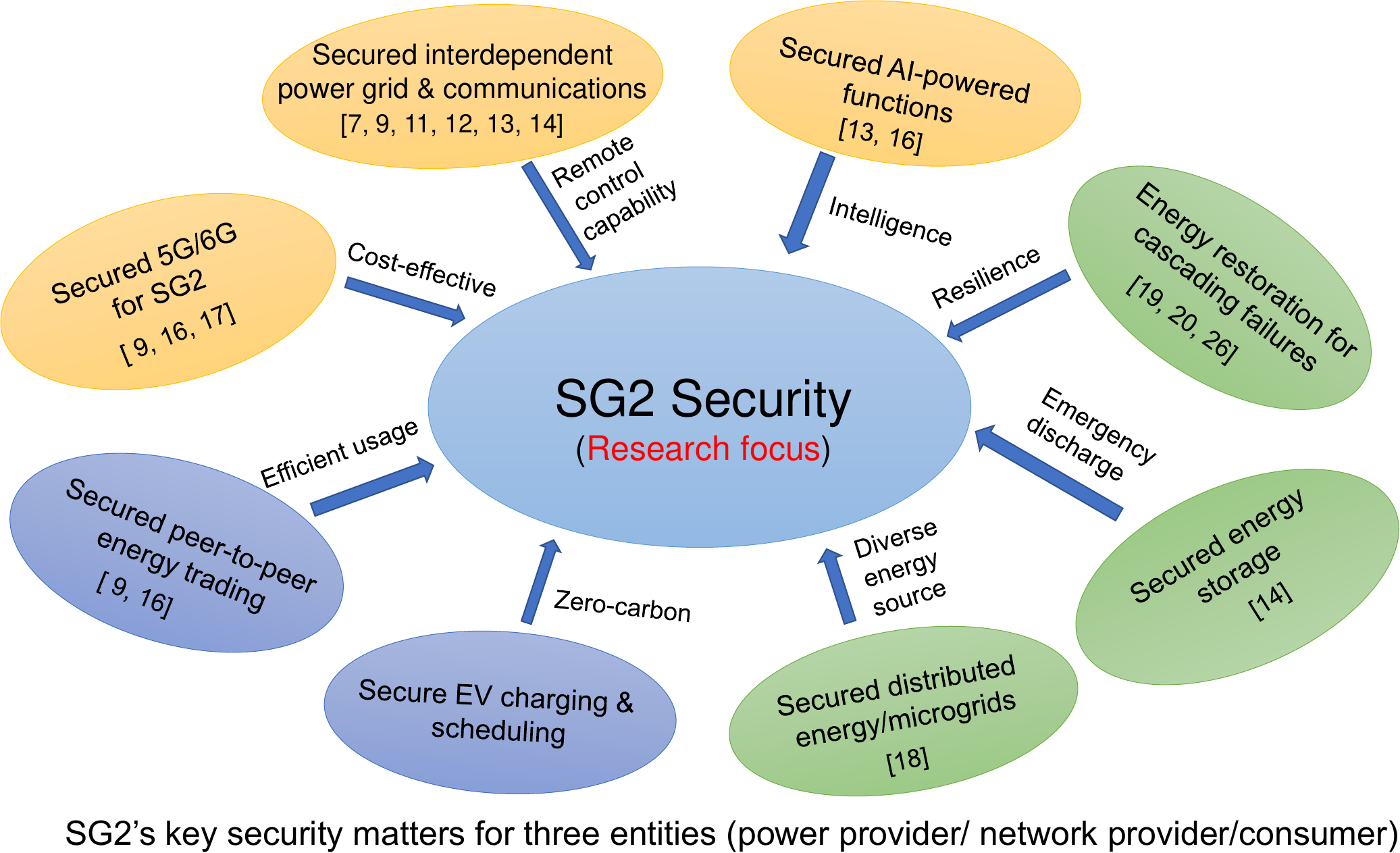}
	\end{center}
	\centering
	\caption{This work addresses energy security principles for SG2 from a view of interdependent power grid communication networks, notably with the introduction of new technologies for three entities (power provider, communication network provider, consumer), such as energy storage, 5G/6G, AI-powered functions, and peer-to-peer energy trading models. }
	\label{fig:research-position}
\end{figure}

Exploring security threats for smart grids has been a hot topic for years but few studies address \ac{SG2}'s security matters in a comprehensive manner. Fig.~\ref{fig:research-position} presents \ac{SG2}'s essential components and security concerns, together with state-of-the-art relevant studies. Accordingly, most articles focus on security threats in the conventional smart grid that features electricity sources and distribution. For example, the authors in \cite{Yan12} provide a comprehensive survey of typical security attacks and vulnerabilities of authentication and security protocols in the conventional energy model. AI and blockchain for conventional smart grids and related security is briefed in \cite{Bose17, zibaeirad2024, AYUBKHAN2023103282}. However, the paper covers few aspects of security protection or \ac{AI} role in specific functions in each layer or from related stakeholders (power generator, communication provider, and consumers). Similarly, the surveys in \cite{Komninos14,Kumar19COMST,GUNDUZ20} cover a narrow scope of security in specific smart grid networks, e.g., the connection between home and grid supply, metering data collection and transmission \cite{SAKHNINI21}.

Recently, the survey papers focus on the security threats in communications among power generation and distribution components of the newer smart grid model with battery storage systems \cite{Ogino23,Amin21,Tala22,Trevizan22}. As technology and infrastructure continue to advance, renewable energy (e.g., wind, solar, geothermal, hydropower) plays an increasingly vital role in the global energy transition toward a cleaner and more sustainable future. Energy storage systems are required to maintain the stability of such distributed sources. Several surveys on the safety of smart inverters \cite{Li2023}, battery storage/swap \cite{Hasan23}, or control systems \cite{Vahidi23} against remote attacks or physical tampering is also presented. On the other hand, the authors in \cite{Khan23,Ghasemi23,Liu22} provide a holistic view of control and communication strategies in multi-energy generation grids or robust models against cascading failure in interdependent power-communication networks. However, the studies did not address the security threats or specific attacks for each entity (power provider, communication network provider, consumer). Unlike prior studies, this work aims to investigate weaknesses in the interdependence of microgrids that heavily rely on distributed energy sources and communication technologies. Additionally, there will be a focus on identifying new risks associated with AI-powered energy control and novel energy trading/storage models. This research will be particularly important as many new small energy sources (e.g., from solar roofs) are integrated into management networks. Overall, the first goal of our work is to provide a comprehensive view of cybersecurity in these new elements, referring to \ac{SG2}'s energy security principles, which have received little attention in the existing literature. 

Besides, many governments considered grid security as a national security matter and proposed measures to improve information security protection \cite{SECenergyStorage,IEC,NIST,ENISA} and resilience strategies in disaster and crisis scenarios \cite{Khan23,Ostergaard21,Kliem20}. Grid security refers here to the consistent and reliable availability of all fuels and electricity sources in a timely, sustainable, and cost-effective manner. For example, several standards for information security in smart grids, such as the framework developed by the \ac{NIST}, have been developed. The US Department of Energy is carrying out the Cybersecurity for Energy Delivery Systems (CEDS) program \cite{ENERGY}, aimed to enhance the security and resilience of the country's energy infrastructure. The \ac{NERC} has established network security standards for the power industry in North America. The \ac{ENISA} has issued guidelines for safeguarding EU's smart grids \cite{ENISA}. The \ac{IEC} has established standards for network security in power systems (IEC 62443, 62351 standards) \cite{IEC}. Therefore, this survey’s second goal is to determine which features have not yet been defined in the security standards and what standards the nationals apply for their energy management systems based on facility availability, deployment cost, and environment compatibility. Further, understanding energy restoration plans for potentially cascading failures of communication technologies is critical to consult a proper model for deploying \ac{SG2}.

\subsection{Review methodology}

Given the difficulties of installing from scratch owing to high costs, \ac{SG2} will likely inherit many control components, existing facilities, and communication infrastructure from the current smart grid. Inspired by this fact, we present possible security concerns in \ac{SG2}, as viewed through the mirrors of two aspects' lessons learned. The first phase involves examining energy security principles, identifying the primary risks to the components of a smart grid system, and assessing the security vulnerabilities in legacy technologies with examples of well-known energy crises and blackout events. Additionally, this involves energy restoration strategies in the event of probable cascade failures resulting from security attacks on communication lines. The rest is to figure out security flaws in emerging technologies and new decentralized energy models that are expected to be the main vehicles towards \ac{SG2}, such as AI-powered energy control functions, battery storage technologies, and the integration of advanced communication technologies for charging stations and distributed renewable energy sources. Security threats are often the motivating factor behind the need to change countermeasure approaches. These attacks typically reveal system faults or protocol issues that were not expected during the design process. Analyzing and learning from attacks like this gives significant insights into the essential security changes for \ac{SG2}, particularly in addressing known weaknesses exploited in prior generations. Finally, this work also addresses security matters from a top-down approach where security threats against the interdependent relationship of stakeholders (power grid operator, communication network provider, consumer) will be assessed and suggested with corresponding defense strategies. The unsolved problems become possible targets for \ac{SG2} improvements, which serve as the foundation for proposing future solutions.

 \begin{figure*}[t]
    		\begin{center}
    			\includegraphics[width=1\linewidth]{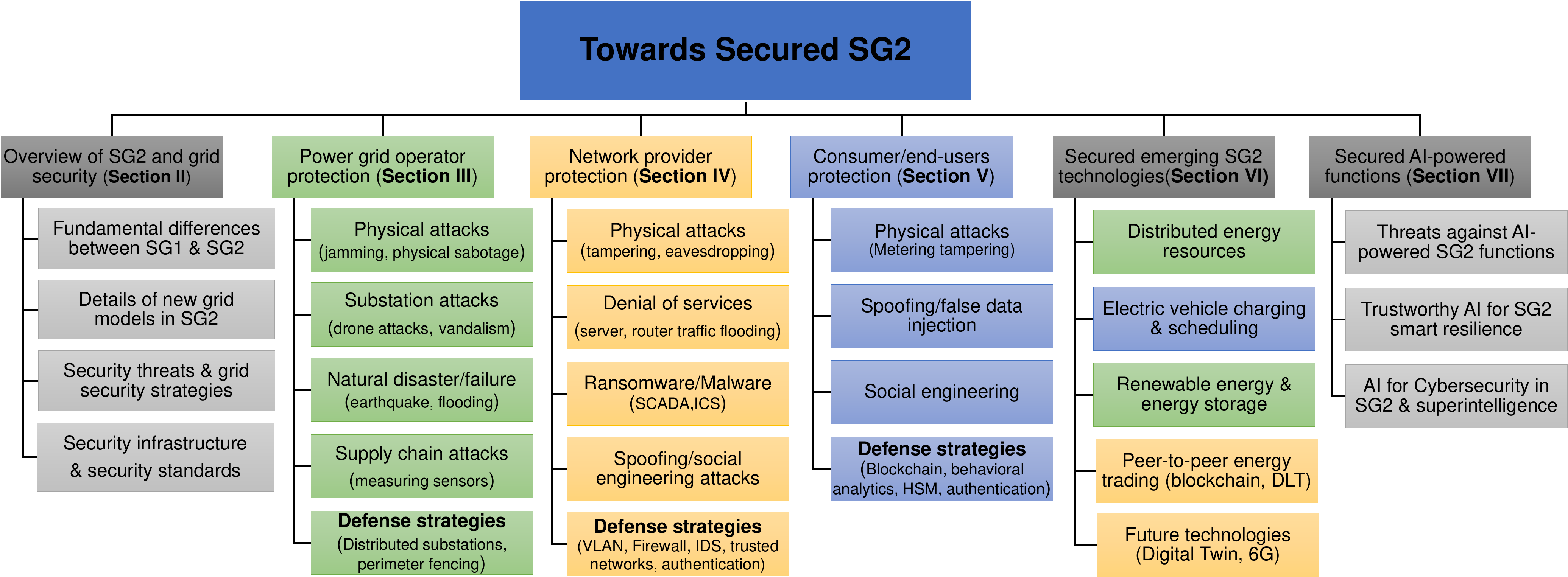}
    		\end{center}
    	\centering
    	 	\caption{The following is a summary of the major findings from our survey on security and protection strategies for Smart Grid 2.0. The decorative colors for technologies match those for the three entities (power provider, communication network provider, and customer) illustrated in the previous figures. }
    	 	\label{fig:survey-structure}
      \end{figure*}

\subsection{Contributions}      
     
 Given the slow transition from legacy to new technologies, it is difficult to predict when \ac{SG2} will be in full operation. However, by drawing a line of relative differences between the current smart grid platform and the expected \ac{SG2} architecture, this research can help the developers and researchers determine the security weaknesses and find the right starting point of \ac{SG2}'s technologies to improve. The primary contributions in this work are summarized as follows. 
 
 \begin{enumerate}
     \item The first attempt to thoroughly investigate the principles of SG2 security for national safety, taking into account the whole perspective of security measures for communication links among power grid operators, communication network providers, and consumers. The study examines the relationship between power grids and communication networks in terms of cascading failures. It identifies potential solutions and necessary improvements for \ac{SG2}, specifically in energy restoration planning and communication isolation.
     
     \item The first attempt to offer a comprehensive perspective on the security risks of \ac{SG2} enabling technologies that need to be adapted to meet the evolving requirements of \ac{SG2}, e.g., blockchain-based energy management/trading, AI-aided grid operations, the networks of electrified transportation systems (EV charging stations) and distributed renewable energy sources. Given that \ac{SG2} follows economic trajectory of the technology evolution, a systematic review of the transition process and potential changes in supply chain management and new communication methods can guide power grid operators and network providers in effectively upgrading their security infrastructure and countermeasure techniques in the future.
     
     \item This study summarized lessons learned from the limitations of current protection implementations in SG1 and the vulnerabilities of SG2 emerging technologies that can aid researchers and developers in determining the problem formulation for further studies. To the best of our knowledge, this survey represents the initial endeavor to comprehensively assess security threat aspects for \ac{SG2}, spanning from vulnerabilities in distributed renewable energy sources to EV charging network architecture, and then AI-powered grid management.
 \end{enumerate}

  \subsection{Structure of the paper}

The rest of this paper is organized as follows.  Section~\ref{sec:age-of-green-energy} briefs the fundamental information about \ac{SG2} architecture, energy security principles, and overall strategies to protect the power grid-communication networks' infrastructure. The security attacks and defense approaches for power providers, communication network providers, and consumer stakeholders are then detailed in Section~\ref{sec:security-physical-layer}, \ref{sec:security-network-layer}, \ref{sec:security-application-layer}, respectively. Section~\ref{sec:security-energy-layer} outlines security risks and some countermeasure techniques in emerging technologies and their role in securing \ac{SG2} is detailed in Section~\ref{sec:security-AI-layer}. Section~\ref{sec:lessons-learned} discusses lessons learned and future research. Section~\ref{sec:conclusion} concludes this paper. Fig.~\ref{fig:survey-structure} summarizes the main points of our survey. The acronyms used in this work are listed as follows.

\renewcommand{\IEEEiedlistdecl}{\IEEEsetlabelwidth{SONET}}
\printacronyms[sort=true]
\renewcommand{\IEEEiedlistdecl}{\relax}

 \section{SG2: Upgrades from SG1 and security concerns from grid security strategies}
\label{sec:age-of-green-energy}

 This section introduces fundamental differences between smart grids (i.e., \ac{SG1}) and \ac{SG2}, key security concerns, and defense strategies. The discussions of \ac{SG2} insecurity and national safety are also presented. 

   \subsection{Overview of fundamental differences between SG1 and SG2}

\begin{figure*}[t]
   	\begin{center}
   		\includegraphics[width=1\linewidth]{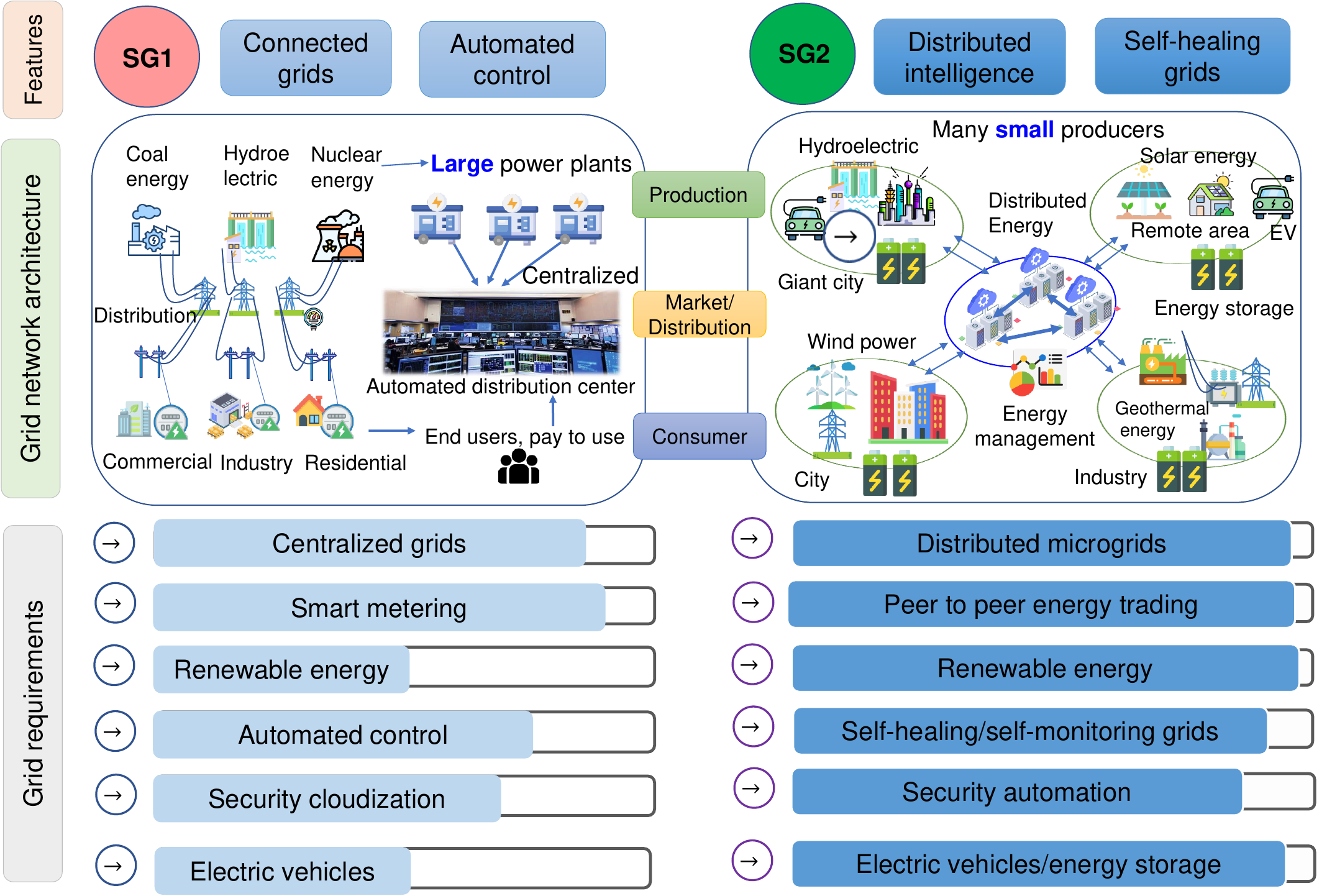}
   	\end{center}
   	\centering
   	\caption{The illustration of comparison between smart grid 1.0 and smart grid 2.0 along with key features, architecture and main requirement differences.}
   	\label{fig:intelligent-energy-network}
   \end{figure*}

   \ac{SG2} is expected to be a comprehensive framework designed to modernize and enhance the usage efficiency, reliability, and sustainability of energy distribution and management in SG1. \ac{SG2} leverages communication technologies and data analytics to transform \ac{SG1} into interconnected and intelligent systems that are capable of optimizing energy production, distribution, and consumer demands at peak periods. Fig.~\ref{fig:intelligent-energy-network} illustrates the fundamental differences between SG1 and SG2. Instead of managing energy distribution from large power plants only as in \ac{SG1}, \ac{SG2}'s architecture will involve both large and small producers in a distributed energy model, where microgrids can automatically be operated to serve city oases and partially managed by AI. In this way, key features in \ac{SG2} are distributed intelligence and self-healing grids, highlighting autonomous energy distribution management. Overall, \ac{SG2} enhances automated control and centralized management model in SG1 with the new capabilities, such as distributed microgrid model, peer-to-peer energy trading, and self-healing/self-monitoring grids. 
  
   \ac{SG2} integrates the goal of reducing carbon emissions toward clean/green energy. For example, to achieve the goal of carbon neutrality by 2050, many countries expect the contribution of solar and renewable energy sources to be up to 40\%  by 2035 or 45\% by 2050 \cite{SOLAR,HouseUK22,China2021}, and a Net-Zero emission target by 2070 \cite{India2023}. As a result, in the next decades, renewable energy sources (hydropower, solar power, wind power, bioenergy, and geothermal energy) will dominate the market. However, given the challenges of far-distance distribution from power plants (e.g., solar farms in the deserts or wind farms at the beach), maintaining the continuity of renewable energy sources, particularly at peak hours, needs the help of advanced technologies, such as energy storage and super grids \cite{EuropeanSuperGrid}.

    \subsection{Details of new energy generation, distribution, and consumption model in SG2} 

\begin{figure*}[t]
    		\begin{center}
    			\includegraphics[width=0.90\linewidth]{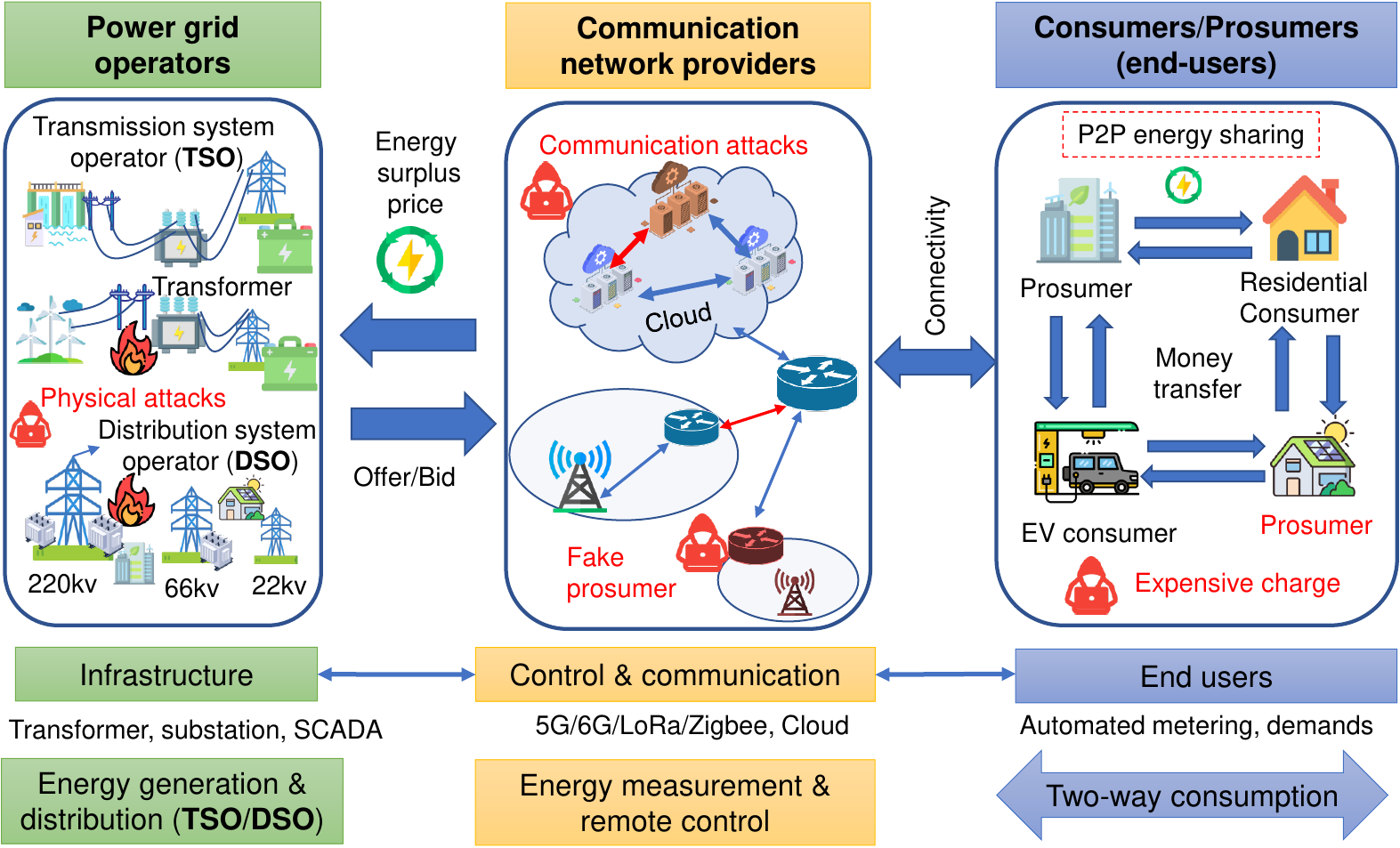}
    		\end{center}
    	\centering
    	 	\caption{The illustration of three stakeholders in SG2: power grid providers, communication network providers, and consumers. Power providers have a complicated infrastructure of power plants, transformers, and substations. Communication network providers include several networking technologies (e.g., Zigbee, LoRa), which provide connectivity to connect sensors from energy generation, transmission, and distribution, to consumers and remote centers.}
    	 	\label{fig:information-flow}
      \end{figure*}

      
     Generally, \ac{SG2} includes three primary components: power grid providers, communication network providers, and consumers, as illustrated in Fig.~\ref{fig:information-flow}. Power grid providers/operators have a complicated infrastructure of power plants, power lines, transformers, and substations. Power grid providers/operators play a role in energy generation and distribution. Network providers provide communication technologies for energy measurement and remote control, i.e., connectivity to synchronize data from sensors at substations or consumers' smart metering devices to management applications, e.g., \ac{SCADA}. The communication infrastructure includes the \ac{WAN}, \ac{NAN}, and \ac{HAN} with networking technologies like LoraWAN/Zigbee \cite{Hasan23}. 
     Generally, the \ac{WAN} is the primary network infrastructure for establishing a connecting backbone. \ac{NAN} and \ac{FAN} are used for connecting customer smart meters, substations, and \ac{WAN}. The \ac{HAN} facilitates connectivity for devices in the house/building/factory \cite{Li2023}. The diversity of connectivity technologies in \ac{HAN} and \ac{NAN} is to maintain flexibility for supply chains in different countries. In this work, we use the general network architecture for research reference.
    

    Regarding specific upgrades from \ac{SG1}, there are two major upgrades in \ac{SG2}: (1) upgrading the way of generating, distributing, trading, and consuming energy in power grid providers/operators (2) upgrading the networking infrastructure and consumer devices. For the first major upgrade, the energy landscape evolves from a one-way distribution model to a dynamic, bidirectional flow of energy and information \cite{Vahidi23, Hasan23}. Decentralized energy resources, such as solar panels and energy storage systems, assume a more pivotal role, enabling consumers to be energy users and contributors to the grid. Consumers evolve from passive users to active participants who can not only monitor their energy consumption in real-time but also contribute excess energy back to the grid (P2P trading), transforming them into prosumers (as illustrated in the right side of Fig.~\ref{fig:information-flow}). Real-time data from the smart sensors at substations enables utilities to optimize energy distribution and  balance demand and supply proactively \cite{ALI2021}. 
    	
    Moreover, to enhance grid resilience, the introduction of microgrids \cite{Olivares14,JASIM2023}, electric vehicle integration, and predictive maintenance ensures a more resilient, reliable, and efficient energy infrastructure. Microgrids emerge as localized energy systems that can operate autonomously or interdependently with the main grid. By enabling localized generation and consumption, microgrids bolster grid resilience, allowing communities to maintain power during disruptions and emergencies. Through data-driven predictive maintenance and AI-powered control algorithms \cite{MACHLEV22, ZHOU2022, LEHNA2023}, grid operators can preemptively identify and address potential faults, minimizing downtime and optimizing system performance. The integration of electric vehicles (EV) also becomes a crucial component in \ac{SG2}, with the ability to intelligently manage EV charging and discharging \cite{Ye22}, not only minimizing grid congestion but also utilizing EV batteries as a form of distributed energy storage \cite{Nina21}. Further, that is the appearance of novel customer-oriented energy management as a service (EMaaS) \cite{Chen15} or a fair demand response with electric vehicles \cite{Chen18}. This evolution necessitates communication and management modernization, a significant upgrade from the conventional smart grid model. 
    
    	
    For the second upgrade type, most upgrades of network infrastructure come from new versions of existing WAN/FAN/LAN platforms (e.g., Zigbee, LTE) to be integrated into the smart metering and control management systems \cite{Hasan23}. Recently, the European Union (EU), China, and private sectors in the United States and other countries proposed a novel communication model to exploit the technical features of the fifth-generation (5G) communications for SG2 \cite{CIGRE,Das22}. By exploiting the broadband capability to collect massive amounts of unstructured data (video, images), 5G can assist energy distributors in improving their energy transmission flow efficiency and delivering personalized services to specific consumers. Implementing cloud offloading, such as optimizing energy production and delivery, is crucial for system security and stability \cite{CIGRE}. These activities may be time- and space-intensive. Cloud and edge computing from 5G may reduce computational burden in the center and allow the system function distributed and resiliently. In the long term, the sixth-generation (6G) networks can be the successor of 5G to continue enhancing the efficiency of the operations and enabling the development of new renewable energy sources \cite{Tariq21}. 6G with non-terrestrial network segments, e.g., satellite communications, can help utilities monitor and manage electricity distribution over larger areas, which is especially beneficial for microgrid communities in rural and remote areas.
    
    
    \subsection{Security concerns in transition networks towards SG2} 
     \label{subsec:potential-upgrade}

     Electricity has always been crucial for maintaining essential societal functions, including electrified transportation, machinery lines, public light, and economic development. The transition trend from fossil fuels to renewable energy in \ac{SG2} also motivates many users to use electric vehicles, which increases pressure on power grid infrastructure. Since society and millions of people depend on massive amounts of electric resources, ensuring a timely, sustainable, and cost-effective electricity supply is a priority. Blackouts over several hours, let alone days, can disrupt crucial services in hospitals and nursing homes \cite{Satter24}, halt manufacturing lines/transportation, and cost millions of dollars in damages \cite{Italia}. For example, in December 2015, some versions of the BlackEnergy and KillDisk malware disrupted the functioning of many substations in Ukraine, resulting in the loss of power for around 225,000 citizens for a few hours \cite{Ukcraine2015}. Generally, electricity security is a national security matter. Besides civil usage, electricity is also vital for military factories and combat control operations. Table~\ref{tab:main-threat-smart-grid} summarizes key security threats and potential protection approaches. The security threats and typical examples are classified by three target stakeholders, as shown in Fig.~\ref{fig:information-flow}, i.e., power grid provider/operator, communication network provider, and consumer. The severity and likelihood aspects are assessed using records of well-known attacks or catastrophic economic losses. The following sections detail the security threats and defense for three stakeholders.

      \begin{table*}[ht]
    	\caption{The main threats, typical examples, their severity level, likelihood and potential protection approaches in \ac{SG2}. }
    	\label{tab:main-threat-smart-grid}
    	\begin{tabular}{|l|l|l|l|l|l|l|}
    		\hline
    		\rowcolor[HTML]{EFEFEF} 
    		\multicolumn{1}{|l|}{\cellcolor[HTML]{EFEFEF}\textbf{Target}} & \multicolumn{1}{|l|}{\textbf{Security threats}} & \multicolumn{1}{|l|}{\textbf{Example}} & \multicolumn{1}{|l|}{\textbf{Severity}} & \multicolumn{1}{|l|}{\textbf{Likelihood}} & \multicolumn{1}{|l|}{\textbf{Protection method}} & \multicolumn{1}{|l|}{\cellcolor[HTML]{EFEFEF}\textbf{Reference}} \\ \hline
    		& Equipment failure & \begin{tabular}[c]{@{}l@{}}$\blacktriangleright$ Power generator\\$\blacktriangleright$ Cooling system \\ $\blacktriangleright$ Actuators/sensors \\ $\blacktriangleright$ RTU/PMU/HTS \\ $\blacktriangleright$ PLC microcontrollers \\$\blacktriangleright$ Transformer \\ $\blacktriangleright$ Switchgear \\ $\blacktriangleright$ Substations\end{tabular} & High & Low & \begin{tabular}[c]{@{}l@{}}$\checkmark$ Regular maintenance \\ $\checkmark$ Emergency backup\\$\checkmark$ Equipment stockpile \end{tabular} &  \begin{tabular}[c]{@{}l@{}}\cite{Hines2008,Islam19, Khan23,Satter24} \\ \cite{Hasan23, Ghasemi23,Havlena2023,TVABuildsResilience} \end{tabular}\\ \cline{2-7} 
    		\multirow{-2}{*}{\begin{tabular}[c]{@{}l@{}}Power grid \\ operator \\ (TSO/DSO) \end{tabular}} & Physical security & \begin{tabular}[c]{@{}l@{}}$\blacktriangleright$ Time synchronization\\$\blacktriangleright$ Substation attacks\\$\blacktriangleright$ Thief and vandalism\\ $\blacktriangleright$ Terrorist/drone attacks\\$\blacktriangleright$ Natural disasters\\$\blacktriangleright$ Supply chain \\$\blacktriangleright$ Aging devices\end{tabular} & High & High & \begin{tabular}[c]{@{}l@{}}$\checkmark$Perimeter fencing \\ (camera/barrier)\\ $\checkmark$ Distributed substations \\ $\checkmark$ Fault detection \\$\checkmark$  Mobile transformers\\ $\checkmark$ Emergency stocking plan \\ $\checkmark$ Diverse supplies \\$\checkmark$ Regular drills \\ $\checkmark$ Legal actions \\(thief/vandalism) \end{tabular} &  \begin{tabular}[c]{@{}l@{}} \cite{SupplychainAttack,MetcalfAttack,EnelAttack,Ukcraine2015} \\ \cite{Hasan23, Ghasemi23,Havlena2023,TVABuildsResilience} \\ \cite{WANG2023108889, YangZhiyuan23,NIST}\end{tabular}\\ \hline
    		& Equipment failure & \begin{tabular}[c]{@{}l@{}}$\blacktriangleright$ Router/modems\\ $\blacktriangleright$ Network cables\\$\blacktriangleright$ Base station/towers \\$\blacktriangleright$ SCADA center\end{tabular} & High & Low & \begin{tabular}[c]{@{}l@{}} $\checkmark$ Backup routing network \\$\checkmark$ Equipment stockpile \\$\checkmark$ Distributed towers \\$\checkmark$ Diverse supplies \end{tabular}&  \begin{tabular}[c]{@{}l@{}}\cite{CopperThief,EnelAttack,ElectricGridSecurity} \end{tabular} \\ \cline{2-7} 
    		& Physical security & \begin{tabular}[c]{@{}l@{}}$\blacktriangleright$ Physical tampering\\$\blacktriangleright$ Link sabotage \\$\blacktriangleright$ Supply chain\end{tabular} & High & Medium & \begin{tabular}[c]{@{}l@{}}$\checkmark$ Blockchain, HSM \\$\checkmark$ Legal actions \\ (Copper thief)\end{tabular} & \cite{LI2023103686, Ukcraine2015, CopperTheftLawsuit} \\ \cline{2-7} 
    		\multirow{-3}{*}{\begin{tabular}[c]{@{}l@{}}Network \\provider \\ (ISP)\end{tabular}} & Cybersecurity & \begin{tabular}[c]{@{}l@{}}$\blacktriangleright$ Denial of services\\$\blacktriangleright$ Ransomware/malware \\$\blacktriangleright$ False data injection \\$\blacktriangleright$ Social engineering \\$\blacktriangleright$ AI-aided attacks\\$\blacktriangleright$ Indirect attacks  \end{tabular} & High & High & \begin{tabular}[c]{@{}l@{}}$\checkmark$ Network slicing \\ $\checkmark$ Trusted/VLAN networks \\$\checkmark$ Encryption/authentication \\$\checkmark$ Endpoint/IDS/IPS \\ $\checkmark$ DMZ/Firewalls/MTD\\$\checkmark$ Blockchain/ledgers \\$\checkmark$ Behavioral analytics \\$\checkmark$ Secure service edge \end{tabular} & \begin{tabular}[c]{@{}l@{}} \cite{DoSAttacks, MalwareFrench,CDW,RansomwareJohannesburg} \\ \cite{Luo2023,Habib2023,Yan12} \\ \cite{5GSmartGrid,TR33.811,CEN-CENELEC-ETSI,Hasan23} \\ \cite{Zhang24, Mohamed21,WANG2023108889,3GPP33501} \\ \cite{ETSI.TS133.501,Nguyen21} \end{tabular} \\ \hline
    		& Equipment failure & Smart metering & Low & Low & \begin{tabular}[c]{@{}l@{}} Equipment stockpile \end{tabular} &  \begin{tabular}[c]{@{}l@{}}\cite{Zheng18,TVABuildsResilience,NIST} \end{tabular} \\ \cline{2-7} 
    		& Physical security &  \begin{tabular}[c]{@{}l@{}} $\blacktriangleright$ Metering tampering \\$\blacktriangleright$ Energy thief \\$\blacktriangleright$ Data exposure \\$\blacktriangleright$ Malware/DoS \\ (industrial zone) \end{tabular}& Medium & High & \begin{tabular}[c]{@{}l@{}}$\checkmark$ Blockchain, HSM \\ $\checkmark$ Abnormal detection \end{tabular}&  \begin{tabular}[c]{@{}l@{}} \cite{MeteringTampering,SHOKRY2022358,Mall2022}  \\ \cite{Bera21}  \end{tabular}\\ \cline{2-7} 
    		\multirow{-3}{*}{Consumer} & Cybersecurity & \begin{tabular}[c]{@{}l@{}}$\blacktriangleright$ Social engineering \\$\blacktriangleright$ False data injection\end{tabular} & Medium & Medium & \begin{tabular}[c]{@{}l@{}}$\checkmark$ Blockchain/ledger \\ $\checkmark$ Behavioral analytics \\$\checkmark$ Encryption/authentication\end{tabular} &  \begin{tabular}[c]{@{}l@{}} \cite{khoei2022comprehensive, Zografopoulos2021,Industroyer,ZHOU2022} \\ \cite{LI20232,MACHLEV22,ietf-quic-http-34}  \end{tabular} \\	
    		\hline
    		& Equipment failure & Communication links & High & High & \begin{tabular}[c]{@{}l@{}} Compatible standards \end{tabular} &  \begin{tabular}[c]{@{}l@{}}\cite{WANG2023108889,TVABuildsResilience} \end{tabular} \\ \cline{2-7} 
    		& Physical security &  \begin{tabular}[c]{@{}l@{}} $\blacktriangleright$ Cable thief \\$\blacktriangleright$ Monitoring fragmentation \\$\blacktriangleright$ Insider attacks \\$\blacktriangleright$ Hijacked hardware  \\$\blacktriangleright$ Malicious supply chain \end{tabular}& Medium & High & \begin{tabular}[c]{@{}l@{}}$\checkmark$ Trusted supply chain \\ $\checkmark$ Power switch plan \\ $\checkmark$ Intrusion detection \\  $\checkmark$ Joint warning \end{tabular}&  \begin{tabular}[c]{@{}l@{}} \cite{Mohamed21,Zhuang21,Habib2023}  \\ \cite{Bera21}  \end{tabular}\\ \cline{2-7} 
    		\multirow{-3}{*}{\begin{tabular}[c]{@{}l@{}} Joint stakeholders \\(ISP/DSO/TSO) \end{tabular}} & Cybersecurity & \begin{tabular}[c]{@{}l@{}}$\blacktriangleright$ DDoS attacks \\$\blacktriangleright$ False data injection  \\$\blacktriangleright$ Impersonate stakeholder \\  $\blacktriangleright$ Ransomware \end{tabular} & Medium & Medium & \begin{tabular}[c]{@{}l@{}} $\checkmark$ Firewall/IDS/IPS \\ $\checkmark$ Joint incident response \\$\checkmark$ Unified/SSO authentication \\ $\checkmark$ Mutual trust protocols\end{tabular} &  \begin{tabular}[c]{@{}l@{}} \cite{Junho19,SAKHNINI21,Park23} \\ \cite{Leite17,Nazemi21,Hassan22}  \end{tabular} \\ \hline
    	\end{tabular}
    \end{table*}

     For power grid operators (the first target in Table~\ref{tab:main-threat-smart-grid}), one of the major security concerns is sudden hardware failures of power generators, actuators, monitoring sensors, and microcontrollers \cite{Khan23}. Although the severity level is high, the failures of these devices are rare because of the stringent standards for lasting quality of supply in smart grid infrastructure \cite{NIST}. The other biggest security concern for power providers is physical security, where there are diverse threats, such as physical sabotage, vandalism, drone attacks (against important substations or high-voltage power lines, e.g., 500kv, 220kv), compromised supply chain, and even natural disasters. Further, since renewable energy sources are used to generate electricity, the stability of these supplies are critical, too. The details of attacks and protection methods are discussed in Section~\ref{sec:security-physical-layer}.       
     
     For network providers (the second target in Table~\ref{tab:main-threat-smart-grid}), major security concerns are physical and cyber attacks. Several typical examples are communication link tampering, ransomware/malware injection to internal computers or \ac{SCADA}, and denial of services against routers and control centers. For example, Enel Group's internal IT network in Europe was temporarily blocked by Snake Ransomware attacks in June 2020, resulting in the disruption of customer service activities \cite{EnelAttack}. The increased use of digital and networking technologies to optimize energy usage and distribution efficiency and provide one-click remote control capability. However, this interconnected model theoretically creates new space for attackers to access the power grids. In \ac{SG2}, the communication technologies help to synchronize measurement sensor data at power providers' facilities (e.g., cooling systems, microcontrollers) into overload monitoring applications. The heavy dependence of control systems on data exchange or remote control convenience is vulnerable to cyberattacks. The attacks target two primary purposes: (1) interrupt the data stream exchange or routing devices; (2) falsify the data \cite{Mohamed21}. Disconnecting the data stream or collecting inaccurate data might result in incorrect decisions on the activation or deactivation of individual electricity lines. Further, an attacker might potentially manipulate the power distribution by redirecting the power supply or disabling certain power lines, resulting in widespread power outages affecting several residences. The interdependence of communication networks and the power grid then requires many protection layers (e.g., a combination of firewall, IDS, VLAN, and advanced authentication) and optimized power distribution networks \cite{Parandehgheibi13}. If several efforts to receive data fail, the power grid may activate its backup network communication systems. These communications (e.g., VPN, leased lines) are theoretically isolated from the Internet and used for emergencies or authorized access privileges. The attacks and defense for the communication network provider are detailed in Section~\ref{sec:security-network-layer}.

     For consumers/end-users (the third target in Table~\ref{tab:main-threat-smart-grid}), major security concerns are tampering attacks on smart metering devices and false data in peer-to-peer energy trading models. The meters, which monitor and report electricity usage, can be vulnerable to physical and cyber attacks. Tampering meters can involve physical alterations to the device to manipulate consumption data, leading to inaccurate billing and potential financial losses for consumers \cite{MeteringTampering}. Additionally, smart two-way communication meters can be compromised through firmware updates that inject false data, leading to spoofed readings that can either inflate or deflate actual energy usage \cite{Mohamed21}. This false data injection can mislead utility companies and disrupt billing accuracy, posing a significant threat to the integrity of energy consumption data \cite{Luo2023,Tran2021}. In the context of peer-to-peer (P2P) energy trading, where consumers trade excess energy directly with others, spoofing and false data injection can have even more severe implications. Malicious actors can falsify transaction data to manipulate energy prices or falsely report energy trades that never occurred \cite{Havlena2023}. This can lead to financial fraud, undermine trust in the P2P trading system, and destabilize the energy market. Addressing these security concerns is essential for maintaining the reliability and trustworthiness of smart grids, particularly as they become more decentralized. The attacks and defense for the consumers are detailed in Section~\ref{sec:security-application-layer}.
     
     \subsection{Specific security threats in novel SG2 networks} 
     \label{subsec:specific-security-threats}
     
    In SG2, the rise of novel decentralized energy markets, private medium-small renewable grid operators, and peer-to-peer energy trading opens up a joint stakeholder model (illustrated in Figure~\ref{fig:information-flow}), where Internet Service Providers (ISPs), Distribution System Operators (DSOs), and Transmission System Operators (TSOs) cooperate to operate the grid and distribute energy to end users. However, new attack surfaces, such as price manipulation and fake energy transfers in peer-to-peer energy exchanges, are predicted to grow quickly. For example, the secondary control functions of microgrids and small operators via communications increase the complexity of network management \cite{WANG2023108889}. When SG2 incorporates large-scale EV and various charging vendor infrastructures, and private grid operators (particularly from renewable power plants, households, and small providers),  fragmentation of grid status monitoring systems or inconsistency of joint security infrastructure among diverse stakeholders open the door for \textbf{insiders} and \textbf{supply chain attacks}. Inconsistent security protocols among these entities make it easier for attackers to perform {Man-in-the-Middle (MitM) attacks} and {fake data injection (FDI)} \cite{Mohamed21,Zhuang21,Habib2023}, leading to incorrect grid balancing or energy billing errors.  
     Attackers exploit the fact that ISPs, DSOs, and TSOs may have separate monitoring and management systems that don't fully communicate with each other. This fragmentation creates blind spots, where an attack on one part of the grid or communication network may go unnoticed by the others, leading to undetected breaches and delayed mitigation actions.   \textbf{A lack of coordination in incident response} can delay action during \textbf{Distributed Denial-of-Service (DDoS) attacks} \cite{DoSAttacks} targeting ISPs, which can disrupt real-time data exchange between distributed energy resources (DERs) and grid operators, leading to grid imbalances \cite{Zografopoulos2023}. 
     Attackers can exploit this delay to intensify distributed attacks, escalate damage, or prolong service disruptions, knowing that each party might expect the other to handle the response.
     
     Lack of cooperation between ISPs, DSOs, and TSOs can allow attacks to propagate from the communication network (e.g.,  \textbf{malware on ISP-controlled systems}) to the grid's operational technology (OT) environment, such as SCADA systems, leading to power disruptions \cite{Industroyer,Italia}. The disjointed approach to cybersecurity between ISPs and grid entities also opens up risks for \textbf{ransomware attacks}, data manipulation, and insider threats. If there is no unified strategy for incident response, ransomware can quickly spread, locking down critical components of both the communication and power grid systems. Further, without unified authentication schemes and mutual trust protocol standards \cite{Mall2022}, attackers can gain unauthorized access to critical grid systems by exploiting the weakest link in the network (e.g., private grid operators or small energy distributors with limited security defense capability). In the absence of cooperation, an \textbf{insider} from one party (e.g., an ISP or ISP tenants) could bypass inadequate security measures of another party (e.g., DSO) to carry out attacks. SG2's significant expansion in IoT devices to provide real-time data for smart management, such as smart meters, home automation systems, and linked appliances, may introduce new vulnerabilities of lightweight security protocols and authorization \cite{Li2023}. Also, due to the growing complexity of supply chain management, malware or \textbf{hijacked hardware components} can be inserted into grids through different stakeholders with a lack of strict plugin-in validation.

    Mitigating the security threats in cooperative grid operations requires a comprehensive solution. The straightforward solution is to deploy \textbf{intrusion detection} and  \textbf{endpoint security systems} across both communication networks and operational technology environments \cite{Junho19,SAKHNINI21}. Further, automated responses to attacks (e.g.,  \textbf{isolating compromised network segments through network slice} of the grid,  \textbf{rerouting power lines} \cite{Park23}) can shorten the response time and help maintain grid stability. Self-healing solutions like  \textbf{automatic local switching plans} \cite{Leite17} to switch fault operations and isolate failures within the shortest time interval can prevent wider disruptions. However, given the complexity of multiple ISPs/TSOs/DSOs, particularly the integration of renewable energy sources (RESs) in power distribution systems (PDSs), there should have \textbf{joint incident response} plans through API or open interfaces, with clearly defined roles and responsibilities. The restoration plan can be based on non-convex non-linear stochastic optimization formulation with joint probabilistic constraints \cite{Nazemi21}. The related parties should share \textbf{playbooks for distributed coordinated attacks}, periodic drills, and communication strategies during an incident.
    
   Implementing platforms for \textbf{real-time data sharing and secured protocols} between DSOs, TSOs, and ISPs can improve situational awareness. The secured protocols (e.g., TLS 1.3-based and quantum-resistant-based  \cite{Alshowkan22}) are essential to provide synchronized views of grid operations, cyber activities, and network health, enabling faster detection and better-informed decision-making. However, these solutions may come up costly and require a trust agreement among stakeholders. At this point, \textbf{blockchain-based or Zero-Trust verification platforms} can be the solutions \cite{Hassan22} to secure communications and maintain the relative trust. For example, by providing a tamper-proof ledger, blockchain can help ensure the integrity of grid-related communications and improve coordination during incident response. Further, by  \textbf{decentralizing energy generation and integrating more DERs} (like solar panels, batteries, and microgrids), TSOs/DSOs can improve their ability to react to vandalism, drone attacks, or natural disasters. In this model, DERs or microgrids can maintain power at a local level while the larger grid is being stabilized. Finally, \textbf{unified authentication} frameworks (e.g., single sign-on) using physical uncloneable functions for cooperative microgrids \cite{Badar2021} can be a good idea to mitigate the unbalanced security capability of each party. 

         \begin{table*}[]
    	\centering
    	\caption{Summary of several security standards used in the power grid systems in some countries.}
    	\label{tab:methods-summary}
    	\begin{adjustbox}{width=1\textwidth}
    		\small   
    		\begin{tabular}{llllllllll}
    			\hline
    			\textbf{\#} & \textbf{Security standard} & \textbf{Features} & \textbf{Countries} & \textbf{Application cover} & \textbf{Latest version} & \textbf{Type}  & \textbf{Issuer}\textcolor{red}{*}  & \textbf{Study} \\ \hline
    			1 & NISTIR 7628 & \begin{tabular}[c]{@{}l@{}} $\blacktriangleright$Describe security controls, risk \\ management, and privacy\end{tabular} & $\checkmark$ US &  \begin{tabular}[c]{@{}l@{}}All components of \\ Smart Grid \end{tabular}&  2014 & Guidelines & NIST &  \cite{NIST} \\
    			2 & NERC CIP & \begin{tabular}[c]{@{}l@{}}$\blacktriangleright$Describe critical infrastructure \\ protection,  incident response, \\ and recovery planning\end{tabular} & $\checkmark$ US & \begin{tabular}[c]{@{}l@{}}Secure industrial \\ critical Systems \end{tabular} & 2021 & Guidelines  & \begin{tabular}[c]{@{}l@{}}NERC \end{tabular} &  \cite{NIST} \\
    			3 & IEEE C47.230 & \begin{tabular}[c]{@{}l@{}}$\blacktriangleright$Standard security requirements \\ for substation automation, \\ protection, and control systems)\end{tabular} & $\checkmark$ Worldwide &  \begin{tabular}[c]{@{}l@{}}Substation \& ICS\end{tabular}&  2018 & Standard & IEEE  & \cite{IEEEC37} \\
    			4 & FIPS 140-3 & \begin{tabular}[c]{@{}l@{}}$\blacktriangleright$Describe  standards\\ for cryptographic modules\end{tabular} & $\checkmark$ US&   \begin{tabular}[c]{@{}l@{}}Hardware modules \end{tabular}& 2022 & Standard & NIST  &  \cite{NIST} \\
    			5 & IEC 61850 & \begin{tabular}[c]{@{}l@{}}$\blacktriangleright$Describe communication protocols \\ \& data models for substation \\ automation systems (including for\\ energy trading)\end{tabular} & $\checkmark$ US & Grid communications &  2016 & Standard &  IEC &  \cite{IEC61850} \\
    			6 & ISA/IEC 62443 & \begin{tabular}[c]{@{}l@{}}$\blacktriangleright$Describe network segmentation, \\ access control, and security \\ monitoring\end{tabular} & $\checkmark$ Worldwide&  \begin{tabular}[c]{@{}l@{}}Industrial automation \\ \& control systems \\ (+energy storage)\end{tabular}  & 2021  & Standard &  ISA/IEC &  \cite{IEC62443}\\
    			7 & IEC 62351 & \begin{tabular}[c]{@{}l@{}}$\blacktriangleright$Describe securing communication \\ protocols used in power systems \\ automation\end{tabular} &  $\checkmark$ Worldwide & \begin{tabular}[c]{@{}l@{}}Network protocols \\ of power systems\end{tabular}& 2023 & Standard & IEC &  \cite{IEC} \\
    			8 & ENISA & \begin{tabular}[c]{@{}l@{}}$\blacktriangleright$Describe emerging threats, \\ risk assessment methodologies, \\ best practices for securing  \\ smart grid deployments\end{tabular} & $\checkmark$ EU & All components & 2019 & Guidelines &  EU& \cite{ENISA} \\
    			9 & ISO/IEC 27001 & \begin{tabular}[c]{@{}l@{}}$\blacktriangleright$Describe information security\\  management risks (including for\\ energy trading)\end{tabular} & \begin{tabular}[c]{@{}l@{}}$\checkmark$ Worldwide \end{tabular}& Information Security & 2022 & Standard & ISO/IEC & \cite{ISOIEC27001} \\
    			10 & ISO/IEC 15408 & \begin{tabular}[c]{@{}l@{}}$\blacktriangleright$Assess the security of smart grid \\ components and systems\end{tabular} & $\checkmark$ Worldwide & Product security& 2022 & Standard & ISO/IEC  & \cite{IEC15408} \\
    			11 & ISO 15118 & \begin{tabular}[c]{@{}l@{}}$\blacktriangleright$Describe security standards \\ for Vehicle-Grid Communications\end{tabular} & $\checkmark$ Worldwide & Product security& 2022 & Standard & ISO  & \cite{ISO15118} \\
    			
    			12 & IEEE 1686 & \begin{tabular}[c]{@{}l@{}}$\blacktriangleright$Describe security standards  for \\ Intelligent electronic devices\end{tabular} & $\checkmark$ Worldwide & Substation security& 2023 & Standard & IEEE & \cite{IEEE1686} \\
    			
    			13 & IEEE 2030 & \begin{tabular}[c]{@{}l@{}}$\blacktriangleright$Guide for the interoperability \\ of energy storage systems \\ with the grid infrastructure\end{tabular} & $\checkmark$ Worldwide & Energy Storage& 2019 & Standard & IEEE & \cite{IEEE2030} \\
    			
    			14 & CEN-CENELEC-ETSI & \begin{tabular}[c]{@{}l@{}} $\blacktriangleright$Describe interoperability, \\ protocols, data exchange formats\end{tabular} & $\checkmark$ EU  & Smart Meters & 2018 &  Framework & EU  & \cite{CEN-CENELEC-ETSI} \\
    			15 & Privacy Act & $\blacktriangleright$Address the privacy of user data & $\checkmark$ Local Law & Regulation & - & Policy & Local country &  \\
    			16 & \begin{tabular}[c]{@{}l@{}}Smart Grid \\ Security Guidelines\end{tabular} & \begin{tabular}[c]{@{}l@{}}$\blacktriangleright$General guidelines\end{tabular} & $\checkmark$ China & Custom Smart Grid & 2012 & Guidelines & SGCC, CEC &  \cite{Yu12} \\
    			17 & \begin{tabular}[c]{@{}l@{}} Smart Grid \\ Security Guidelines\end{tabular} & \begin{tabular}[c]{@{}l@{}}$\blacktriangleright$General guidelines\end{tabular} & $\checkmark$ Japan & Custom Smart Grid & 2019 & Guidelines & METI &  \cite{METI} \\
    			18 & JISEC Framework & \begin{tabular}[c]{@{}l@{}}$\blacktriangleright$The Japan information security \\ evaluation and certification, \\ certifies the security of IT products \\ \& systems\end{tabular} & $\checkmark$ Japan & Product security & 2023 & Framework  & CCRA &  \cite{JISEC}\\
    			19 & \begin{tabular}[c]{@{}l@{}}Industrial Control\\ Security Guidelines\end{tabular} & \begin{tabular}[c]{@{}l@{}}$\blacktriangleright$Security measures, vulnerability \\ management, access controls, and \\ incident response\end{tabular} & $\checkmark$ Worldwide & ICS Security  & 2015 &  Guidelines & NIST & \cite{SP80082} \\
    			20 & \begin{tabular}[c]{@{}l@{}}Cyber security \\in Power Sector\end{tabular} & $\blacktriangleright$Guidelines for grid security & $\checkmark$ India& Custom Smart Grid & 2022 & Guidelines  & CERC   & \cite{IndiaPowerGridSecurityPower}  \\ \hline
    		\end{tabular}
    	\end{adjustbox}
    	\begin{tablenotes}
    		\item \textcolor{red}{*}  NIST: National Institute of Standards and Technology; NERC: North American Electric Reliability Corporation; ISO: International Organization for Standardization; ISA: International Standards on Auditing; IEC: International Electrotechnical Commission; CERC: Central Electricity Regulatory Commission; EU: European Union; METI: Ministry of Economy, Trade and Industry; CCRA: Common Criteria Recognition Arrangement;  SGCC, CEC: State Grid Corporation of China, China Electricity Council.
    	\end{tablenotes}
    \end{table*}

     \subsection{Discrepancy of security standards used in the grid sector and security threats of remote control models in \ac{SG2}}
	\label{sec:standard}

      Securing the power grid is a global concern, and various countries have established security standards and implementations to safeguard their energy infrastructures. Unlike civil applications, security standards for \ac{SG2} are based on \ac{ICS} and \ac{SCADA} standards (e.g., IEC 62443-3-2 and NIST SP 800-82), which are well-designed for large-scale and critical infrastructure. Theoretically, the \ac{SCADA} for the integrated security operation center or in energy management systems is isolated by industrial \ac{DMZ} and protected with many protection layers of firewalls, anomaly detection, and \ac{IPS} platforms from remote access.  Meanwhile, the \ac{WAN} and \ac{LAN} in substations are bounded by security perimeter solutions, e.g., \ac{IPS}, \ac{DPI}. For multiple networking technologies (e.g., 5G, Zigbee, LoRaWAN), the protection is implemented from the devices and gateways to the core networks (\ac{WAN}).

      Each nation also uses the \ac{ICS} security architecture in smart grids differently. For example, the US North American Electric Reliability Corporation (NERC) implements critical infrastructure protection requirements via its CIP framework \cite{NIST}. This framework covers physical and cybersecurity security for power generating, transmission, and distribution systems. To maintain grid dependability and resilience, NERC CIP guidelines include strong access controls, incident response plans, and frequent security evaluations. The European Network for Cyber Security (ENCS) collaborates with energy companies to establish cybersecurity requirements tailored to the energy sector \cite{ENISA}. Japan prioritizes energy security but follows local standards. China's National Energy Administration (NEA) \cite{Shu17} has outlined their own measures for energy network security, addressing aspects like data protection, risk assessment, and response plans. The Central Electricity Regulatory Commission (CERC) in India has introduced their own guidelines for cybersecurity in the power sector \cite{IndiaPowerGridSecurityPower}. 

      Generally, the countries have developed their comprehensive security standards and implementations to protect their power grids. Most standards and guidelines are aligned with international best practices (e.g., ISO/IEC 27001). However, given the national security, many implementations are adapting to their unique energy landscapes to follow local laws and create efficient energy ecosystems (appropriate to the country's available resource capacity and supply chains). Due to many reasons (sloppy implementations, employee mistakes, poor security design), the power grids in many countries may not be entirely equipped with the best protection mechanisms as expected in the security standards \cite{Nguyen21}. This is why the attackers may successfully launch an attack against the power grids, even with many well-known attacks listed in this paper.
      
      Further, the growing centralization and remote control capabilities of \ac{SG2}, especially for large consumers, offer substantial grid security vulnerabilities. The foreign-made devices have raised concerns about potential vulnerabilities that could allow unauthorized access and control. Imagine that installation technicians can access websites that control large numbers of inverters, creating a scenario where the grid could be destabilized with a single command. This underscores a broader problem of security and control in grid-connected devices, regardless of their origin. To prevent the one-click control attacks or dependence of a single vendor, there are several defense approaches: (1) network slicing, VLAN, and firewalls isolate communication networks; (2)  strict authentication and remote access permission with secure protocol/information security standards and strong oversight of a balance responsible party (BRP); (3) decentralizing control and communication network systems, (4) diverse supply chains through trusted international partners. Finally, we believe, to maintain the good of all these technical issues, human resource training and regular stress tests are also important too. However, all these protective mechanisms may be too expensive for certain stakeholders (power grid providers, communication network providers, consumers). To mitigate the damage, the approaches (1) and (2) can be affordable to implement. The following sections detail the security attacks and threats in general that can impact the platforms in many countries. Based on these threat examples and lessons learned, we also highlight the importance of collaboration among stakeholders and local laws to enhance \ac{SG2}'s security.
     

\section{Security threats and protection models for power grid providers/operators}
\label{sec:security-physical-layer}

 As stated earlier and summarized in Table~\ref{tab:main-threat-smart-grid}, major security concerns for power grid operators come from hardware failures and physical security. Since the grid infrastructure plays a critical role to nation energy security, equipment used in power plants is often subject to strict maintenance and quality management cycles. However, given the complexity of the smart grid operations and aging devices, the grid operators are vulnerable to many security matters and also the target of many adversaries \cite{Tala22,Habib2023}. This section summarizes typical security threats against the power providers. Security protection mechanisms are then presented. Finally, we summarize the remaining challenges of protection models.


     \subsection{Typical security concerns against power grid providers}

    Security threats to electricity suppliers come from both inside and outside. The \textbf{inside security concerns} refer to problems in the internal grid control networks or broken devices. For example, disgruntled employees or those with malicious intent can sabotage operations or leak sensitive information. Human error is another significant internal threat, with mistakes by employees potentially leading to operational disruptions or security breaches \cite{Yan12}. \textbf{Outside factors} might include unforeseen events (natural disasters), physical attacks (drones, terrorism, vandalism), and cyber attacks (malware/ransomware) by adversaries. Further, failures in software, hardware, and older equipment cause less resistance to attacks.
    
    Several security threats with specific damages are summarized as follows. 
   



      \textcircled{\raisebox{-0.9pt}{1}} \textbf{Time synchronization attacks}: This attack threatens the precise operation of phasor measurement units (PMUs) in grid operators by disrupting their time synchronization processes \cite{Moussa19,Vahidi2024}. Since PMUs rely on accurate time-stamping from Global Navigation Satellite Systems (GNSS) like GPS, any interference with these signals can compromise the reliability of synchrophasor measurements. The attack can cause significant disruptions in phase angle monitoring (PAM) by corrupting total vector error (TVE), leading to inaccurate phase angle calculations and erroneous power transfer dynamics. These attack types can also trigger false alarms, unwarranted trip commands, and degrade voltage stability controllers, compromising the overall reliability and protection of the smart grid.
        
\begin{figure}[t]
         	\begin{center}
         		\includegraphics[width=1\linewidth]{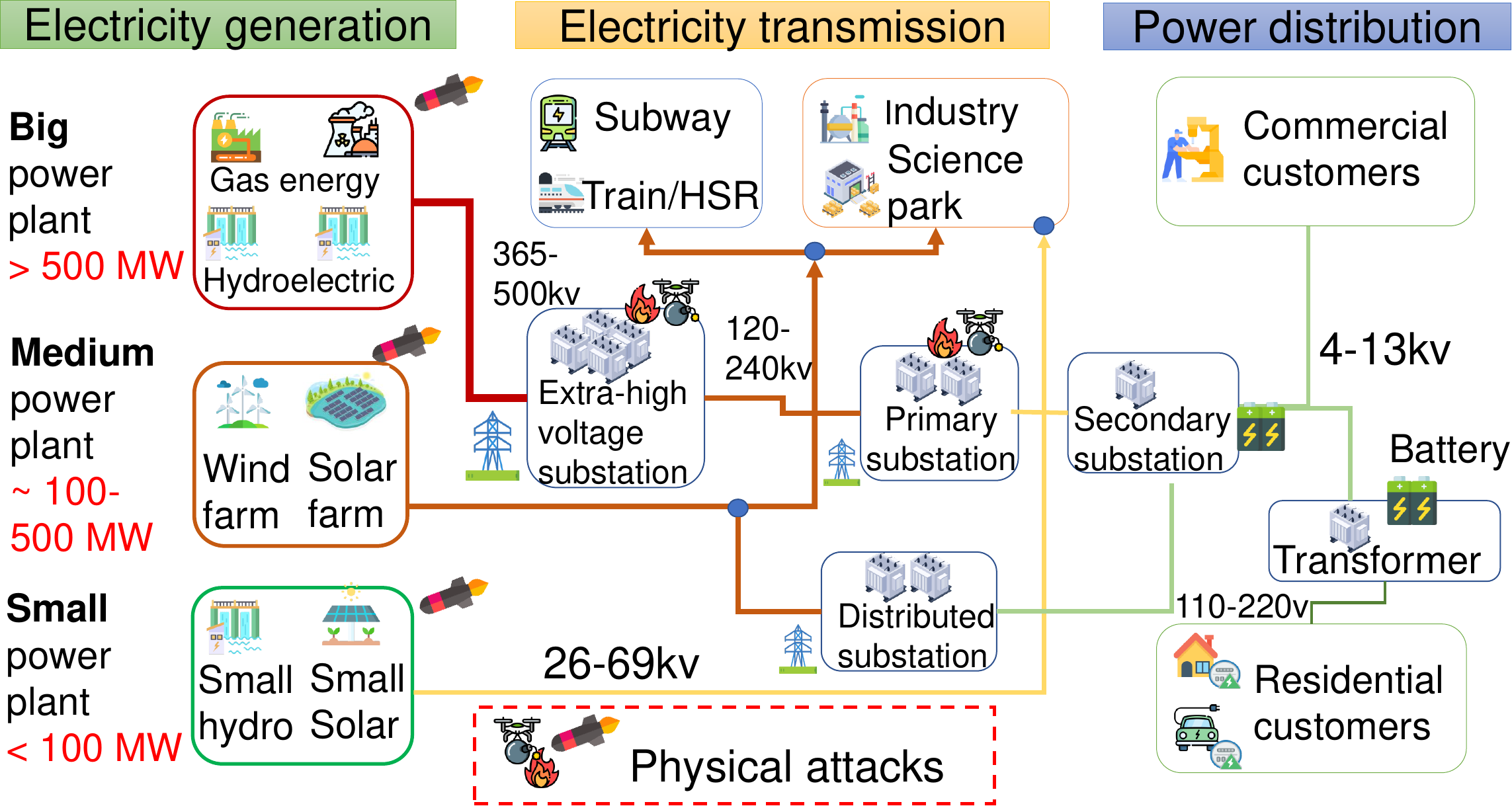}
         	\end{center}
         	\centering
         	\caption{The illustration of an electric distribution model to different consumer types, from public transportation, industry clients, medium-small commercial consumers, and residential users. Failures of extra-high voltage or primary/secondary substations can cause devastating consequences, i.e., large blackouts. They are also the targets of many physical attacks.}
         	\label{fig:electric-distribution-model}
         \end{figure}

       \textcircled{\raisebox{-0.9pt}{2}} \textbf{Equipment failure}: The failures can come from aging microcontrollers and electricity circuits in operation systems of power plants \cite{Islam19, Khan23}. According to Statista, between 2000 and 2023, the large states in the US (e.g., Texas, California) has an average of 251 blackouts in the 23-year period due to aging broken devices or equipment failures from thunderstorms, wind, hurricanes, or just a brownout during a heat wave. The failures cause substantial direct economic costs, e.g., insurance claims up to $3$ billion USD \cite{Hines2008}. Fig.~\ref{fig:electric-distribution-model} illustrates an electric distribution model for different consumer types. Accordingly, failures of extra-high voltage or primary/secondary substations can cause devastating consequences, i.e., large blackouts. Besides, they also generate other substantial damages, including metro/subway passenger delays and emergency vehicle stops owing to traffic signal failures in urbanized cities. The amount, duration, location, and time of day of a blackout affect its societal effect. Furthermore, equipment failures due to aging infrastructure or technical malfunctions can compromise the reliability of the electric supply. For example, the large blackout in the northeastern United States in 2003 was partly attributed to aging infrastructure and system failures \cite{Blackout2003}.  
         
         
\begin{figure}[t]
    		\begin{center}
    			\includegraphics[width=1\linewidth]{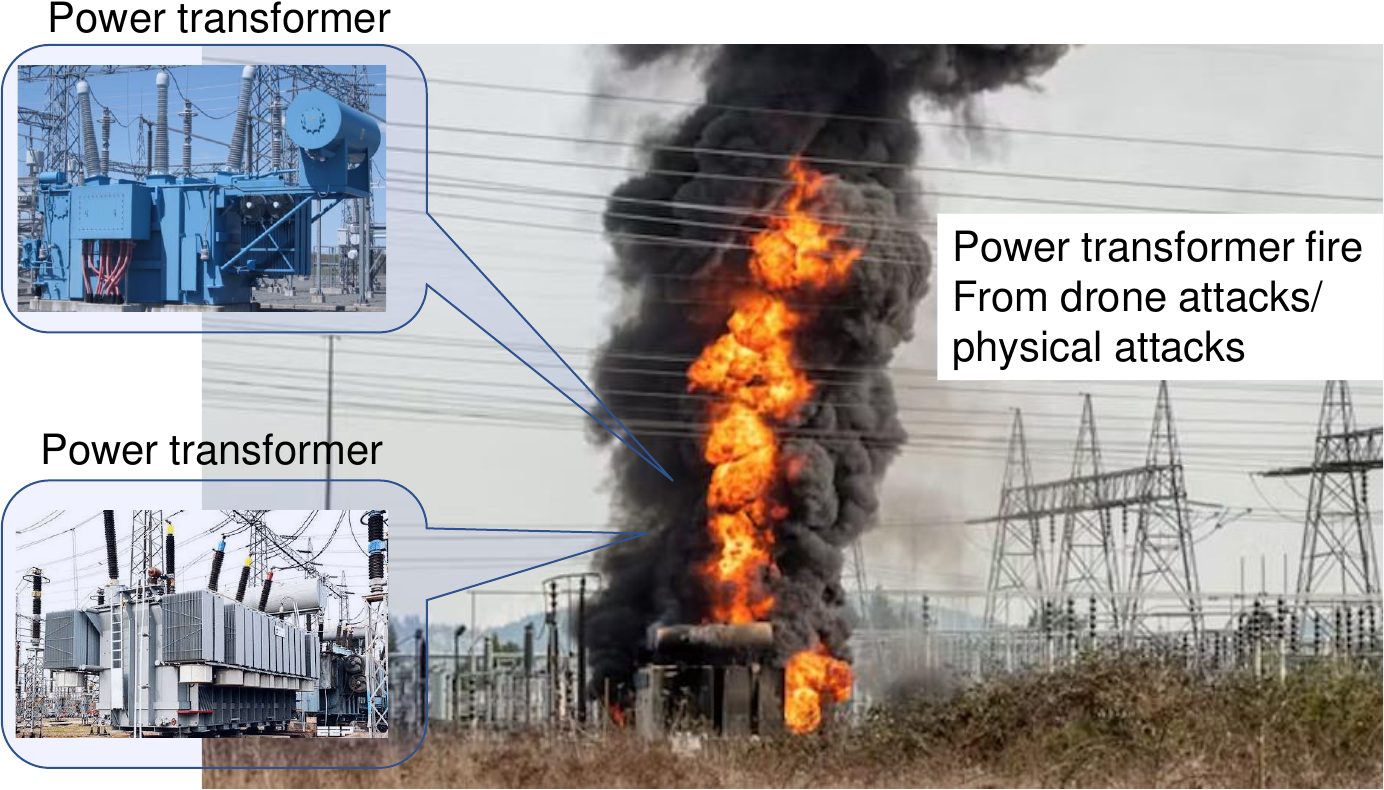}
    		\end{center}
    	\centering
    	 	\caption{The illustration of a substation transformer attack by drones/vehicles or vandalism that causes fire and electric outage.}
    	 	\label{fig:PHY-power-transformation-fire}
      \end{figure}

        \textcircled{\raisebox{-0.9pt}{3}} \textbf{Physical/substation attacks}: In this attack, attackers might tamper with equipment, disable security measures, or even manipulate operational settings, leading to widespread disruptions. For instance, in 2013, attackers cut fiber optic cables and used sniper rifles to damage transformers at a PG\&E substation in Metcalf \cite{MetcalfAttack}, California, causing significant damage up to $15$ million USD worth of repairs and raising concerns about an unprecedented and sophisticated terrorism attack on an electric grid substation with military-style weapons \cite{CaliforniaAttack}. Another emerging threat is manipulating energy storage systems, such as batteries, essential for grid stability and load balancing. Malicious actors could exploit vulnerabilities in these systems to manipulate energy flow, leading to grid instability or even overloading components \cite{He2016StorageAttack}. For example, the 2015 cyber attack on Ukraine's power grid involved hackers gaining access to control systems and changing operational settings, leading to power outages for hundreds of thousands of people \cite{Ukcraine2015Hack}. The other threat is the attack from cruise missiles or military drones that target essential power transformers (extra-high voltage or primary/secondary substations as shown in Fig.~\ref{fig:electric-distribution-model}) to cause fire and power outages over large areas, as illustrated in Fig.~\ref{fig:PHY-power-transformation-fire}. Another variant is to ram vehicles into substations or transmission lines, which can cause significant damage and outages. For example, an attack on two power substations left more than 40,000 people without power in Moore County, US, on December 3, 2022 \cite{SubstationAttacks}.

       \textcircled{\raisebox{-0.9pt}{4}} \textbf{Supply chain attacks}: The attacker intentionally can introduce compromised or malicious hardware (remote monitoring sensors \cite{Miller12}) into the infrastructure, which can later open a backdoor for physical intrusion or remote control \cite{SupplychainAttack}. This type of attack can be particularly insidious, as it may go undetected until the compromised hardware is activated. Theft of critical infrastructure equipment, such as heating monitoring sensors in power plant's cooling systems or copper cables  \cite{ElectricGridSecurity}, is a significant concern. We can imagine how devastating it would be for emergency infrastructures if power systems went down owing to a loss of control at the management center from stolen devices. Further, through recent trade wars, pandemic/conflict crisis, and the decarbonization goals toward 100\% clean electricity, a severe shortage of key commodities, such as key minerals/materials (rare earth elements) and gas supplies (in gas-electric power plants), can create an inflationary cost environment for companies, resulting in increased costs of transformers, wire and cables, batteries, and solar panels. Also, the clean energy transition will require a large expansion of transmission and distribution infrastructure, including new and end-of-life line replacements, power transformers, high-voltage direct current systems, digital relays, and smart inverters \cite{SupplychainAttack}. Many of basic supplies rely on a few nations, leaving them susceptible to interruption.

      In general, physical and supply chain attacks can have severe implications for grid stability, functionality, and even public safety. However, these attack types often require the presence of the attackers near the attack areas, limiting the attacker's capability if the grid infrastructure is physically well-protected (e.g., by building walls). Finally, insider threats cannot be overlooked. Human errors or individuals with authorized access to the physical infrastructure could intentionally misuse their privileges to compromise system integrity. 
     As \ac{SG2} becomes more reliant on digital controls and interconnected devices, the risk of cyber-physical attacks grows \cite{Hasan23}. Major security concerns in \ac{SG2} will be severe shortages of key minerals and materials in the supply chain to produce solar panels and electric vehicles or the risks of long-range cruise missiles/drones targeting critical transformers and substations, given the popularity of UAV/drones everywhere. The volatile movements in the geo-political landscape towards a new multipolar order, coupled with changes in global economic dynamics, especially concerns about trade wars, are prompting many nations to establish a roster of reliable partners for sources of essential components in \ac{SG2} as detailed below.

   \subsection{Specific threats for new power grid models in SG2}
   
   Unlike in SG1, the energy market in SG2 may open for thousands of grid operators, including private sector and small-scale renewable energy producers. In this distributed energy resources (DERs), solar panels and wind turbines in small producers with limited security capability are often remotely controlled and may lack robust security mechanisms \cite{SOLAR,Ostergaard21}, making them vulnerable to \textbf{brute-force cyber attacks} with default credentials as in current IoT devices. Further, the
   	growing use of IoT devices in SG2 by smaller grid operators
   	increases the risks, where the IoT devices have inadequate ability (e.g., no enterprise firewall or intrusion prevention system) to protect themselves from being attacked. If those devices are hacked, then the extensive devices can be used to attack the core network (e.g., DDoS) or disturb the electricity flow and information exchange. Note that, smaller operators may not invest adequately in infrastructure or maintenance or even fail to fully comply with technical grid codes and operational standards (due to lack of resources, knowledge, or intentional disregard), leading to substandard equipment or outdated technologies in grid-connected systems. With many small operators involved in energy trading, attackers could target exploiting vulnerabilities of the integrity of financial transactions (e.g., through \textbf{51\% attacks} or \textbf{consensus mechanism manipulation} \cite{Chowdhury22,Kaur21}) and automated energy bids. This could lead to significant financial losses for specific operators if the collusion of a portion of the network or the incorrect pricing (due to false data attacks \cite{Mohamed21} during synchronization or energy trading) occurs for a certain time. Small DSO operators, including small private sector participants, might have  \textbf{disgruntled employees or contractors} who could abuse their access in energy biding systems and grid operation to compromise the grid to initiate local cascading grid failures. In extreme cases, grid operators in a competitive market, especially those relying on intermittent renewable energy sources, may not always align their energy generation with grid demand \cite{Hasan23} but price bidding ( \textbf{selling only when prices are favorable}). Some might \textbf{intentionally withhold or oversupply energy} to manipulate market prices or gain an advantage in energy trading markets, particularly in crisis situations and peak demand periods. The time synchronization attacks \cite{Moussa19} can be a threat in SG2. This could stress or destabilize the grid, leading to outages. In the contrary, several operators, especially those using solar and wind, may produce excess energy during periods of low demand without proper coordination with grid operators. This  \textbf{overproduction} can lead to voltage rises, grid congestion, and potential damage to grid infrastructure, especially if energy storage systems are not in place to absorb the excess.

  \subsection{Security protection models for power grid operators}

    There are various countermeasures to mitigate the magnitude of security threats against power grid providers. This subsection and the protection method column in Table~I summarizes several typical models.
    
\begin{figure}[t]
    	\begin{center}
    		\includegraphics[width=1\linewidth]{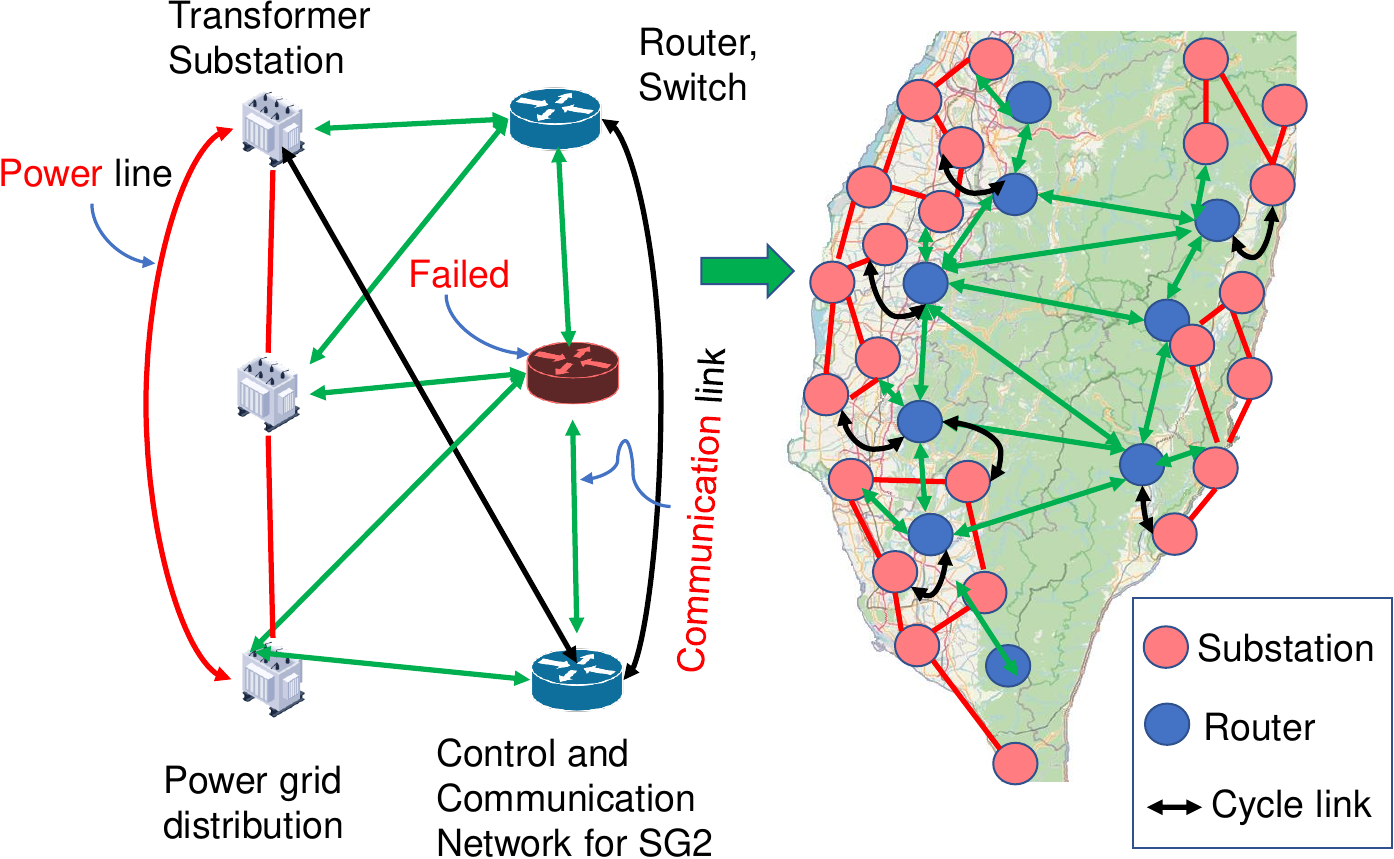}
    	\end{center}
    	\centering
    	\caption{The illustration of a power grid model in Taiwan and a design for connection lines of grid control systems to prevent cascading failures. Links or optimal coupling between the nearest neighbors are created to guarantee that no failure substation/router will cause the failure of the whole communication control network or a blackout of the whole grid.}
    	\label{fig:cascading-failure}
    \end{figure}

   To mitigate equipment failures stemming from aging micro-controllers and electricity circuits, power plants can employ several key strategies, such as regular maintenance and timely replacement of outdated components \cite{DigitalisationProtection}. Additionally, implementing redundant systems and automated failover mechanisms ensures that operations continue seamlessly in the event of equipment failure. For example, in 2013, the Tennessee Valley Authority upgraded its grid infrastructure in a way, that if one line or component fails, the system automatically reroutes power through alternate pathways, preventing outages and maintaining continuous operation \cite{TVABuildsResilience}. The implementation also includes special circuits and fiber networks to prevent electromagnetic pulse attacks and geomagnetic disturbance \cite{mate2021analyzing}. Fig.~\ref{fig:cascading-failure} illustrates a study model on mitigating the minimum number of node failures to prevent cascading failures. Accordingly, the authors use graph structure to model equipment failures or offline routers and suggest a star-topology-based heuristic algorithm to find a near-optimal routing solution. Recent studies on the flow dynamics in both the grid and the failure rollover reinforced that increasing the power/communication link coupling is beneficial to the resilience against cascading failures \cite{Ghasemi23,Khan23}. The utilization of advanced data analytics and AI models, along with datasets containing information on equipment failure rate and natural events (e.g., earthquake/heavy rain/thunderstorm risks), enables real-time monitoring and predictive maintenance \cite{Hasan23}. This enhances the capability to detect early signs of failure and facilitate quick repairs. For internal grid operations, developing misbehavior detectors to check abnormal operational data (e.g., unscheduled power on/off key devices, suddenly increased voltage levels) can also mitigate insider attacks \cite{Havlena2023}.

    For physical attacks, primary solutions are detection intrusion, equipment tampering detection, video monitoring, camera analytics, and using new cut/climb/ram-resistant fences.  Table~\ref{tab:main-threat-smart-grid} in the previous section summarizes several prospective protection solutions against physical communication links in the grid operators. The most common idea is to equip the metering devices and substation components with perimeter fencing (camera/barrier), \ac{HSM}, and card-based entry systems for on-site security personnel. \ac{HSM} can provision cryptographic keys for critical functions such as encryption, decryption, and authentication \cite{Sommerhalder2023} in relevant metering devices. The authentication methods help limit access to authorized individuals only and prevent physical access/tampering since any physical manipulation of the devices (e.g., firmware/operational settings modification) will be detected by \ac{HSM}'s tamper-evident and intrusion-resistant safeguarding capability. For transformer-targeted attacks, additional spares inventory is critical to speed emergency replacements \cite{TVABuildsResilience}. Mobile transformer fleets and mobile switch houses with racks and portable truck-mounted GIC switchgear can offer the greatest flexibility for restoration in the case of a series of missile attacks and substation failures. 
    
   The most challenging task is to avoid substation attacks, given the diversity of attack sources (e.g., intentional sabotage by trucks and flammable substances). Multiple physical measures are required to protect this critical infrastructure. For example, besides perimeter fencing around the substation with physical barriers (concrete walls) to restrict unauthorized access, implementing access control measures such as gates, locks, card readers, and biometric systems is also an efficient method to ensure that only authorized personnel can enter the substation \cite{NERCReliabilityGuidelines,Islam19}. The sensor alarms can be equipped to trigger shout alarms or notifications to security personnel once the substation is intruded. Installing robust fencing, anti-climbing devices, barriers around substations, or high-definition cameras with night vision capabilities at strategic locations around substations can deter theft and provide evidence for investigations \cite{PhysicalSecurity}. Besides the strict regulations and increasing penalties for copper theft, educating the public about the impact of the theft activities on critical infrastructure and encouraging them to report suspicious activities can enhance community vigilance \cite{ElectricGridSecurity}.
    	
    For complicated attacks (e.g., by remote bombs and cruise missiles), building backup substations/distributed microgrids or anti-drone technologies and missile defense systems can be a way to mitigate the damage of physical attacks and ensure the substation can continue operating in case of an attack or equipment failure \cite{WANG2023108889}. Accordingly, instead of constructing 5-10 major power plants with great production capacity and a substantial portion of the energy balance, there should be dozens to hundreds of small power plants spread throughout the nation that can power a city if one fails. However, to implement this decentralized electric management model, the AI-based automated operation capability and distributed security management are critical. Section\ref{sec:security-AI-layer} clarifies several security concerns in AI and distributed security management. The other way is to build risk-aggregated substation testing cases to assess potential security matters for future defense \cite{YangZhiyuan23}. If a local station is flooded, the mobile transformer or backup one from the emergency stocking plan can be used. 
    
\begin{figure}[t]
    	\begin{center}
    		\includegraphics[width=1\linewidth]{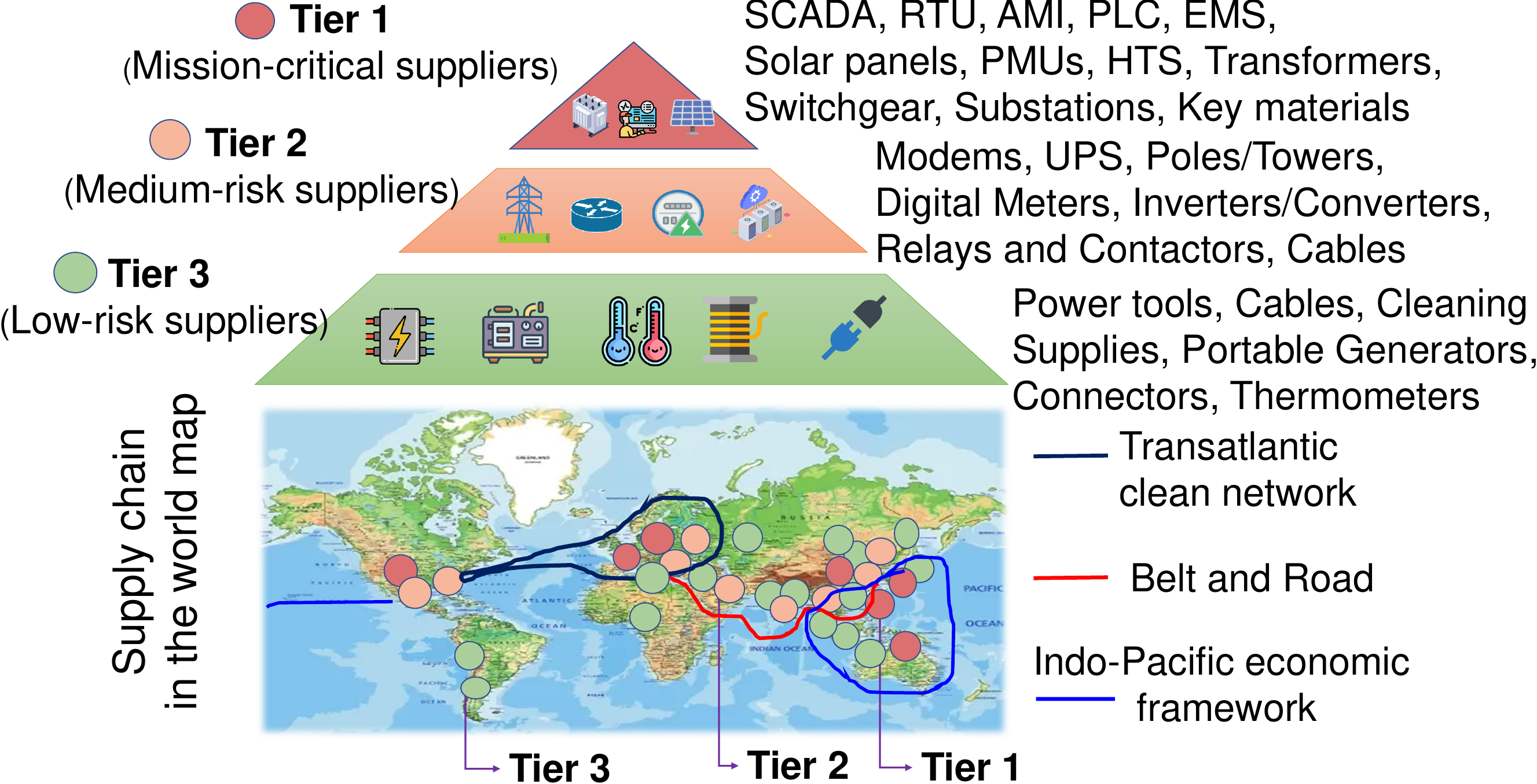}
    	\end{center}
    	\centering
    	\caption{The hierarchy model of suppliers by their important role in the supply chain's resilience and sustainability. Tier 1's suppliers are critical to the supply chain while tier 2 and tier 3 companies are medium-to-low risk suppliers. There are three typical supply chain model examples through trade initiatives: Transatlantic Clean Network \cite{TCN} (now a part under U.S.-E.U. Trade and Technology Council supervision), Belt and Road Initiative \cite{BRI}, Indo-Pacific Economic Framework (IPEF) \cite{IPEF}. }
    	\label{fig:supply-chain}
    \end{figure}

\begin{figure}[t]
    	\begin{center}
    		\includegraphics[width=1\linewidth]{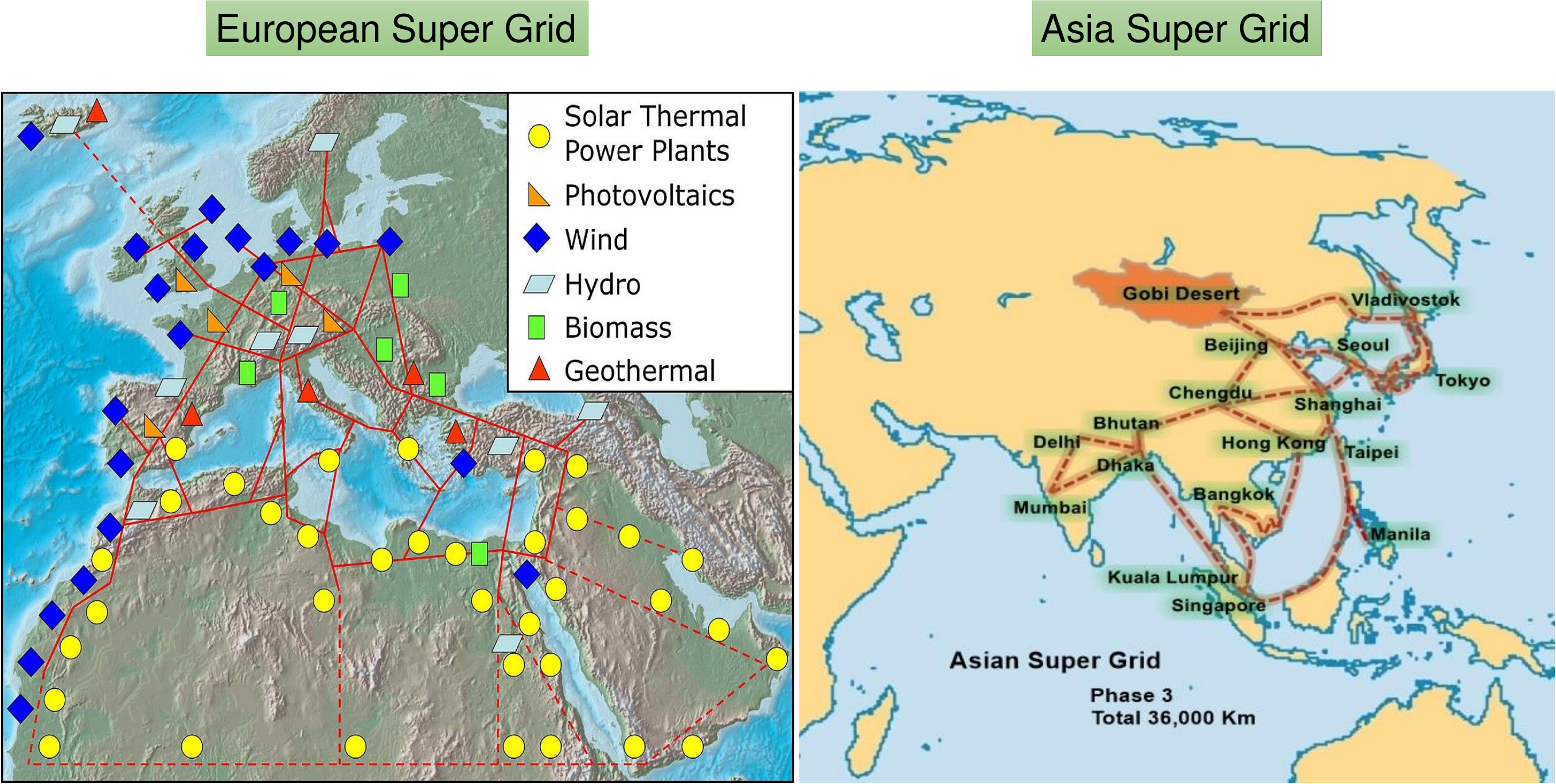}
    	\end{center}
    	\centering
    	\caption{Two conceptual plans for maintaining sustainable energy supplies: a super grid linking renewable sources across North Africa, the Middle East, and Europe \cite{EuropeanSuperGrid}; a super grid to establish an power transmission network connecting China, South Korea, Mongolia, Russia, Japan, and India\cite{AsianSuperGrid}.}
    	\label{fig:super-grid}
    \end{figure}

    To mitigate supply chain attacks, materials/fuel strategic reserves (gasoline, diesel, aviation fuel and propane, copper, rare earth elements) are necessary in certain conditions where post-storm, pandemic, and conflicts may restrict the supply access \cite{SupplychainAttack}. Establishing a varied supply chain consisting of a range of smart grid components and services, organized in a hierarchical list of trustworthy partners, as shown in Fig.~\ref{fig:supply-chain}, may also provide a clear strategy for determining which suppliers should be given priority in order to enhance collaboration and reduce reliance on a single country. The cooperation can upgrade to build intercontinental super grids linking renewable sources, e.g., across North Africa, the Middle East, and Europe \cite{EuropeanSuperGrid} or connecting China, South Korea, Taiwan, Mongolia, Russia, Japan, and India in Asian\cite{AsianSuperGrid} as illustrated in Fig.~\ref{fig:super-grid}. However, national security concerns, country rules, the safety of power transmission routes (e.g., underground), geopolitical and environmental sustainability \cite{RiskSuperGrid}, and expensive capital costs of ultra-high voltage (UHV) power lines (up to 756 or 1100kv) and transformers (up to 800 metric tons) limit this global super grid venture to market demand. Several countries like the US and China are considering building a better distribution system to connect renewable sources in remote areas, e.g., a West-to-East grid in the US, hydropower plants in Mongolia, and solar/wind power plants in the Gobi Desert to the coastal regions in China. 
    
    To mitigate the specific threats against the DERs model with thousands of small producers or microgrids, a multi-layer protection model should be applied. First, government agencies may require regulatory standards and periodic inspections to \textbf{ensure small operators comply with grid codes, hardware integrity, security audits, and operational standards} \cite{NIST}. Further, a  \textbf{real-time monitoring and demand-response system} to ensure that energy generation from private operators aligns with grid demand can be critical. The system can provide price signals and grid conditions to operators to avoid overproduction or underproduction. Any \textbf{violation or abnormal demand-response activity} or energy withholding/oversupply during peak demand can face severe penalties.  \textbf{Local power switch optimization} for grid balancing in energy production from small operators \cite{Leite17} can be helpful, particularly during periods of high or low demand.

   Finally, according to the NIST guidelines \cite{NIST}, saving electricity and building emergency backup power lines can be a high degree of priority if most electricity infrastructure is often damaged by unplanned extreme events or wars. The resiliency in this case requires well-planned information to the local communities and integration of the grid infrastructure with restoration plans in emergency situations. Further, electricity restoration for critical infrastructures (hospitals, nuclear plants) can be then prioritized. Regular national drills and employee training for countering security threats and power outage scenarios are extremely helpful in guaranteeing grid security.

 \subsection{Remaining challenges}
    
   There are many remaining challenges to preventing security threats against grid operators in \ac{SG2}. One of the biggest challenges is to protect substations, transformers, and critical transmission lines against intentional physical attacks, such as vandalism, sabotage, or theft of critical components. Since those attacks can come from insiders, civilians, or enemies, predicting ``when and where the attacks potentially occur'' remains a complicated technical matter. Equipping anti-drone and sabotage activities for many substations may cause significant capital costs and operational expenditures. Second, many existing power grid infrastructure components are legacy systems with limited or outdated security features (probably up to five or ten years old). Retrofitting or securing legacy systems can be costly and complex. Third, the state-sponsored attack groups may launch cyber-physical attacks (e.g., Stuxnet \cite{Stuxnet}) that exploit vulnerabilities in programmable logic controllers of \ac{SG2} to damage SCADA and PLC systems of power plants. Fourth, many utilities in small operators face resource constraints, including budget limitations, that may hinder their ability to equip \ac{HSM} for all metering devices or backup components (e.g., transformers). Fifth, supply chain vulnerabilities can introduce security risks, as compromised or counterfeit components may be integrated into the grid infrastructure, potentially enabling attacks or equipment failures. Finally, detecting and responding to physical security incidents in real time can be challenging. A delayed response can lead to significant damage or electric disruptions for many households.

\section{Security threats and protection models for communication network providers}
\label{sec:security-network-layer}

 Communication technologies, such as wired and fiber networks, help enable real-time data exchange between various components of the grid, such as power plants, substations, microgrids, and consumer devices. This facilitates enhanced monitoring and control, allowing for better management of energy flow and quicker response to issues like outages. In \ac{SG2}, communication networks support the integration of decentralized renewable energy sources and demand response programs, thereby increasing grid reliability and resilience. However, the increasing dependence of \ac{SG2} on communication technologies poses severe security threats to the reliability and integrity of energy management and distribution systems. Generally, security attacks on communication network providers target vulnerabilities of transmission and security protocols (e.g., wireless, TLS, TCP/IP protocol). This section summarizes typical security threats against the network providers in SG1 vs \ac{SG2} and protection mechanisms. 
 
   \begin{figure*}[t]
 	\begin{center}
 		\includegraphics[width=1\linewidth]{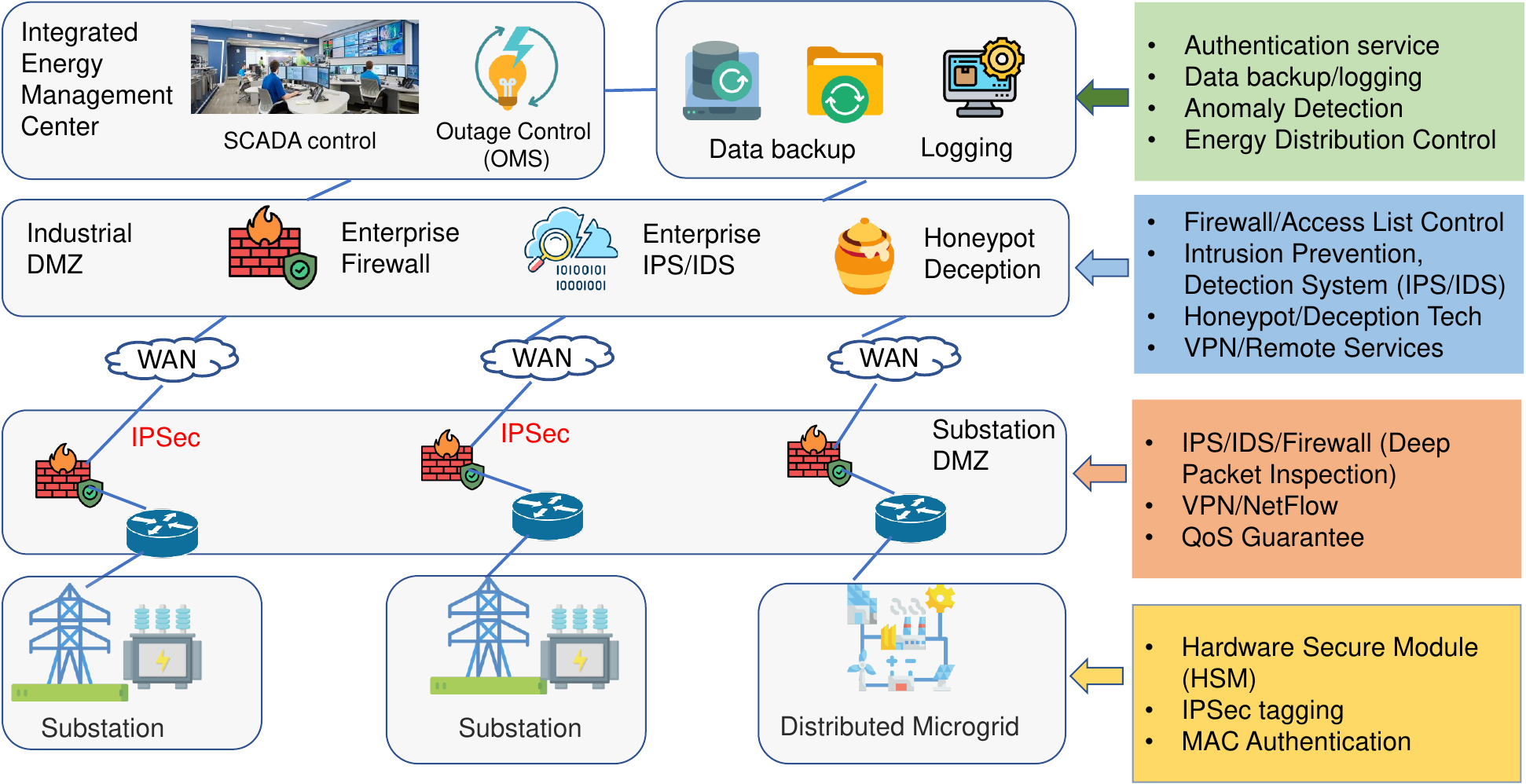}
 	\end{center}
 	\centering
 	\caption{An example of security implementation for smart grid where networks are connected and protected by industrial DMZ, enterprise firewalls, and IPS/IDS platforms.}
 	\label{fig:ics-security-for-power-grid}
 \end{figure*}

 \subsection{Typical security attacks against network providers}
 
  Security threats to network providers mostly come from many vulnerabilities of Internet technologies. Given the diverse range of security attacks on the Internet, we summarize some of the most severe threats that have been recorded via specific damage events and discuss several hints on \ac{SG2}. 

   \textcircled{\raisebox{-0.9pt}{1}} \textbf{Denial-of-service (DoS)}: This attack can overwhelm the communication network infrastructure with a massive volume of traffic, rendering it unavailable for legitimate users and hindering critical communications. Various well-known DoS attacks at the network layer can be low-rate HTTP requests or DNS/NTP amplification. DoS attacks can disrupt the grid's operation by causing power outages or blackouts or disrupting communication between smart grid devices. For example, in March 2019, a DDoS attack hit electrical system operations in Kern County, California, and Converse County, Wyoming, USA and temporarily disrupting the online services \cite{DoSAttacks}. DDoS attacks can lead to a chain breakdown of indirect problems such as inaccurate readings or preventing critical information exchange and real-time grid management \cite{Yan12,Hasan23}.

    \textcircled{\raisebox{-0.9pt}{2}} \textbf{Malware/Ransomware attacks}: Malware is one of the most popular source threats against smart grid infrastructure. Their target is to disrupt or disable grid operations and steal sensitive data. For example, in June 2018, Ingerop company in France was hacked by malware and lost 65 GB of data with 11,000 files from planned nuclear-waste dump projects, details of the Fessenheim nuclear power plant, and personal details of more than a thousand Ingerop employees \cite{MalwareFrench}. Malware can be installed on smart grid devices in various ways, such as phishing emails, infected websites, or USB drives. Ransomware is the other variant of malware. Ransomware encrypts the victim's data and demands a ransom payment in order to decrypt it. The Colonial Pipeline ransomware attack in May 2021 was the largest US energy infrastructure attack \cite{CDW}. The hack gang entered Colonial Pipeline Co.'s network using a stolen VPN password of an employee from a previous data theft. The hackers took 100 GB of data in two hours and threatened to release it. Colonial paid a 75-bitcoin ransom ($4.4$ million USD \cite{CDW}) and restored operations five days later. In a separate incident, Ryuk ransomware encrypted Volue's front-end platforms, impacting 2,000 customers in 44 countries \cite{CDW}. Ransomware attacked City Power, a major electricity supplier in Johannesburg, South Africa, leaving more than a quarter of a million people without power \cite{RansomwareJohannesburg}. Ransomware is also used in many recent attacks, e.g., Enel Group \cite{EnelAttack}.

    \textcircled{\raisebox{-0.9pt}{3}} \textbf{False/spoofing data injection attacks}:  This attack introduces fraudulent information into the network, such as false packets/untrusted MAC sensor nodes, further undermining data integrity and leading to erroneous energy distribution or energy theft \cite{Luo2023,Habib2023}. False/spoofing data may include inaccurate measurements, improper timestamps, or incorrect sender IDs \cite{HOTSEC08}. Further, the attacker can eavesdrop on sensitive data transmissions and manipulate control signals, potentially leading to incorrect grid operations, energy theft, or even physical damage \cite{Jhala21}. The attacker can advance the attacks by impersonating the metering devices' IP addresses/MAC identities to pretend to be legitimate nodes and compromise the authenticity of the communication, enabling unauthorized access and data breaches \cite{Tala22}.

    \textcircled{\raisebox{-0.9pt}{4}} \textbf{Communication link sabotage attacks}: Attackers destroy the communication cables to disrupt the connectivity between smart meters and the grid management centers. On the other hand, copper theft in communication cables is a significant concern. The copper theft cost Lumen, a global communications service provider, up to $500,000$ USD to fix business landlines \cite{CopperThief}. According to S\&P Global Market Intelligence, copper demand will triple by 2035, leading to supply shortages in 2025 \cite{CopperThief}. Copper phone lines and wire thefts have skyrocketed as criminals may sell their hauls for cash. The organized theft rings may target mobile towers and landlines.

     \textcircled{\raisebox{-0.9pt}{5}} \textbf{Indirect attacks}: Besides the above direct attacks (which damage the infrastructure), the hacker can launch the indirect attacks, e.g., manipulate the network routing paths of data, causing data to be directed to unauthorized destinations, potentially exposing sensitive information or causing disruption. Generally, the attack goal is often used to inject malware and ransomware into the grid's internal networks. This attack type has many variants, e.g., routing information spoofing, alteration or replay, blackhole and selective forwarding attacks, sinkhole attacks, and Sybil attacks \cite{Yan12,Tala22}. Another variant is social engineering/phishing which exploits human psychology to trick users into taking actions that are harmful to the grid. For example, an attacker might pose as a customer service representative from an electric distributor and ask users to provide their personal information to update billing information; otherwise, the electricity will be cut off. The hacker can bait users to use discount services and download fake software \cite{khoei2022comprehensive, Zografopoulos2021}. Finally, the remote terminal units in decentralized energy resources (DER) controllers or SICAM substation automation systems can be the targets of stack buffer overflow or firmware vulnerability exploitation \cite{Tala22}.

      Unlike the attacks against the power grid operators, the attacker can remotely launch cyberattacks against the network providers. Besides, individuals in trusted private networks for smart grids could misuse their network access privileges to compromise network security in the other network segments. In \ac{SG2}, the communication technologies are expected to expand for connecting multi-energy generation grids \cite{Khan23}. Major security concerns against this network model in \ac{SG2} will be still the malware/ransomware and blackmail attacks. However, as cyberwarfare becomes integrated into national security strategies and cyberattacks have become a lucrative money-making industry, these attacks are expected to be considerably intensified and, in some instances, supported by governments. Further, networking technologies for specified \ac{SG2} generation, such as renewable energy sources and distributed energy resources, are the new targets.

 \subsection{Specific threats for new communication models in SG2}
 
In SG2, thousand of grid operators and millions of consumers may connect and rely on the connectivity provided by network operators. The complexity of maintaining the connectivity for such a sophisticated grid exponentially increases and building a leased line or dedicated network infrastructure for individual grids can be a challenge. Further, the real-time data exchange and high communication traffic could potentially lead to network congestion, latency, or insufficient bandwidth, affecting the timely management of grid stability. The other threat is poorly synchronized communication among thousands of operators that could lead to mismatches in grid data, causing imbalances in energy distribution. In such complicated and diverse communications, five attacked identified in the previous subsection (e.g., DoS, communication link sabotage \cite{CopperThief}, false data injection \cite{Luo2023,Habib2023}, ransomware \cite{CDW}) are expected to be more intensive. Potential other attacks can be routing attacks (blackhole) and Sybil attacks \cite{Yan12,Tala22} (fake multiple grid operators from a single identity to create false coordination). The attacks could result in imbalanced grid operations or unreliable energy trading. Finally, the use of lightweight communication protocols for IoT-enabled SG2 like MQTT \cite{Abir2021,AKKAD2023} may introduce security gaps (without using TLS protocols) if they are not properly secured from data tampering and transaction interception \cite{Abir2021}.

\subsection{Protection technologies for network providers in \ac{SG2}}

 Securing the communication networks requires a comprehensive approach of multiple protection models at different network layers, as illustrated in Fig.~\ref{fig:ics-security-for-power-grid}. Ideally, for each essential component in \ac{SG2}, the protection models involve a DMZ (e.g., substation/industrial DMZ). The DMZ is integrated with many advanced protection techniques, e.g., IPS/IDS, authentication, honeypot, at ingress/egress routers/gateways.  Table~\ref{tab:network-security-changes} summarizes several key protection models (the fourth column) for five security attacks above. An approach, e.g., IDS/IPS, can be used for prevent various threats.

\begin{figure}[t]
    		\begin{center}
    			\includegraphics[width=1\linewidth]{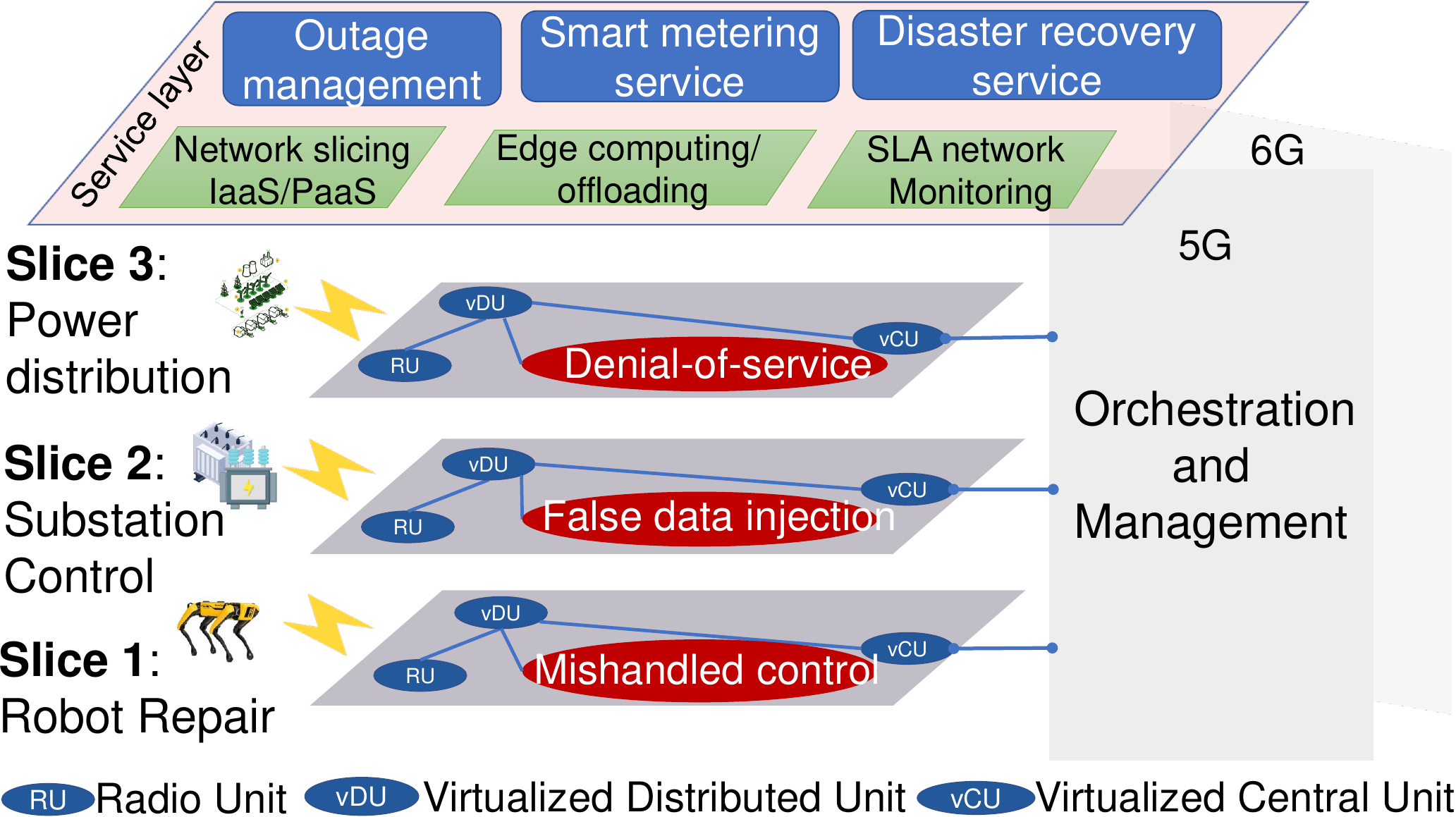}
    		\end{center}
    	\centering
    	 	\caption{Illustration of a 5G network slicing concept for \ac{SG2} \cite{5GSmartGrid} where each service, depending on their security requirements, can be dedicated in a slice to protect the communication secrecy. For example, communications between substations and the remote center or between power distribution control sensors and the center.}
    	 	\label{fig:5g-slicing-IEN}
      \end{figure}

 \textcircled{\raisebox{-0.9pt}{1}} \textbf{Network slicing}: In this paradigm, \ac{SG2} communications are technically separated from civil networks \cite{TR33.811}. According to this approach, any possible security interference/breach that happens in one slice will not affect the others. Ideally, \ac{SG2} communications, particularly \ac{SCADA} and substation networks, are isolated from the Internet through private communication links or VPN connections. Fig.~\ref{fig:5g-slicing-IEN} illustrates a 5G network slicing concept for \ac{SG2} where different grid services are expected to be isolated based on their service priority and response time requirement. Accordingly, end-to-end network slicing for \ac{SG2} \cite{5GSmartGrid} is a promising technique. It is vital in network slicing to define isolation characteristics for \ac{SG2} individually, specify KPI criteria, and enforce them. A high-volume DoS attack, on the other hand, may make this exceedingly difficult to contain. 
 Defining isolation properties for each slice, as well as defining and enforcing KPI standards in smart grid environments, is crucial. A network slice manager, such as Network Slice Management Function (NSMF) \cite{TR33.811}, is expected to be in charge of abstract virtual network functions. 

  \textcircled{\raisebox{-0.9pt}{2}} \textbf{Virtual local area networks (VLAN)}: Similar to network slicing, however, instead of applying for wide area communication networks, VLANs help isolate different types of traffic in a single physical communication network infrastructure. For instance, in a \ac{SG2}, VLAN 10, VLAN 20, VLAN 30, and VLAN 40 are assigned for the communication networks of power generation plants, substation control systems, administrative offices, and backup/recovery systems, respectively. The isolated communication networks logically prevent unauthorized access against unassigned communication network areas, limiting the impact of potential cyberattacks \cite{Zhang24}. For example, control teams in power distribution and staff for equipment/communication link maintenance can be placed on different VLANs, ensuring that only authorized personnel can access critical operational systems. However, while VLANs can offer some security benefits by isolating sensitive data and limiting the spread of broadcast domains, the complexity of configuring large-scale VLANs can bring up new threats. For example, the assigned network segments may access each other if there is a misconfiguration  \cite{Zhang24}. In this way, relying solely on VLANs for security can be misleading. For robust network security, VLANs should be used in conjunction with other security measures such as authentication, IDS/IPS, and strong access control policies.

      
  \textcircled{\raisebox{-0.9pt}{3}} \textbf{Secure control and communication protocols}: For network access control, stronger authentication between energy distributors and the consumers is required to prevent spoofing attacks and man-in-the-middle attacks. For example, 5G \ac{AKA} and future 6G \ac{AKA} \cite{Nguyen21} will be core services to unify the authentication for mobile users and IoT metering devices, although the metering subscriber identity (i.e., National Meter Identifier) must likely synchronize with mobile identifiers (eSIM). In this way, users can use one unified identity with a single point of access to manage all electricity service fees, car charging, battery swap, or P2P electricity trading. Users benefit from this single sign-on (SSO) capability, allowing them to access multiple systems with one set of credentials, thereby improving convenience and reducing password fatigue. Unified authentication generally facilitates interoperability among different energy providers, enabling the integration of multi-energy generation grids and optimizing cost in peak time. This interoperability is crucial as the smart grid expands, enabling efficient load sharing of new energy sources. This unified model also helps simplify the administration of user credentials and the authentication protocol upgrading process for all devices. 
  
  However, a private \ac{PKI} infrastructure for \ac{SG2} authentication \cite{CEN-CENELEC-ETSI} can still be necessary to utilize secure communications for special services (defense industry). Further, cryptographic algorithms in secure communication protocols such as Internet Protocol Security (IPsec) and Transport Layer Security (TLS) must be adopted to prevent unauthorized interceptors from accessing communications. Toward the age of quantum computing, adopting quantum-safe cryptographic schemes (e.g., extending the key length of AES-128 to AES-256/AES-512) \cite{ETSIQuantum}, quantum-resistant algorithms (e.g., NTRU/AES) or Quantum Key Distribution schemes \cite{Lakshmi23,Kong2022} can be a benefit for communications between SCADA and substations. End-to-end encryption and decreasing the cost of protection (e.g., energy consumption, deployment cost) are critical challenges for next-generation communication security, which will serve as the foundation for \ac{SG2} communications. Because metering data transfer between substations might be enormous, adopting end-to-end encryption may be prohibitively expensive. Currently, 3GPP and standards organizations advocate the optional use of security measures based on their capacity to fulfill necessary services. If \ac{SG2} needs mandatory end-to-end encryption, it is unclear how to implement this requirement over the multiple communication subnetworks, given their co-existence of legacy and modern technologies.

   \textcircled{\raisebox{-0.9pt}{4}} \textbf{Advanced endpoint security and intrusion detection systems}:  As shown in Fig.~\ref{fig:ics-security-for-power-grid}, endpoints play a guard role in protecting communication gateways and the borders of the core network infrastructure of the smart grid (i.e., substation DMZs). To guard against external threats, \ac{SG2} core networks have historically required DMZ and control nodes. To prohibit unwanted traffic from core network parts, a security gateway may examine bi-directional traffic against the operation rules. The types of security gateways include IDS, service-oriented architecture API protection, antivirus programs, VPN, and so on. In a traditional smart grid, as illustrated in Fig.~\ref{fig:ics-security-for-power-grid}, security gateways located at substation DMZ and Industrial DMS are responsible for inspecting all traffic between Internet and Industrial Industrial Control System (ICS) \cite{3GPP33501,ETSI.TS133.501}. Such gateways will need to enhance their capacity in \ac{SG2} significantly. Many experts anticipate that the enhanced capabilities for the next-generation endpoint security will include (1) in-line deep packet inspection (DPI), (2) encrypted traffic inspection, (3) next-generation firewall, (4) next-generation intrusion detection/prevention, and (5) unified identification \cite{Nguyen21}. Current AI-powered engines will need significant improvements in detection capabilities, such as higher ability for online traffic training and less impact from imbalanced datasets, as well as resilience for defending varied communication protocols. As suggested by artificial general intelligence, one viable option is to increase the generative learning capabilities of deep learning models \cite{ALDWEESH2020105124,SILVER2021103535}. 

  \begin{table*}[ht]
 	\caption{Comparison of Security threats and protection models for the communication network provider in SG1 vs SG2}
 	\label{tab:network-security-changes}
 	\begin{adjustbox}{width=1\textwidth}
 		\begin{tabular}{|l|l|l|ll|l|l|l|l|}
 			\hline
 			\rowcolor[HTML]{EFEFEF} 
 			\cellcolor[HTML]{EFEFEF} & \cellcolor[HTML]{EFEFEF} & \cellcolor[HTML]{EFEFEF} & \multicolumn{2}{c|}{\cellcolor[HTML]{EFEFEF}\textbf{Protection model}} & \cellcolor[HTML]{EFEFEF} & \cellcolor[HTML]{EFEFEF} & \cellcolor[HTML]{EFEFEF} & \cellcolor[HTML]{EFEFEF} \\ \cline{4-5}
 			\rowcolor[HTML]{EFEFEF} 
 			\multirow{-2}{*}{\cellcolor[HTML]{EFEFEF}\textbf{Security threats}} & \multirow{-2}{*}{\cellcolor[HTML]{EFEFEF}\textbf{Severity}} & \multirow{-2}{*}{\cellcolor[HTML]{EFEFEF}\textbf{Likelihood}} & \multicolumn{1}{l|}{\cellcolor[HTML]{EFEFEF}\textbf{SG1}} & \textbf{SG2} & \multirow{-2}{*}{\cellcolor[HTML]{EFEFEF}\textbf{Efficiency}} & \multirow{-2}{*}{\cellcolor[HTML]{EFEFEF}\textbf{Deployment cost}} & \multirow{-2}{*}{\cellcolor[HTML]{EFEFEF}\textbf{Open issues}} & \multirow{-2}{*}{\cellcolor[HTML]{EFEFEF}\textbf{Reference}} \\ \hline
 			&  &  & \multicolumn{1}{l|}{IDS/IPS/MTD} & \begin{tabular}[c]{@{}l@{}}Intelligent IDS/IPS/MTD\\ (Superintelligence)\end{tabular} & High & Low & \begin{tabular}[c]{@{}l@{}}- High complexity\\ - Deal with encrypted data\\ - Comprehensive dataset\\ for SG2\end{tabular} &  \cite{Zhang23} \\ \cline{4-9} 
 			\multirow{-2}{*}{\begin{tabular}[c]{@{}l@{}}Denial of service\\ (HTTP, NTP, DNS)\end{tabular}} & \multirow{-2}{*}{High} & \multirow{-2}{*}{High} & \multicolumn{1}{l|}{} & Distributed microgrid & High & Medium & - Complicated management & \cite{Nguyen21} \\ \hline
 			&  &  & \multicolumn{1}{l|}{IDS/IPS} & \begin{tabular}[c]{@{}l@{}}Intelligent IDS/IPS\\ (new AI models)\end{tabular} & High & Low & \begin{tabular}[c]{@{}l@{}}- Adversarial attacks\\ - Deal with fragment \\ malware\\ - Lack of rigorous datasets\end{tabular} &  \cite{MalwareFrench} \\ \cline{4-9} 
 			&  &  & \multicolumn{1}{l|}{Layer 3 VLAN} & Layer 3/7 VLAN & Medium & Low & \begin{tabular}[c]{@{}l@{}}- Misconfiguration\\ - Complicated management\\ - Limited VLAN IDs\end{tabular} &  \cite{Zhang24} \\ \cline{4-9} 
 			&  &  & \multicolumn{1}{l|}{Network isolation} & Deep slicing & Medium & High & \begin{tabular}[c]{@{}l@{}}- High cost\\ - Scalability\end{tabular} & \cite{TR33.811,5GSmartGrid} \\ \cline{4-9} 
 			\multirow{-4}{*}[20pt]{Malware/ransomware} & \multirow{-4}{*}{High} & \multirow{-4}{*}{High} & \multicolumn{1}{l|}{Firewall/DPI} & Intelligent Firewall/DPI & High & Low & - Adversarial attacks &  \cite{ALDWEESH2020105124,SILVER2021103535}\\ \hline
 			&  &  & \multicolumn{1}{l|}{Abnormal detection} & \begin{tabular}[c]{@{}l@{}}Intelligent IDS/IPS\\ (new AI models)\end{tabular} & High & Low & \begin{tabular}[c]{@{}l@{}}- Comprehensive dataset \\ for SG2\end{tabular} & \cite{Hasan23} \\ \cline{4-9} 
 			&  &  & \multicolumn{1}{l|}{Authentication} & Unified authentication & High & Low & \begin{tabular}[c]{@{}l@{}}- Single point of failure \\ - Provider acceptance\end{tabular} &  \cite{Nguyen21} \\ \cline{4-9} 
 			\multirow{-3}{*}{\begin{tabular}[c]{@{}l@{}}False data injection\\ Impersonation\end{tabular}} & \multirow{-3}{*}{High} & \multirow{-3}{*}{Medium} & \multicolumn{1}{l|}{\begin{tabular}[c]{@{}l@{}}Encryption\\  (TLS 1.3)\end{tabular}} & \begin{tabular}[c]{@{}l@{}}Blockchain \\ (Quantum-safe TLS)\\ Trusted networks\end{tabular} & High & Medium & \begin{tabular}[c]{@{}l@{}}- High computation, scalability\\ - Zero-Trust implementation\end{tabular} & \cite{Khan23,Habib2023} \\ \hline
 			& High & Medium & \multicolumn{1}{l|}{\begin{tabular}[c]{@{}l@{}}Educate public\\ on theft consequence\end{tabular}} & \begin{tabular}[c]{@{}l@{}}Educate public \\ on theft consequence\end{tabular} & Medium & High & \begin{tabular}[c]{@{}l@{}}- Lack of awareness on\\ high severity, sustained efforts \\ with many partnerships.\end{tabular} &   \cite{CopperTheftLawsuit}\\ \cline{2-9} 
 			\multirow{-2}{*}[10pt]{\begin{tabular}[c]{@{}l@{}}Communication\\ link sabotage \\ (copper theft)\end{tabular}} &  &  & \multicolumn{1}{l|}{Legal actions} & Legal actions & High & High & \begin{tabular}[c]{@{}l@{}}- Detection and damage proof\\ - Available law provision\end{tabular} & \cite{CopperThief,CopperTheftLawsuit} \\ \hline
 			Social engineering & High & High & \multicolumn{1}{l|}{\begin{tabular}[c]{@{}l@{}}Educate public\\ Identity verification\end{tabular}} & \begin{tabular}[c]{@{}l@{}}Educate public/\\ Identity verification\end{tabular} & Medium & Medium & \begin{tabular}[c]{@{}l@{}}- Fast changing phishing tactic\\ - Sustained efforts to teach\end{tabular} &  \cite{khoei2022comprehensive, Zografopoulos2021}\\ \hline
 			\begin{tabular}[c]{@{}l@{}}Indirect attacks\\ (jamming/Sybil)\end{tabular} & Medium & Low & \multicolumn{1}{l|}{IDS/IPS} & \begin{tabular}[c]{@{}l@{}}Intelligent IDS/IPS\\ (physical layer)\end{tabular} & Medium & Low & \begin{tabular}[c]{@{}l@{}}- High complexity\\ - Unclear efficiency\end{tabular} &  \cite{Hasan23} \\ \hline
 		\end{tabular}
 	\end{adjustbox}
 	\begin{tablenotes}
 		\item \textbf{Severity/Likelihood}: \textbf{High} (occurred and caused damage/finance loss); \textbf{Medium} (occurred but the damage is small; \textbf{Low} (threats tested in research lab))
 		\item \textbf{Deployment cost}: \textbf{High} (require many extra devices and efforts to implement); \textbf{Medium} (require several extra devices or efforts); \textbf{Low} (Easy to deploy)
 	\end{tablenotes}
 \end{table*}

    \textcircled{\raisebox{-0.9pt}{5}} \textbf{Deep packet inspection and next-generation firewall}: Along with Endpoint Security Gateways, Deep Packet Inspection (DPI) and Firewall are vital components to protect substation DMZs, as illustrated in Fig.~\ref{fig:ics-security-for-power-grid}. DPI technologies provide advanced packet analysis capability (e.g., analysis of packet headers and protocol fields of source data, sending frequency, network protocols, malicious payload patterns) for distributed microgrids and substations \cite{Hasan23}. This capability is particularly useful, given the diversity of the management models in \ac{SG2} microgrids. On the other hand, stateful firewalls offer additional features like the smart metering application or buffer overflow attack detection, fragmented worm or intrusion detection, and access list control to prevent vulnerability penetration or keep unauthorized users out, particularly in \ac{SG2} microgrids \cite{Mohamed21,WANG2023108889}. The primary challenge is that many next-generation security designs are still only concepts. For example, many AI-driven DPI and firewall technologies also struggle with privacy preservation in training/learning, scalable online learning, false alarms, and vulnerability to many adversarial attacks. 
    
     \textcircled{\raisebox{-0.9pt}{6}} \textbf{Physical attack defense solutions}: Unlike substation protection, affordable and effective protection (e.g., camera, fencing) against copper theft and communication cable sabotage attacks is a challenge for communication network providers since communication cables are often placed in many residence areas. One of the most effective methods is to educate the public about the impact of copper theft on critical infrastructure and encouraging them to report suspicious activities can enhance community vigilance \cite{ElectricGridSecurity}. The other is to build backup lines to prevent cascading failures, as introduced in Section~\ref{sec:security-physical-layer}.B and Fig.~\ref{fig:cascading-failure}. In the worst case, several companies consider pursuing legal action against thefts and scrap metal businesses that knowingly purchase stolen copper. For example, Bell Canada has taken legal action against individuals stealing copper wire from their communication networks. The thefts typically involve cutting down telecommunication cables from poles and selling the copper for cash, with each incident taking 10 to 12 hours to repair and significantly impacting service reliability \cite{CopperTheftLawsuit}.    

    \textcircled{\raisebox{-0.9pt}{7}} \textbf{Emerging defense solutions:} Software-defined Networking is deploying widely over the Internet. By separating the packet forwarding (data plane) and routing (control plane) operations, SDWAN attempts to increase network control performance and intelligence. In \ac{SG2}, SDWAN is expected to be utilized for power utility distribution automation \cite{Jakaria21,Al-Rubaye19} . DoS/DDoS attacks and insider adversaries are the most severe risks to SDN/SD-WAN. The most common mitigation strategies are to exploit AI models to classify abnormal traffic in \ac{IDS} \cite{Tala22} and \ac{MTD} \cite{Zhang23}. If an anomaly is discovered, the detection system may instruct the SDN controller to rewrite the data plane (programmable switches) to reduce the magnitude of the attack \cite{Cunningham23}. A promising solution is to use MTD-based solutions by masking or altering critical network features (e.g., true IP addresses) on a regular basis to avoid DDoS attacks and reconnaissance scans. Secure access service-edge (SASE) architecture \cite{WOOD20206} can be used to provide protection for SDWAN in  \ac{SG2}. Another promising technology is substation virtualization \cite{Robert22}, where critical grid management applications are virtualized into the cloud or edge servers. Network functions and control applications can be optimized and deployed quickly at the stability and scalability advantage of the cloud. 
    
    In a complex network to connect thousands of grid operators, to prevent cyberattacks, all strategies above should be included properly, from enforcing strong encryption and mutual authentication to secure data and identities; deploying IDS/IPS/DPI/DMZ \cite{Mohamed21,WANG2023108889} to detect and mitigate threats like DDoS; implementing network segmentation (VLAN, VPN) and slicing \cite{5GSmartGrid} to isolate and protect grid communication, adopting blockchain node for each grid operator, and ensuring regular security updates for IoT devices \cite{IEC}. Additionally, other classical methods, such as security audits, adaptive synchronization protocols \cite{3GPP33501,ETSI.TS133.501}, and multiple communication links \cite{Khan23}, can still enhance overall grid resilience.

\subsection{Remaining challenges}

  Table~\ref{tab:network-security-changes} summarizes the potential changes of security threats and defense solutions against the communication network providers in the smart grid and \ac{SG2}. One of the biggest remaining challenges for security matters for communication technologies is standardizing and implementing commercial post-quantum cryptographic schemes. This information security model will impact not only the security communication protocols in \ac{SG2} but also the Internet, given its essential role in every communication. Furthermore, emerging technologies such as deep slicing and blockchain technology have the potential to be game changers, but their complexity and high energy consumption may make widespread adoption difficult. Furthermore, if the technologies' overhead computation and security vulnerabilities are not addressed, they will most likely be disregarded. However, seamless upgrades for many networking technologies at the same time pose significant dangers. One such option is to include AI in the update automation \cite{NISTResponsibleAI} (e.g., federated learning-based automated firmware/protocol version upgrade systems for thousands of routers, switches, and firewalls in the networks of distributed grids and substations). However, given the numerous known adversarial attack concerns of AI-based models (e.g., evasion attacks, data poisoning, model poisoning attacks, and label-flipping attacks in federated learning \cite{NISTResponsibleAI, alsulaimawi2024}), it is questionable if AI can considerably enhance the situation or, on the contrary, it makes the poisoned updates spreading faster. In this case, the integration of blockchain for protecting the integrity of AI models and trustable AI may help.

\section{Security threats and protection models for consumer/end-users}
\label{sec:security-application-layer}

Consumers in \ac{SG2} means individuals, households, public infrastructure, or commercial groups consume electricity generated by the grid. Generally, since the consumer types are diverse, security defense capabilities vary. In the traditional power grid, given that there is no major financial reward if the attack is successful, hackers may be less motivated to target people or families consumers for money than for retribution or other wicked purposes. Commercial electric consumers, such as public transportation and industrial complexes, are the favorite hacking targets. However, in SG2, with the rapid change to distributed generation, heating of residential homes with heat pumps, need for charging EVs and UAVs, home batteries, households are rapidly becoming the ideal targets. For example, sudden charges or discharge of energy from millions of connected households and EVs from rerouting energy flow or malicious energy demands of compromised vehicles can be a threat to the stability of grid operations. Generally, the small consumers are not professionally secured with enterprise firewalls and intrusion prevention systems to prevent attacks.  In this case, remote home batteries and connected EV cars can become targets for malware, potentially being turned into zombie systems to carry out malicious tasks under the control of a remote hacker. This section summarizes several typical threats and protection models for consumers.

\subsection{Typical security attacks against the consumers}

Security threats to consumers mostly come from vulnerabilities of smart metering devices and internal information technology (IT) networks (of commercial consumers). Based on the literature, we summarize several typical attacks and threats as follows.

\begin{figure}[t]
	\begin{center}
		\includegraphics[width=1\linewidth]{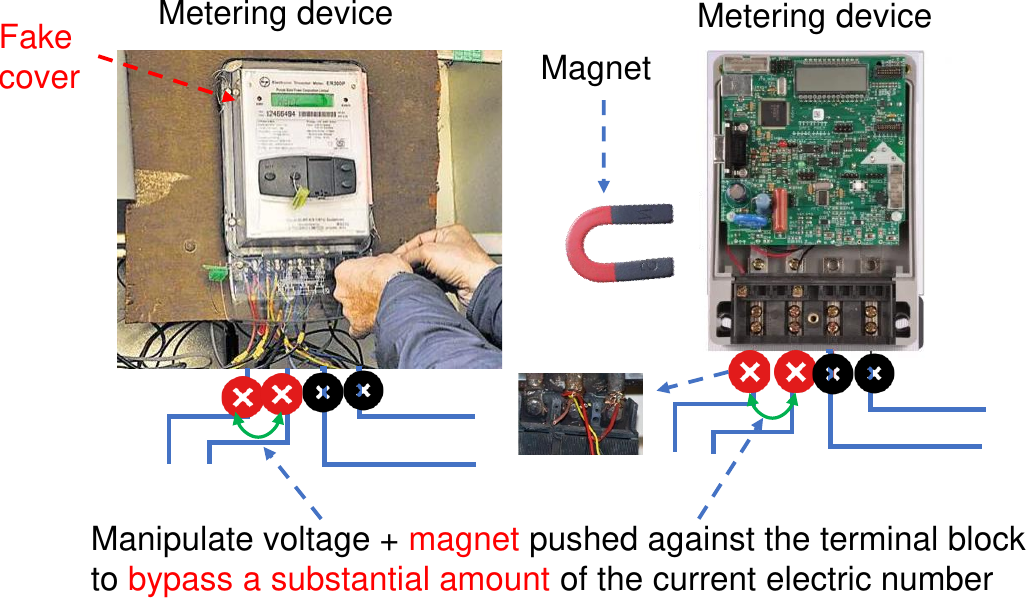}
	\end{center}
	\centering
	\caption{The illustration of a fake cover and smart meter tampering attack where magnets and voltage interference can bypass energy numbers by the metering device \cite{MeteringTampering}. }
	\label{fig:PHY-tampering}
\end{figure}

 \textcircled{\raisebox{-0.9pt}{1}} \textbf{Metering device interference/Energy theft}: Attackers destroy the communication cables to disrupt the connectivity between smart metering sensors and the grid management centers. These sabotage attacks can lead to sensor/link failures, prevent the \ac{SG2} from responding to operational demands, and potentially cause economic loss or electric disruption. The other variants can be time synchronization attacks or smart meter tampering attacks to bypass counting and electricity theft \cite{Zheng18} and cut out power bills \cite{SHOKRY2022358}. As illustrated in Fig.~\ref{fig:PHY-tampering}, the attacker can illegally use metal objects (magnets) to push against the terminal block and bypass a substantial percentage of power bills \cite{MeteringTampering}. Generally, a smart electricity meter calculates energy consumption by measuring voltage between the input and neutral lines, and current across a shunt between the input and output lines. The total energy consumed is derived from the sum of the voltage-current products over time. This tampering method includes reversing the line connections, which can result in negative power readings, and using metal objects to bypass the current sensor, thus reducing the recorded power. Another method involves applying magnets to affect transformers and current sensors, leading to lower power readings. The attacker can also use a fake cover to mislead electric readers. The physical attacks against a meter can be changing code or retrieving a pre-shared key (PSK) (e.g., in LoRaWAN-based) to clone the meter \cite{Nguyen21}. 

 \textcircled{\raisebox{-0.9pt}{2}} \textbf{Social engineering}: Social engineering attacks are designed to exploit human psychology to trick users into taking actions that are harmful to the grid. For example, an attacker might pose as a customer service representative from an electric distributor and ask users to provide their personal information to update billing information; otherwise, the electricity will be cut off. Also, an attacker impersonates a maintenance worker or technician from the utility company and contacts users, claiming they need remote access to the user's smart meter for maintenance purposes. The hacker can also bait users into using discount services and download fake software \cite{khoei2022comprehensive, Zografopoulos2021}. All personal details can be used to launch secondary attacks, e.g., using the stolen credentials to log into the consumer's smart meter account, manipulate energy usage data, or gain access to broader network segments within \ac{SG2}.

 \textcircled{\raisebox{-0.9pt}{3}} \textbf{Malware attacks}: Malware attacks mostly target commercial consumers. For example, Industroyer is a sophisticated malware designed specifically for industrial control systems (ICS) in industry complexes \cite{Industroyer}. Industroyer has multiple components, including a backdoor, data wiper, and payloads that control electric circuit breakers, causing power outages by repeatedly opening and closing breakers. The malware communicates with its command and control servers over the Tor network, making it difficult to detect and mitigate. Some variants from Industroyer exploited CVE-2015-5374 vulnerabilities to launch a denial-of-service (DoS) and cause Siemens SIPROTEC relays unusable \cite{Industroyer}. Malware can be installed in the consumers' IT networks in various ways, such as phishing emails of online electric bills or from compromised smart metering devices. Note that if consumers' smart meter devices do not often support two-way communications or direct data write capabilities, spreading malware via compromised smart metering devices is challenging.

    Security threats against consumers in \ac{SG2} are a significant concern, particularly due to the varying levels of protection among different consumer groups. Individual and household consumers often utilize less well-protected methods (e.g., firewalls, IDS, barriers), making them more vulnerable to attacks. The attack damage can be reduced if the consumer devices do not support bidirectional communications as in SG1. Devices that only support one-way communication frequently send data to a remote server but do not receive commands or updates from it, limiting the potential for malware to propagate or for attackers to manipulate the devices directly. For commercial consumers, such as industrial complexes and science parks, the stakes are higher. These entities typically rely on more sophisticated metering devices that support bidirectional communications. While this capability allows for more efficient and responsive local energy management, it also opens up more vectors for potential attacks, e.g., malware/ransomware, DoS. The attacks can cause the consumer's IT systems substantial operational disruptions and safety hazards \cite{McDaniel09}. 

 \subsection{Specific threats for the new consumer models in SG2}
 
  Unlike in SG1, consumers in SG2 can also play the role of the grid providers, i.e., both consume and produce energy. With the rise of DERs, such as solar panels, wind turbines, home energy storage systems \cite{Trevizan22}, and electric vehicles (e.g., battery-based), consumers can generate their own electricity and feed surplus energy back into the grid. Big prosumers (e.g., a factory with solar/wind turbine farms) can adjust their energy usage in response to grid demands, helping balance load and maintain grid stability during peak periods or energy shortages (e.g., for households in extreme weather or crisis situations). A well-coordinated plan with many prosumers (e.g., big prosumers. millions of EVs, home portable electric generators) can reduce grid congestion, provide local energy solutions for households, and improve grid resilience (against physical attacks, grid outage) \cite{Nazemi21}. However, \textbf{compromised consumers and prosumers} could also target demand response systems, \textbf{artificially inflating or reducing consumer demand signals} to disrupt grid stability or manipulate energy prices for financial gain \cite{HUEROSBARRIOS2022100620}.  As consumers in SG2 gain more control over energy production (e.g., via solar panels) and consumption, uncoordinated energy flow or production can cause grid imbalances if not properly managed or synchronized with the grid. Besides, \textbf{energy theft and fraud} through altering consumption data, inaccurate billing, or fraudulent energy trading practices, can become common, given the difficulty of managing thousands of contracted prosumers. Further information on the specific threats in energy trading models are detailed in Section~\ref{subsec:energy-trading}. 
 
 \subsection{Specific security protection models for the consumers}

   In essence, protecting the consumer requires several comprehensive techniques. For individual consumers, physical security measures such as tamper-evident seals and secure meter enclosures can help prevent unauthorized access. Regular inspections and monitoring can detect any signs of tampering early. For commercial consumers, \textbf{multiple protection layers} may require authentication, data encryption, application security protocols, firewalls, service identity access management, operation/kernel systems reinforcement, abnormal behavior detection, and so on. For example, industrial consumers may need an IT team and should be educated about recognizing phishing attempts and the importance of not sharing sensitive information. Implementing \textbf{multi-factor authentication (MFA)} can add an additional layer of security. The data transmission and secure communications in smart electricity-related tasks should follow IEC 62056-21, an international standard for reading utility meters \cite{Mall2022}. 
   	
\begin{figure}[t]
    		\begin{center}
    			\includegraphics[width=1\linewidth]{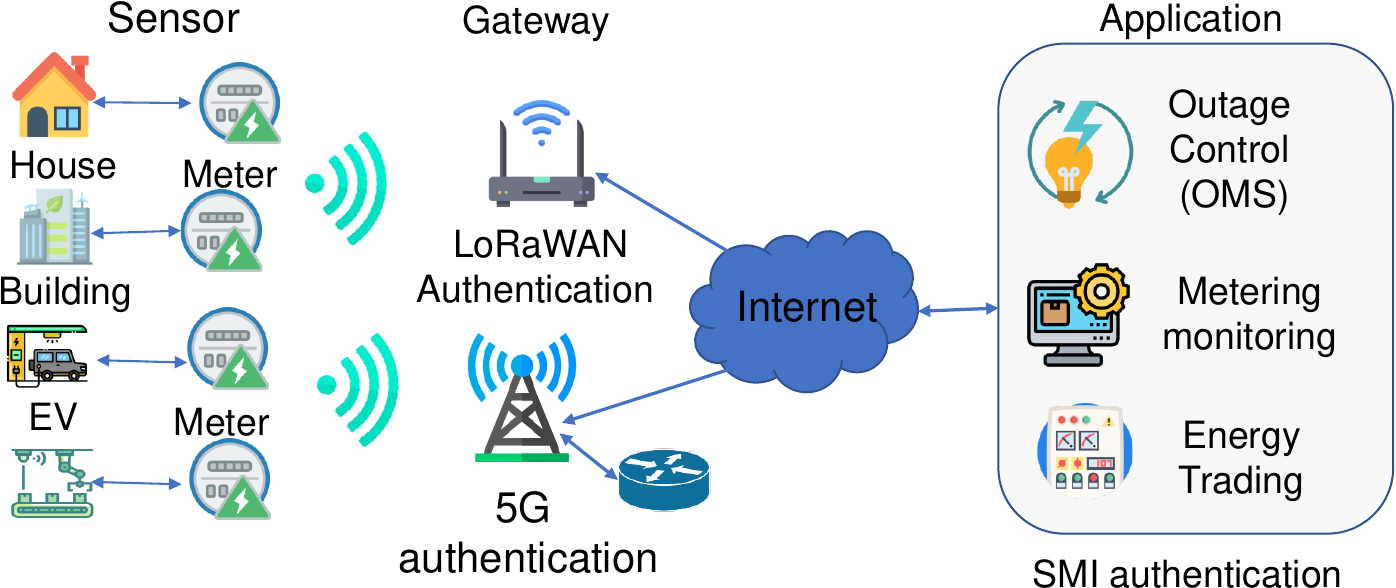}
    		\end{center}
    	\centering
    	 	\caption{The illustration of multiple authentication mechanisms for carrying data from smart metering in smart grids. The authentication infrastructure will be tailored to the communication technologies equipped for meter devices. }
    	 	\label{fig:lora-authentication}
      \end{figure}

   The connections between commercial customers' smart meters and the grid supplier can be authenticated through the PKI or custom \textbf{authentication mechanisms} of network carriers, e.g., LoRaWAN gateway, as illustrated in Fig.~\ref{fig:lora-authentication}. The credentials used in authentication can vary, such as pre-shared keys/tokens/certificates. For 5G-based smart grids, the authentication is based on 5G PKI architecture. However, each smart meter or end device is required to be equipped with a SIM card or eSIM (which may cause a high cost). Authentication, among other services in smart grids, e.g., camera surveillance and sensor monitoring, is also followed by custom models \cite{Badar2021}. To prevent energy theft from consumers, several emerging technologies have been promising to be used in smart grids, e.g., blockchain \cite{Bera21,Mollah2021,MONIRUZZAMAN2023109111} or zero-sum peer-to-peer transaction settlement verification \cite{Jia22}. This can create an immutable and transparent record of energy transactions and prevents fraudulent alterations to consumption records and ensures accountability for energy usage. For secure communications, \textbf{encrypting all data transmissions} can protect against interception. Implementing secure communication protocols such as transport layer security (TLS) can further enhance security. In \ac{SG2}, AES can be upgraded with a new key length, e.g., 256 bits. WiFi or Ethernet-based HAN may still rely on TLS 1.3 \cite{Nguyen21}.

   \textbf{Firewalls, intrusion detection systems (IDS), and abnormal behavior detection} can help detect and mitigate DoS attacks and malware/ransomware targeting commercial consumers' communication networks. Implementing machine learning algorithms to analyze smart meter data for anomalies and alerting authorities are the hot topics currently. Further, with the trend toward virtualization of networking technologies, the security-as-a-service (SECaaS) paradigm will be crucial. In this paradigm, service prosumers or energy distributors may contract with Internet security providers to do all security services (malware detection, cyber attack prevention). The other is the enhancement of AI models to allow proactive learning and response against numerous risks \cite{ZHOU2022,LI20232,MACHLEV22}. To prevent imbalances in the network of large-scale prosumers, distributed cooperative grid synchronization mechanisms  \cite{LI2024109624,Zhong2024} for smart energy management will be essential to coordinate and maintain seamless handover between energy production and consumption. Smart meters can be equipped with tamper detection sensors \cite{Sommerhalder2023} that alert utility companies in real-time if any attempt is made to physically or digitally interfere with the meter. By continuously monitoring energy consumption patterns with advanced data analytics and machine learning, utilities can detect anomalies that indicate potential theft. Sudden drops or spikes in usage that deviate from benign patterns can trigger further investigation.

   \subsection{Remaining challenges}

   For consumers in SG2, security attacks as energy thief or malware/ransomware/DoS attacks become more comments, given the complexity of managing bidirectional communications in smart meters. Among these, metering sensor interference or artificial energy price inflation attacks are specific ones for the integration of consumers and prosumers in SG2. The unique difference is, in the context of \ac{SG2}, successful attacks could allow the attacker to manipulate energy consumption patterns or intentional interference on consumer/prosumer devices, impacting power bills. Securing these consumer systems requires both physical deployment with boxes, robust authentication mechanisms or cooperative grid synchronization mechanisms, and IDS at connected multi-source networks. Furthermore, unified authentication and end-to-end data encryption techniques can be critical but have not yet been implemented in \ac{SG2}. The hierarchy protection by service priority or the relevance of data transfer should be properly constructed. In the next part, we cover the security risks and protection of many upcoming technologies that are projected to be prevalent in \ac{SG2}, such as distributed energy resources and peer-to-peer energy trading.

\section{Security attacks and protection models in SG2 enabling technologies}
\label{sec:security-energy-layer}

\ac{SG2} will involve a series of enhancements from the conventional smart grid model, e.g.,  the efficiency, reliability, and sustainability of energy distribution and management. This section summarizes the security problems and corresponding solutions of enabling technologies that are expected to be popular in \ac{SG2}: distributed energy resources (DERs), peer-to-peer energy trading, and energy storage technologies.

\subsection{Security threats and defense for DERs}

DERs denote small-scale or micro units of energy generation connected to the electricity grid at the distribution level. DERs can be operated automatically or through a remote cloud center \cite{WangKuan23}. This application concept revolutionizes the smart grid by permitting decentralized, community-generated energy from energy resources near electricity usage locations. DERs minimize power system augmentation, consumer costs, and emissions. To maintain energy stability without long-distance transmission power lines, DERs depend on rooftop solar panels, small wind turbines, small hydroelectric generating, and battery storage as more people avoid living near toxic power plants. However, to maintain the distributed microgrids, DERs rely much on the communication network systems. Therefore, one major threat in DERs is \textbf{cybersecurity attacks}, which can exploit \textbf{vulnerabilities in TCP/IP-based protocols}, as summarized in Section~IV. Further, due to low cost, many local grids of DERs may not be equipped with advanced security protection, e.g., substation DMZ, enterprise firewalls \cite{Zografopoulos2023}. According to \cite{Jhala21}, \textbf{data injection and physical layer attacks} in DERs may be less challenging to launch than in well-protected large-scale smart grids. Further, many DER components (e.g., metering devices and authentication gateways) may come from various suppliers, and some may sacrifice security features for low cost and simplicity. If these compromised components make their way into a larger interconnected DER, they could be used as entry points for attackers to infiltrate the larger energy network or initiate disruptions. Finally, \textbf{the lack of standardization in security practices across various DER technologies} increases the risk of these assets being exploited by malicious actors \cite{Chatzimiltis23}.

 Securing the physical infrastructure of DERs is vital for \ac{SG2}. With the strict requirement on cost, deploying physical access controls, advanced surveillance, and tamper-evident measures can be difficult. As summarized in Table~\ref{tab:emerging-technology-changes}(the first application domain), preventing physical compromises by wall fences and integrating HSM modules on key components of DER is then the most economical protection solution. Further, the elements connected to the Internet should be equipped with \textbf{basic firewalls} (e.g., access list control, limit the control from dedicated IPs, or \textbf{enable peer-to-peer authentication} \cite{Chang23}). Incorporating \textbf{lightweight anomaly detection and intrusion prevention systems} in every DER gateway \cite{Jhala21} can proactively identify suspicious activities or unauthorized access attempts. IDS/IPS technologies are still the main forces to alert grid operators to potential security breaches swiftly, enabling rapid isolation and damage mitigation. Additionally, \textbf{maintaining purchased devices from reliable sources}, \textbf{regular stress tests on security}, and updates to the software and firmware of DER devices help identify vulnerabilities and ensure that security measures stay current against evolving threats. Finally, building solutions for cascading failures of several DERs as introduced in Section~III (protecting power providers) and the study \cite{Valdez20} are essential. 

\subsection{Security threats and defense in new energy trading models}
\label{subsec:energy-trading}

     A significant difference between \ac{SG2} and its predecessor is the appearance of a new pro-consumer energy trading model, where the customer can be both an energy surplus supplier (i.e., sell energy surplus to energy distributors) and a consumer \cite{ALI2021}. As stated in \cite{Chang23,Feng23} and \cite{Morstyn19}, the peer-to-peer (P2P) energy management system, also known as bilateral contract networks, can facilitate the coordination of prosumers for higher efficiency and flexibility. Generally, the main energy trading techniques can be divided into four categories: game theory \cite{Anoh20}, auction theory \cite{Kang17, AlSkaif22, Charithri21}, and constrained optimization \cite{Paudel19}. For example, the Stackelberg game theory method was proposed to coordinate prosumers \cite{Chang23}. To control the demand response issues in residential homes, an hour-ahead and intraday P2P technique were used \cite{Liu20}. A multiclass P2P energy management platform was designed to coordinate prosumers with diverse preferences for energy source/destination \cite{Morstyn19,Wang23}. A consortium blockchain solution can be used for localized P2P power trading, with energy price determined by an auction process, while privacy and transaction security were enhanced \cite{Kang17, AlSkaif22,Gai19}. One of the primary benefits of blockchain is enhanced transparency and trust. Blockchain's decentralized and public ledger ensures that all transactions and contract executions are visible and verifiable by all network participants, which fosters trust among parties who do not need to know or trust each other. An example of blockchain technology for peer-to-peer (P2P) energy trading is the Brooklyn Microgrid project \cite{MENGELKAMP2018870}. This initiative was designed to allow residents in Brooklyn, New York, to generate, buy, and sell energy directly to one another using a blockchain-based platform.
     
     However, the major security threats in P2P electricity trading are \textbf{Byzantine attacks} and \textbf{vulnerabilities of blockchain techniques}. Table~\ref{tab:security-blockchain} summarizes several typical security attacks and protection models for blockchain-based \ac{SG2} models. For example, 51\% attacks or Byzantine attacks introduce misleading or noisy input into several nodes of distributed energy management systems, resulting in nonconvergence of transaction models. For the blockchain-based energy trading models, \textbf{vulnerabilities of the blockchain architecture} itself are the most concerning. Smart contracts, for example, are often self-executing contracts that are maintained on a blockchain. They are used to automate the execution of peer-to-peer energy trading contracts. However, smart contracts can be complex to implement and configure correctly. This can lead to replay attacks or software flaws (e.g., mishandled exceptions, unhandled errors \cite{Atzei17}) that can be exploited by attackers to steal energy or overestimate the transmitted energy surplus (siphon funds) \cite{Huang19}. For instance, the Parity wallet bug in 2017 caused by a coding error led to the freezing of $280$ million worth of digital coin Ether, illustrating how even small mistakes can have severe financial repercussions \cite{Accidentalbug}. The consequence will be devastating if the unvetted mistake occurs for the blockchain network of millions of consumers.
     	
     Blockchain vulnerabilities, further, include incorrect arithmetic operations, unchecked external calls, poor randomness generation, and lack of formal verification \cite{ZHOU2023103555}. Particularly, the immutability of smart contracts will have a negative impact if errors or vulnerabilities are discovered after deployment. For example, when a vulnerability is discovered after deployment, rectifying it can be challenging to maintain due to the long chain of dependent transactions. One prominent example is the Decentralized Autonomous Organization (DAO) hack in 2016 \cite{Shabani22}, where a vulnerability in the DAO's smart contract code was exploited, resulting in the theft of approximately $60$ million worth of Ether. Due to the immutable nature of the Ethereum blockchain, the developers could not simply alter the contract to fix the vulnerability or reverse the fraudulent transactions. This incident led to a contentious hard fork in the Ethereum blockchain, splitting it into two separate chains: Ethereum (ETH) and Ethereum Classic (ETC). This example illustrates that while immutability ensures data integrity, it can also make addressing unforeseen issues and errors exceedingly difficult, sometimes requiring drastic measures such as a hard fork. Also, P2P energy trading systems are vulnerable to a variety of \textbf{conventional cybersecurity attacks}, such as \textbf{denial-of-service attacks} \cite{Chowdhury22}, and \textbf{phishing attacks} \cite{Archana19}.

     \begin{figure}[t]
    		\begin{center}
    			\includegraphics[width=1\linewidth]{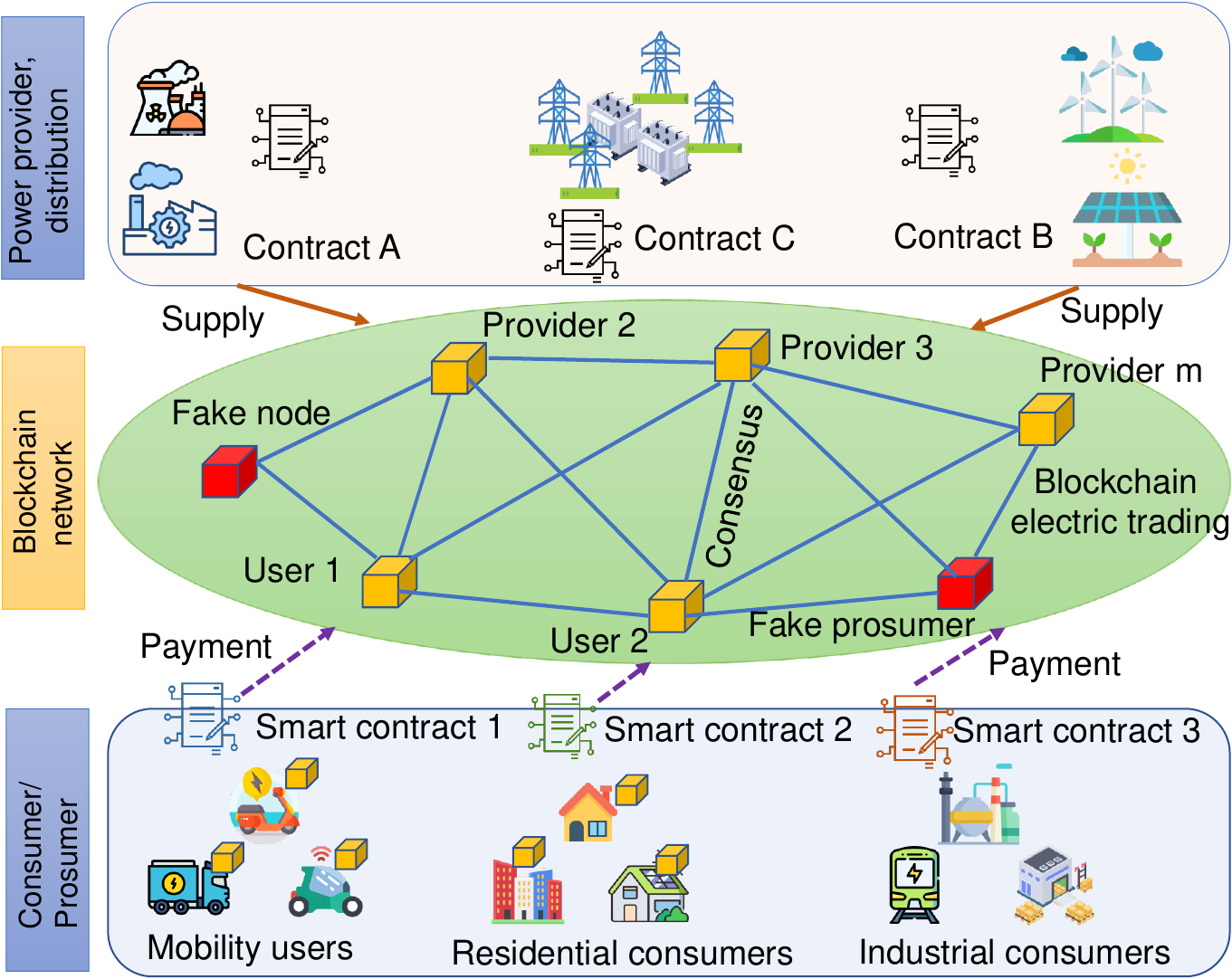}
    		\end{center}
    	\centering
    	 	\caption{Illustration of security threats in peer-to-peer energy trading via a blockchain network where the attackers can create fake prosumers to steal energy or overestimate the transmitted energy surplus (siphon funds).}
    	 	\label{fig:energy-trading-security}
      \end{figure}

     \textbf{Privacy concerns} also arise in new energy trading models. While blockchain and other decentralized technologies can enhance transparency and traceability, they can also \textbf{expose sensitive consumer data to the public domain}. Unauthorized access to participant information, energy consumption patterns, and financial details could lead to privacy breaches and misuse of personal data \cite{Chen20}. For example, the attacker can follow the energy consumption patterns to identify the consumers' habits and target them with phishing attacks \cite{Gai19}. If charging stations are put in medical clinic parks, businesses, and hospitals, sensitive information about EV drivers, such as their driving habits and visited places, can be revealed \cite{Baza21}. 

     \textbf{Collusion} is a major security concern in game-theory-based and auction-theory-based P2P energy trading models. For example, a group of malicious prosumers (consumers who also produce energy) collude to manipulate the energy market. They agree to coordinate their bidding strategies in order to drive up the price of energy to their advantage on a particular day, resulting in higher prices for consumers and lower profits for other energy producers \cite{MONIRUZZAMAN2023109111}. Sybil attacks also pose high risks to P2P energy trading. In the Sybil attack, an attacker creates multiple fake identities to gain undue influence in a system. In the context of P2P energy trading, the Sybil attacker could use their fake identities to manipulate the energy market (drive up the price due to fake high demands), launch collusion attacks, or steal funds from other participants \cite{Chowdhury22}.

     For protection solutions, as summarized in Table~\ref{tab:emerging-technology-changes} (the second application domain), to prevent the negative consequences of immutability matters and vulnerabilities on blockchain-based energy networks, smart contract audit and bug-free verification are essential. For example, the researchers in \cite{Bhargavan16} propose a formal verification method for smart contracts. Accordingly, it is a rigorous mathematical process that can be used to prove that a software system satisfies its specifications. This can help to identify and fix any vulnerabilities in smart contracts before they can be exploited by attackers. Further, robust governance frameworks and smart contract controls that allow for flexibility and error correction, e.g., multi-signature protocols for approving changes \cite{Shabani22}, establishing legal and technical mechanisms for dispute resolution. Additionally, privacy-preserving technologies, such as zero-knowledge proofs, can help protect sensitive participant information while maintaining the benefits of decentralized energy trading \cite{Jiang20,Baza21}. Besides, using reputation-based networks with mutual verification \cite{Bhargavan16} or block alliance consensus mechanisms \cite{Yingsen23} to identify and prevent malicious actors can be a promising approach. Maintaining energy pricing regulations and the eligibility of participants via additional authentication can mitigate the magnitude of the attack.



     \begin{figure}[t]
    	\begin{center}
    		\includegraphics[width=0.9\linewidth]{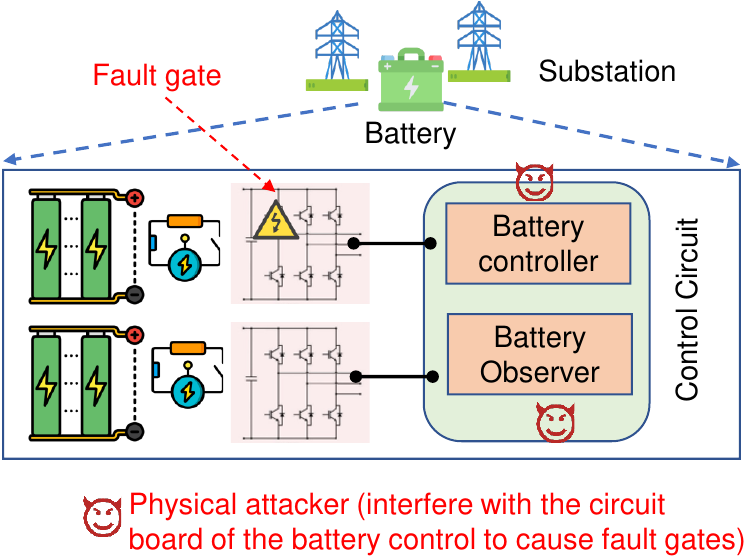}
    	\end{center}
    	\centering
    	\caption{Illustration of energy storage technologies and potential security attacks where the attackers can launch cyber-physical attacks against sensors or external devices of cyber system controls. Their goal is to create abnormal on/off gates that cause faults in key components of smart grids (power converter/transformers).}
    	\label{fig:eds}
    \end{figure}

      \begin{figure}[t]
    		\begin{center}
    			\includegraphics[width=1\linewidth]{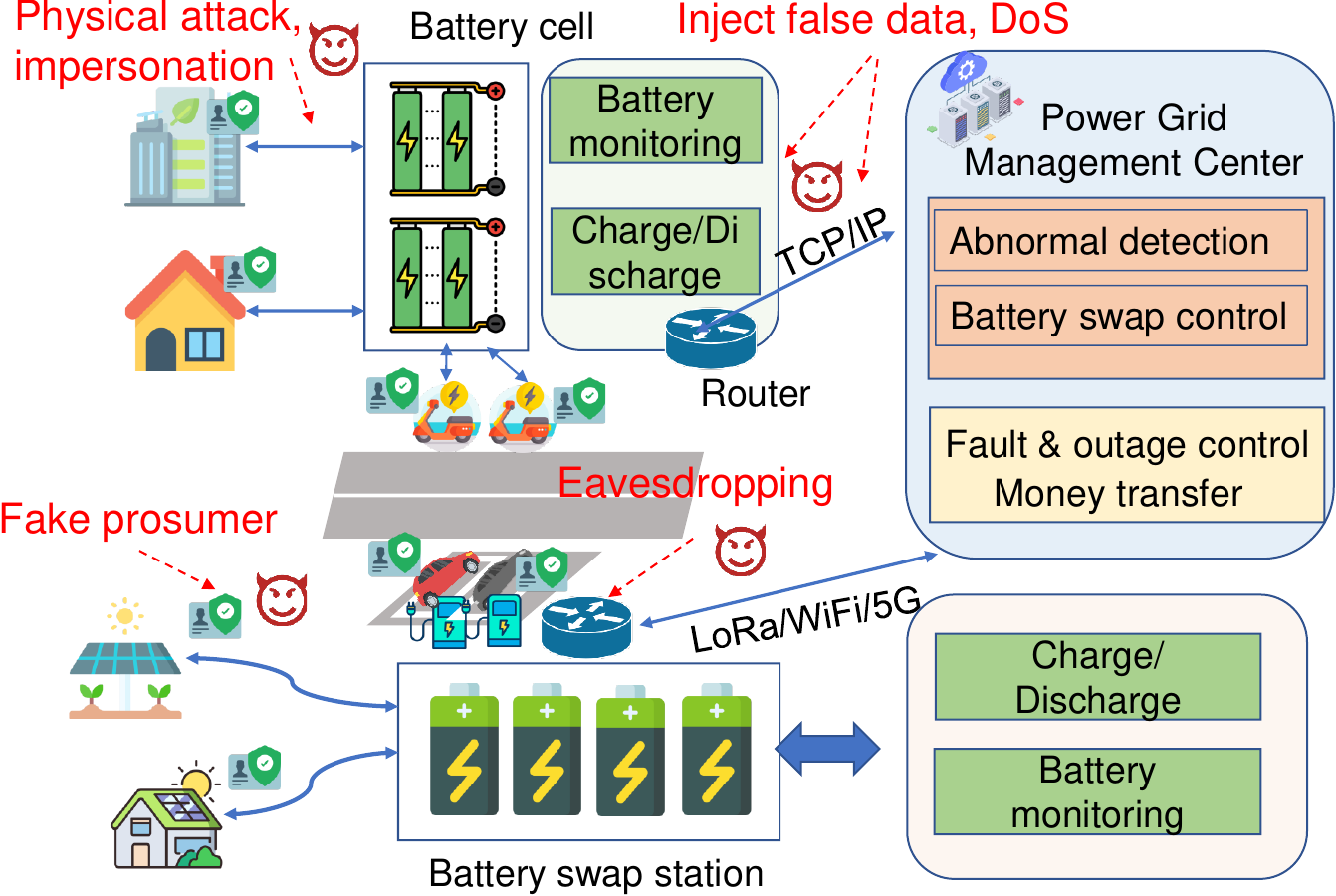}
    		\end{center}
    	\centering
    	 	\caption{Illustration of security attacks against battery storage technologies where TCP/IP-based communications between battery cell pack management and integrated energy management center can be intercepted and injected false data for incorrect control commands, e.g., disable battery charge/discharge functions, or to be fake prosumers. The attacks against EV charging systems are detailed in Fig.~\ref{fig:ev-charging} and Section~\ref{subsec:ev-charging}.}
    	 	\label{fig:battery-attack-1}
      \end{figure}

    In summary, protecting new energy trading models within \ac{SG2} requires a holistic approach that combines technological measures, cryptographic techniques, and regulatory compliance. Addressing these matters still requires time and these topics have not yet been well-explored in the vision of \ac{SG2}.

           \begin{table*}[t]
    	\caption{Summary of security threats, protection models, and open issues in potential blockchain-based \ac{SG2} models}
    	\label{tab:security-blockchain}
    	\begin{adjustbox}{width=1\textwidth}
    	\begin{tabular}{|l|l|l|l|l|l|l|l|l|}
    		\hline
    		\rowcolor[HTML]{EFEFEF} 
    		\textbf{Security threat} & \textbf{Blockchain/Damage} & \textbf{Severity} & \textbf{Likelihood} & \textbf{Potential examples in SG} & \textbf{Protection model} & \textbf{Efficiency} & \textbf{Open issues} & \textbf{Reference} \\ \hline
    		&  &  &  &  & Proof of Stake & Medium & \begin{tabular}[c]{@{}l@{}}A concentration of power among\\  a few wealthy stakeholders or\\ high-energy consumption\end{tabular} &  \cite{Conti18} \\ \cline{6-9} 
    		&  &  &  &  & \begin{tabular}[c]{@{}l@{}}Periodic checkpoints, \\ abnormal detection\end{tabular} & High & \begin{tabular}[c]{@{}l@{}}Require substantial changes to \\ the consensus mechanism \&\\ model for behavior monitoring\end{tabular} &  \cite{Conti18} \\ \cline{6-9} 
    		\multirow{-3}{*}{51\% attacks} & \multirow{-3}{*}{\begin{tabular}[c]{@{}l@{}}- Bitcoin fork (2018)\\ - Loss: \$18 million \\ worth of BTG\end{tabular}} & \multirow{-3}{*}{High} & \multirow{-3}{*}{High} & \multirow{-3}{*}[35pt]{\begin{tabular}[c]{@{}l@{}}A coalition of DER operators\\ or prosumers gains control \\ over more than 51\% of the \\ total energy generation \\ capacity could manipulate \\ the energy supply, creating \\ artificial shortages or \\ surpluses\end{tabular}} & Strong governance & Medium & \begin{tabular}[c]{@{}l@{}}Voter apathy can lead \\ concentration of power among\\  a few active participants\end{tabular} &  \cite{Chowdhury22}\\ \hline
    		&  &  &  &  & Smart contract audit & Medium & \begin{tabular}[c]{@{}l@{}}Depending on the efficiency \\ of audit model and verification \\ engines.\end{tabular} &  \cite{ZHOU2023103555} \\ \cline{6-9} 
    		\multirow{-2}{*}{DAO attacks} & \multirow{-2}{*}{\begin{tabular}[c]{@{}l@{}}- Ethereum (2016)\\ - Loss: \$60 million \\ worth of Ether\end{tabular}} & \multirow{-2}{*}{High} & \multirow{-2}{*}{High} & \multirow{-2}{*}[22pt]{\begin{tabular}[c]{@{}l@{}}Attacker can withdraw \\ funds before the contract \\ updates the balance, resulting \\ in financial loss and \\disruption of grid\end{tabular}} & \begin{tabular}[c]{@{}l@{}}Multi-signature wallets \\ Hard fork \end{tabular}& Medium & \begin{tabular}[c]{@{}l@{}}Complexity in setup and use \\ that requires coordination \\ among multiple parties\end{tabular} &  \cite{Shabani22} \\ \hline
    		FDI attacks & - Bitcoin (2016) & High & Medium & \begin{tabular}[c]{@{}l@{}}Attacker can manipulate \\ energy consumption figures, \\ incorrect energy production \\ data from DERs, or \\ falsified transactions\end{tabular} & \begin{tabular}[c]{@{}l@{}}Fault Tolerance \\Proof of Stake \\ End-to-end encryption\end{tabular} & High & \begin{tabular}[c]{@{}l@{}}Low scalability, complexity in \\ configuration, voter apathy, \\ complexity in key exchange, \\ computational overhead\end{tabular} &  \cite{Jafari23} \\ \hline
    		Internal fraud & \begin{tabular}[c]{@{}l@{}}- Bitcoin (2014)\\ - Loss: \$450 million \\ at the time\end{tabular} & High & High & \begin{tabular}[c]{@{}l@{}}A utility company employee \\ might alter blockchain entries \\ to record false energy \\ consumption data, \\ resulting in lower bills \\ for certain customers\end{tabular} & \begin{tabular}[c]{@{}l@{}}Smart contract audit\\ Role-based access \end{tabular} & Medium & \begin{tabular}[c]{@{}l@{}}Depending on the efficiency\\  of the audit model, verification \\ engines, and access control.\end{tabular} &   \cite{Conti18} \\ \hline
    		\begin{tabular}[c]{@{}l@{}}Parity wallet\\ (Software bug)\end{tabular} & \begin{tabular}[c]{@{}l@{}}- Ethereum (2017)\\ - Loss: \$30 million \\ in Ether\end{tabular} & High & High & \begin{tabular}[c]{@{}l@{}}Attacker can could drain \\ funds allocated for energy\\  transactions, leading to \\ financial losses and \\ operational disruptions\end{tabular} & Smart contract audit & High & \begin{tabular}[c]{@{}l@{}}- High complicated blockchain\\ comes with potential bugs\end{tabular} &  \cite{Accidentalbug} \\ \hline
    	\end{tabular}
    	\end{adjustbox}
    \end{table*}

     \subsection{Security threats and defense in energy storage technologies}

     As we mentioned earlier, battery energy storage systems (BESS) are core components of DERs and future \ac{SG2} by enabling grid stability through the integration of renewable sources and discharging at proper times. However, these technologies also introduce specific security threats. One significant security concern is the potential for \textbf{battery-related incidents or fault gates of energy coordination}, including thermal runaway, leakage, or explosions from the physical attacks \cite{SECenergyStorage, Trevizan22} as illustrated in Fig.~\ref{fig:eds}. Fig.~\ref{fig:battery-attack-1} illustrates four detailed cases of \textbf{physical attacks (e.g., fake prosumer, impersonation)} against signal exchanges between sensors and estimation observers to mislead the voltage/power-line controllers (giving the wrong on-off gate decision). This can lead to incorrect energy flow adjustments or even shutting down a critical power line for various consumers and rivals. Note that BESS control functions rely heavily on telemetry capabilities such as load frequency control, damping oscillations, and power imbalance measurement. Any disconnect from faulty communication networks might result in a loss of control of BESS functions. Besides, manufacturing defects, poor maintenance, or external factors such as temperature variations can trigger the incidents \cite{Hasan23}. In recent years, lithium-ion-based BESS cell failures that resulted in thermal runaway and explosion accidents have also raised serious concerns, e.g., the BESS explosion in Surprise Arizona, US, 2019 \cite{CLOSE2024422}.
     


        \begin{table*}[]
      	\caption{Comparison of Security threats and defense in emerging technologies for SG1 vs SG2}
      	\label{tab:emerging-technology-changes}
      	\begin{adjustbox}{width=1\textwidth}
      		\begin{tabular}{|l|l|l|l|ll|l|l|l|}
      			\hline
      			\rowcolor[HTML]{C0C0C0} 
      			\cellcolor[HTML]{C0C0C0} & \cellcolor[HTML]{C0C0C0} & \cellcolor[HTML]{C0C0C0} & \cellcolor[HTML]{C0C0C0} & \multicolumn{2}{c|}{\cellcolor[HTML]{C0C0C0}\textbf{Protection models}} & \cellcolor[HTML]{C0C0C0} & \cellcolor[HTML]{C0C0C0} & \cellcolor[HTML]{C0C0C0} \\ \cline{5-6}
      			\rowcolor[HTML]{C0C0C0} 
      			\multirow{-2}{*}{\cellcolor[HTML]{C0C0C0}\textbf{Application domain}} & \multirow{-2}{*}{\cellcolor[HTML]{C0C0C0}\textbf{\begin{tabular}[c]{@{}l@{}}Security threats\\ (attack target)\end{tabular}}} & \multirow{-2}{*}{\cellcolor[HTML]{C0C0C0}\textbf{Severity}} & \multirow{-2}{*}{\cellcolor[HTML]{C0C0C0}\textbf{Likelihood}} & \multicolumn{1}{c|}{\cellcolor[HTML]{C0C0C0}\textbf{SG1}} & \multicolumn{1}{c|}{\cellcolor[HTML]{C0C0C0}\textbf{SG2}} & \multirow{-2}{*}{\cellcolor[HTML]{C0C0C0}\textbf{Efficiency}} & \multirow{-2}{*}{\cellcolor[HTML]{C0C0C0}\textbf{Open issues}} & \multirow{-2}{*}{\cellcolor[HTML]{C0C0C0}\textbf{References}} \\ \hline
      			& \begin{tabular}[c]{@{}l@{}}False data injection\\ (data collection)\end{tabular} & High & Medium & \multicolumn{1}{l|}{\begin{tabular}[c]{@{}l@{}}Encryption\\ IDS/IPS\end{tabular}} & \begin{tabular}[c]{@{}l@{}}Quantum-safe Encryption\\ Intelligent IDS/IPS\end{tabular} & High & \begin{tabular}[c]{@{}l@{}}Different security\\ capability of the sources\end{tabular} & \cite{Wang23} \\ \cline{2-9} 
      			\multirow{-2}{*}{\begin{tabular}[c]{@{}l@{}}Distributed\\ energy \\ resource\end{tabular}} & \begin{tabular}[c]{@{}l@{}}Compromised source\\ (federated learning)\end{tabular} & Medium & \begin{tabular}[c]{@{}l@{}}Low\\ (insider)\end{tabular} & \multicolumn{1}{l|}{\begin{tabular}[c]{@{}l@{}}HSM, IDS\\ Authentication\end{tabular}} & \begin{tabular}[c]{@{}l@{}}HSM, IDS\\ Authentication\\ Blockchain\end{tabular} & Medium & \begin{tabular}[c]{@{}l@{}}Unreliable supply\\ chain\end{tabular} & \cite{Chang23} \\ \hline
      			& \begin{tabular}[c]{@{}l@{}}DoS attacks\\ (Network infrastructure)\end{tabular} & Medium & \begin{tabular}[c]{@{}l@{}}Low \\ (distributed\\ model)\end{tabular} & \multicolumn{1}{l|}{\begin{tabular}[c]{@{}l@{}}IDS/IPS\\ Load balance\end{tabular}} & Distributed ledger & Medium & \begin{tabular}[c]{@{}l@{}}Real-time response,\\ scalability\end{tabular} &  \cite{Chowdhury22}  \\ \cline{2-9} 
      			& \begin{tabular}[c]{@{}l@{}}Impersonation/Phishing\\ (Fake prosumer)\end{tabular} & High & High & \multicolumn{1}{l|}{\begin{tabular}[c]{@{}l@{}}Identity \\ authentication\end{tabular}} & \begin{tabular}[c]{@{}l@{}}Smart contract audit\\ Behavior analysis\end{tabular} & High & Scalability of verification & \cite{Vahidi23}  \\ \cline{2-9} 
      			& \begin{tabular}[c]{@{}l@{}}51\% attacks\\ (blockchain-based\\ power grids)\end{tabular} & Medium & \begin{tabular}[c]{@{}l@{}}Low\\ (require large \\ collusion)\end{tabular} & \multicolumn{1}{l|}{\begin{tabular}[c]{@{}l@{}}Proof of stake\\ Consensus\\ mechanism\end{tabular}} & \begin{tabular}[c]{@{}l@{}}Periodic checkpoint\\ Strong governance\end{tabular} & High & \begin{tabular}[c]{@{}l@{}}Transparent nature of\\ blockchain\end{tabular} &  \cite{Mollah2021} \\ \cline{2-9} 
      			& \begin{tabular}[c]{@{}l@{}}Collusion attacks\\ (Auction theory-based\\ energy trading models)\end{tabular} & High & \begin{tabular}[c]{@{}l@{}}Low\\ (Insiders, \\ fraud)\end{tabular} & \multicolumn{1}{l|}{\begin{tabular}[c]{@{}l@{}}Anomaly\\ detection\end{tabular}} & \begin{tabular}[c]{@{}l@{}}Anomaly detection\\ Blockchain\end{tabular} & High & \begin{tabular}[c]{@{}l@{}}Coordinated behaviors\\ of insider attackers\end{tabular} & \cite{MONIRUZZAMAN2023109111} \\ \cline{2-9} 
      			\multirow{-5}{*}{\begin{tabular}[c]{@{}l@{}}P2P\\ energy\\ trading\end{tabular}} & \begin{tabular}[c]{@{}l@{}}Byzantine attacks\\ (Electric controllers)\end{tabular} & Medium & Medium & \multicolumn{1}{l|}{IDS/IPS} & Intelligent IDS/IPS & Medium & \begin{tabular}[c]{@{}l@{}}Complex to detect in\\ a large scale grid\end{tabular} &  \begin{tabular}[c]{@{}l@{}}\cite{Huang19} \\ \cite{Chang23} \end{tabular} \\ \hline
      			& \begin{tabular}[c]{@{}l@{}}Physical attacks\\ Fuel leakage\\ (Battery fault, explosion)\end{tabular} & High & High & \multicolumn{1}{l|}{Fencing/Barrier} & \begin{tabular}[c]{@{}l@{}}Battery swap\\ Fencing/barrier\\ Legal action\end{tabular} & Medium & Economical solutions &  \begin{tabular}[c]{@{}l@{}}\cite{SECenergyStorage}\\ \cite{Trevizan22}\end{tabular}\\ \cline{2-9} 
      			& \begin{tabular}[c]{@{}l@{}}Spoofing attacks\\ (Battery operations)\end{tabular} & Medium & \begin{tabular}[c]{@{}l@{}}Low\\ (Insider)\end{tabular} & \multicolumn{1}{l|}{Encryption} & \begin{tabular}[c]{@{}l@{}}Anomaly detection\\ Blockchain\end{tabular} & Medium & Insider attacks & \cite{Hasan23}  \\ \cline{2-9} 
      			& \begin{tabular}[c]{@{}l@{}}Firmware update\\ tampering\end{tabular} & High & \begin{tabular}[c]{@{}l@{}}Low\\ (compromised)\end{tabular} & \multicolumn{1}{l|}{\begin{tabular}[c]{@{}l@{}}HSM, secure\\ boot\end{tabular}} & \begin{tabular}[c]{@{}l@{}}End-to-end encryption\\ HSM, blockchain\end{tabular} & High & Affordable solutions & \cite{Hasan23} \\ \cline{2-9} 
      			& \begin{tabular}[c]{@{}l@{}}Supply chain\\ (Lithium-ion battery)\end{tabular} & High & High & \multicolumn{1}{l|}{\begin{tabular}[c]{@{}l@{}}Trusted vendor\\ list\end{tabular}} & \begin{tabular}[c]{@{}l@{}}Blockchain-based \\ management\end{tabular} & Medium & \begin{tabular}[c]{@{}l@{}}Limit of options, \\ supply quality\end{tabular} & \begin{tabular}[c]{@{}l@{}}\cite{Duman22}\\ \cite{SupplychainAttack}\end{tabular}   \\ \cline{2-9} 
      			\multirow{-5}{*}{\begin{tabular}[c]{@{}l@{}}Energy \\ storage\\ systems\end{tabular}} & \begin{tabular}[c]{@{}l@{}}Adversarial attacks\\ (AI-powered function\end{tabular} & High & Medium & \multicolumn{1}{l|}{\begin{tabular}[c]{@{}l@{}}No mechanism\\ (yet be applied)\end{tabular}} & \begin{tabular}[c]{@{}l@{}}Adversarial training,\\ Robust AI models\end{tabular} & High & Diverse attacks & \cite{Jhala21} \\ \hline
      			& \begin{tabular}[c]{@{}l@{}}Data breaches\\ (Payment, personal data)\end{tabular} & Medium & \begin{tabular}[c]{@{}l@{}}Low\\ (encryption)\end{tabular} & \multicolumn{1}{l|}{\begin{tabular}[c]{@{}l@{}}OCPP 1.0\\ TLS 1.3\end{tabular}} & \begin{tabular}[c]{@{}l@{}}OCPP 2.0\\ End-to-end encryption\\ Quantum-safe encryption\end{tabular} & High & High cost &  \cite{Hasan23} \\ \cline{2-9} 
      			& \begin{tabular}[c]{@{}l@{}}Identity theft\\ (Charge card)\end{tabular} & High & Medium & \multicolumn{1}{l|}{\begin{tabular}[c]{@{}l@{}}Two-factor\\ authentication\end{tabular}} & Biometric authentication & High & \begin{tabular}[c]{@{}l@{}}User-friendly \\ authentication\end{tabular} &  \cite{Hasan23} \\ \cline{2-9} 
      			& \begin{tabular}[c]{@{}l@{}}Physical sabotage\\ (CSMS links)\end{tabular} & High & High & \multicolumn{1}{l|}{No mechanism} & Anomaly detection & Medium & \begin{tabular}[c]{@{}l@{}}Long links exist in\\ residential areas\end{tabular} & \cite{Johnson22} \\ \cline{2-9} 
      			& \begin{tabular}[c]{@{}l@{}}Collusion attacks\\ (Imbalanced grid)\end{tabular} & High & \begin{tabular}[c]{@{}l@{}}Low\\ (large \\ collusion\end{tabular} & \multicolumn{1}{l|}{No mechanism} & \begin{tabular}[c]{@{}l@{}}Blockchain\\ Anomaly detection\\ Trusted networks\end{tabular} & High & \begin{tabular}[c]{@{}l@{}}Collusion of\\ insider attacks\end{tabular} & \cite{He2016StorageAttack} \\ \cline{2-9} 
      			\multirow{-5}{*}{\begin{tabular}[c]{@{}l@{}}EV charging\\ and scheduling\end{tabular}} & \begin{tabular}[c]{@{}l@{}}Adversarial attacks\\ (pricing management)\end{tabular} & High & Low & \multicolumn{1}{l|}{No mechanism} & \begin{tabular}[c]{@{}l@{}}Adversarial training\\ Robust AI models\end{tabular} & Medium & Diverse attacks &  \cite{Ren24} \\ \hline
      			& \begin{tabular}[c]{@{}l@{}}False data injection\\ (Measurement sensor)\end{tabular} & High & Medium & \multicolumn{1}{l|}{Anomaly detection} & \begin{tabular}[c]{@{}l@{}}Intelligent anomaly\\ detection, blockchain\end{tabular} & High & \begin{tabular}[c]{@{}l@{}}Real time detection\\ Diverse data\end{tabular} &  \cite{SIFAT2023} \\ \cline{2-9} 
      			\multirow{-2}{*}{Digital twin} & \begin{tabular}[c]{@{}l@{}}DoS attacks\\ (Data transformation)\end{tabular} & High & High & \multicolumn{1}{l|}{CDN/Cloudflare} & \begin{tabular}[c]{@{}l@{}}Distributed management\\ systems\end{tabular} & Medium & Many variant attacks & \cite{SIFAT2023} \\ \hline
      			& \begin{tabular}[c]{@{}l@{}} Signal jamming \\ Energy depletion \\ Pilot contamination\end{tabular} & - & - & \multicolumn{1}{l|}{-} & \begin{tabular}[c]{@{}l@{}}Multiple models as \\ summarized in Table III\end{tabular} & - & - &  \\ \cline{2-9} 
      			\multirow{-2}{*}{\begin{tabular}[c]{@{}l@{}}5G/6G\\ for SG2\end{tabular}} & \begin{tabular}[c]{@{}l@{}}Adversarial attacks\\ (AI-powered functions)\end{tabular} & High & \begin{tabular}[c]{@{}l@{}}Low\\ (black-box\\ attacks)\end{tabular} & \multicolumn{1}{l|}{\begin{tabular}[c]{@{}l@{}}Adversarial training\\ Data santinization\\ Model hardening\end{tabular}} & \begin{tabular}[c]{@{}l@{}}Adversarial training \\ Data santinization \\ Model hardening\\ (enhancement)\end{tabular} & Medium & \begin{tabular}[c]{@{}l@{}}High cost of training\\ and data preprocessing\end{tabular} & \begin{tabular}[c]{@{}l@{}}\cite{Takiddin23}\\ \cite{Huang23}\\ \cite{WangYu23}\\ \cite{catak2021adversarial}\end{tabular} \\ \hline
      		\end{tabular}
      	\end{adjustbox}
      \end{table*}

      Moreover, the increasing reliance on \textbf{lithium-ion batteries}, which dominate the energy storage landscape, raises concerns about \textbf{supply chain vulnerabilities} \cite{Duman22}. As the demand for these batteries grows, so does the potential for compromised components entering the market. Malicious actors could insert \textbf{counterfeit or tampered components into battery systems}, leading to reduced performance, safety risks, or even intentional battery failure. Attackers may take advantage of vulnerabilities introduced into devices prior to shipping or during \textbf{firmware upgrades}. Fig.~\ref{fig:battery-attack-1} illustrates physical attacks on IoT gateways to disrupt data exchange between battery cell packs and remote controllers or consumers. Accordingly, embedded devices and communication modules are highly sensitive to physical attacks, e.g., side-channel attacks. BESS near substations are complicated systems with hundreds of gateway controllers and circuit breakers. Because of this complexity, the operator may buy components from many vendors. Even if most of those suppliers are dependable, an operator may have to use lesser-known vendors due to a sudden shortage of particular components. Even dependable providers might be exposed to supply chain hijacking. For example, in the SolarWinds security issue, malicious code is introduced into the developer's update channel, which is subsequently distributed to all impacted devices \cite{Duman22}. Another attack version is that a malicious attack may send a bogus signal to consumers telling them to charge/discharge their devices when they should not, resulting in network frequency degradation. DoS attacks on the central controller in the event of centralized management may result in the loss of energy control by BESS and the instability of power operations \cite{REZAEIMOZAFAR2022112573}.

      Regarding protection models for BESS, the first and foremost is to install IDS/IPS that can analyze the abnormal charge/discharge behavior from BESS \cite{Hasan23}. Supply chain verification, end-to-end encrypted communications, P2P authentication, and source traceability are also essential. Besides, physical protection like fencing and buildings may help, as summarized in Table~\ref{tab:emerging-technology-changes} (the third application domain). There is a lack of studies in this field, given the difficulty of building a testing or simulation platform. A summary of holistic protection approaches to address the causes of failure in BESS, e.g., full BESS system installation testing, hazard and gas chromatography, and composition testing, can be found at \cite{CLOSE2024422}.

     \subsection{Security threats and defense for electric vehicle charging and scheduling in \ac{SG2}}
     \label{subsec:ev-charging}

    Electric vehicles (EVs) are a crucial part of the green transition to zero-carbon alternatives. EVs help reduce greenhouse gas emissions and reliance on fossil fuels, contributing to cleaner air and a more sustainable environment. The adoption of EVs is supported by advancements in battery swap technology and renewable energy integration. For example, electric scooters now account for 12\% of all scooter sales in Taiwan, a figure that continues to grow with the convenience of battery swapping technology and the high density of charging stations \cite{Gogoro}. In China, EVs comprised 60\% of new car sales in 2023, while in Europe and the US, the figures were around 25\% and 10\%, respectively \cite{EVOutlook}. By 2024, electric automobiles may have a market share of up to 45\% in China, 25\% in Europe, and over 11\% in the US \cite{EVOutlook}.
     

    \begin{figure}[t]
    	\begin{center}
    		\includegraphics[width=1\linewidth]{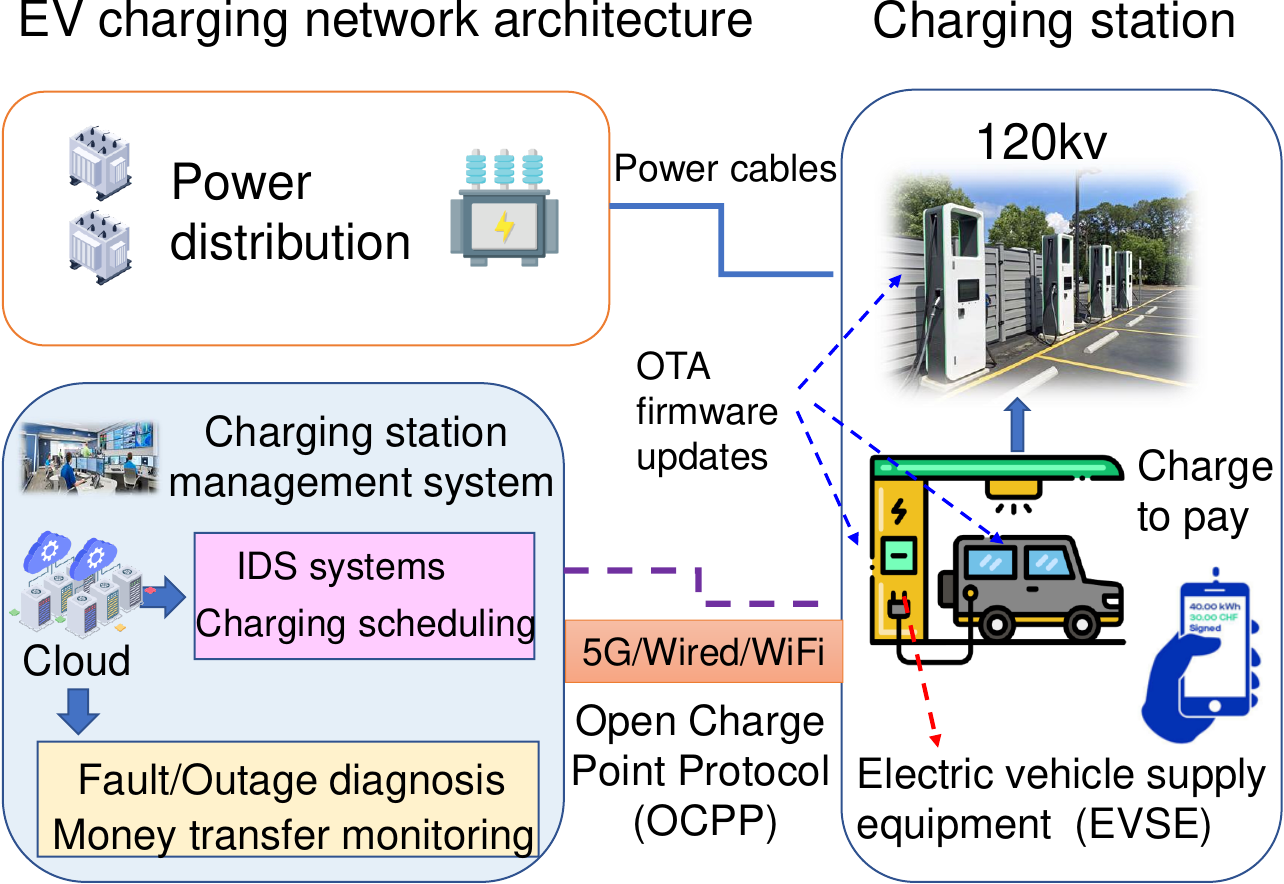}
    	\end{center}
    	\centering
    	\caption{Illustration of an EV charging network architecture and its core components, e.g., electric vehicle supply equipment (EVSE), charging station, charging station management system (CSMS), and network connection between EV and CSMS using Open Charge Point Protocol (OCPP).}
    	\label{fig:ev-charging}
    \end{figure}
     
     A prominent security threat in EV charging systems is the potential for data breaches. EV charging stations collect and transmit a significant amount of user data, including payment information and personal details. If the EVs or charging stations use old generation OCPP or their firmware is not updated frequently, attackers can intercept communication links between the EV, the electric vehicle supply equipment (EVSE), the charging station, and the charging station management system (CSMS) (illustrated in Fig.~\ref{fig:ev-charging}), leading to identity theft or financial fraud \cite{Hasan23}. For instance, researchers in Sandia National Laboratories, US, highlight that several public EV charging networks lacked encryption or some plug-and-charge functions may abort charging sessions by disrupting the PLC communications. This makes it easy for attackers to steal credentials or influence charging sessions via the EV-to-EVSE connection \cite{Johnson22}. Lastly, the integration of distributed energy resources (DER) and demand response systems (i.e., P2P energy trading) into \ac{SG2} may introduce new attack vectors. Hackers could potentially manipulate the scheduling algorithms for EV charging, causing imbalances in the grid. For example, a coordinated attack could instruct numerous EVs to charge simultaneously, creating a sudden spike in demand that the grid cannot handle, leading to blackouts. In 2020, a study \cite{GridReport2020} by the National Renewable Energy Laboratory (NREL) demonstrated how malicious actors could exploit demand response signals to destabilize the grid.

     To defend against these threats, as summarized in Table~\ref{tab:emerging-technology-changes} (the fourth application domain), implementing \textbf{advanced anomaly detection} is crucial \cite{Garofalaki22}. These systems can monitor network traffic for unusual patterns indicative of a DDoS attack, allowing for immediate countermeasures to be deployed. Additionally, adopting more \textbf{secure versions of communication protocols, such as OCPP 2.0}, which includes enhanced security features like encrypted communication and certificate-based authentication, can mitigate the risks of such attacks. Regularly \textbf{patching the software of charging stations} is also essential to protect against known vulnerabilities \cite{Garofalaki22}. To safeguard against data breaches, robust encryption protocols must be employed for all data transmissions between EVs, charging stations, and the central grid. Furthermore, implementing multi-factor authentication (MFA) can add an additional layer of security, ensuring that only authorized users can access sensitive systems and data. Regular security audits and compliance with data protection regulations such as the General Data Protection Regulation (GDPR) can also help ensure that data handling practices meet the highest security standards. Finally, adopting a zero-trust architecture, where every transaction and communication within the grid is verified, can significantly reduce the risk of unauthorized manipulations from firmware tampering. Implementing AI/ML algorithms to predict electric demand more effectively can enhance the resilience of the grid against threats. IDS may use these techniques to identify charging abnormalities and react fast to avoid grid imbalances. The AI/ML solutions for \ac{SG2} technologies are detailed in Section~\ref{sec:security-application-layer}.

    \subsection{Security threats and defense in future technologies for \ac{SG2}}
     
     Digital twin is expected to be one of the most promising future technologies to appear in \ac{SG2} \cite{SIFAT2023}. This technology maintains a virtual world of real-world physical products, systems, or processes through updates from real-time data over time. The straightforward advantage of the digital twin is that it provides a robust and affordable environment for simulation, integration, testing, monitoring, and maintenance, improving efficiency and decision-making. As a result, technological advancements in the monitoring and control functions of energy distribution systems may evolve faster than in the conventional smart grid. However, as summarized in \cite{Bazmohammadi21} and \cite{Jafari23}, a huge volume of data and real-time analysis requirements are two of the major concerns for any security system. Table~\ref{tab:emerging-technology-changes} summarizes two typical attacks in digital twins for SG2 (the fifth application domain). For example, the attacker can launch a false data injection that creates inaccurate representations of the grid state, potentially leading to misguided decisions. Further, DoS attacks target disrupting the real-time functionality of AI/ML models in real-to-virtual transformation functions, degrading the grid's responsiveness \cite{Jafari23}. Since large-scale data processing and real-time reaction are also challenges for many current defense technologies, these attacks are still open issues to address. Distributed IDS or edge-assisted detection systems can be promising approaches that have yet to be explored, particularly in \ac{DERs}, battery storage technologies, and electric vehicle integration environments. 

     Upgrading network infrastructure towards 5G and further 6G networks is expected to be the second significant change in \ac{SG2}, as discussed in Section~\ref{subsec:potential-upgrade}. However, these networks have their security problems that remain concerns. As pointed out in the study \cite{Nguyen21} and summarized in Table~\ref{tab:emerging-technology-changes} (the sixth application domain), signaling DoS attacks/jamming attacks at the physical layer, energy depletion attacks, or exploiting no integrity protection of the user data plane to launch impersonation attacks at the network layer, deep fake/biometric authentication data leakage at the application layer are typical security risks. Furthermore, new risks to 5G/6G-enabled technologies have emerged, such as pilot contamination attacks against massive MIMO-based networks or adversarial attacks against AI-aided network operations (resource allocation, slicing, service offloading, and semantic communications) \cite{Nguyen21}. Explainable AI, quantum-safe communications, distributed ledgers, and differential privacy can significantly reduce the severity of attacks and personal data breaches. However, the research topics of these technologies specified for smart grid environments have not yet been well explored. Given the importance of AI in many features of new smart grid generation, the following section details the research efforts on security attacks and defense in AI-powered functions, including references to smart grid issues.

\section{Security attacks and protection models in AI-enabled power grid management}
\label{sec:security-AI-layer}

AI-empowered energy management represents an essential space for \ac{SG2}'s autonomous and self-healing features, enabling grid distribution optimization and usage efficiency. However, these systems are not immune to security threats. The major concerns are \textbf{data poisoning, adversarial attacks, and model interference}. In the context of \ac{SG2} management, these attacks could lead to inaccurate load forecasts, improper demand response decisions, or faulty optimization strategies, potentially causing grid instability. The following subsections outline several typical threats and defense methods against AI-powered power grid management for \ac{SG2}.

\subsection {Security threats against AI-enabled power grid management in \ac{SG2}}

\begin{figure*}[t]
	\begin{center}
		\includegraphics[width=1\linewidth]{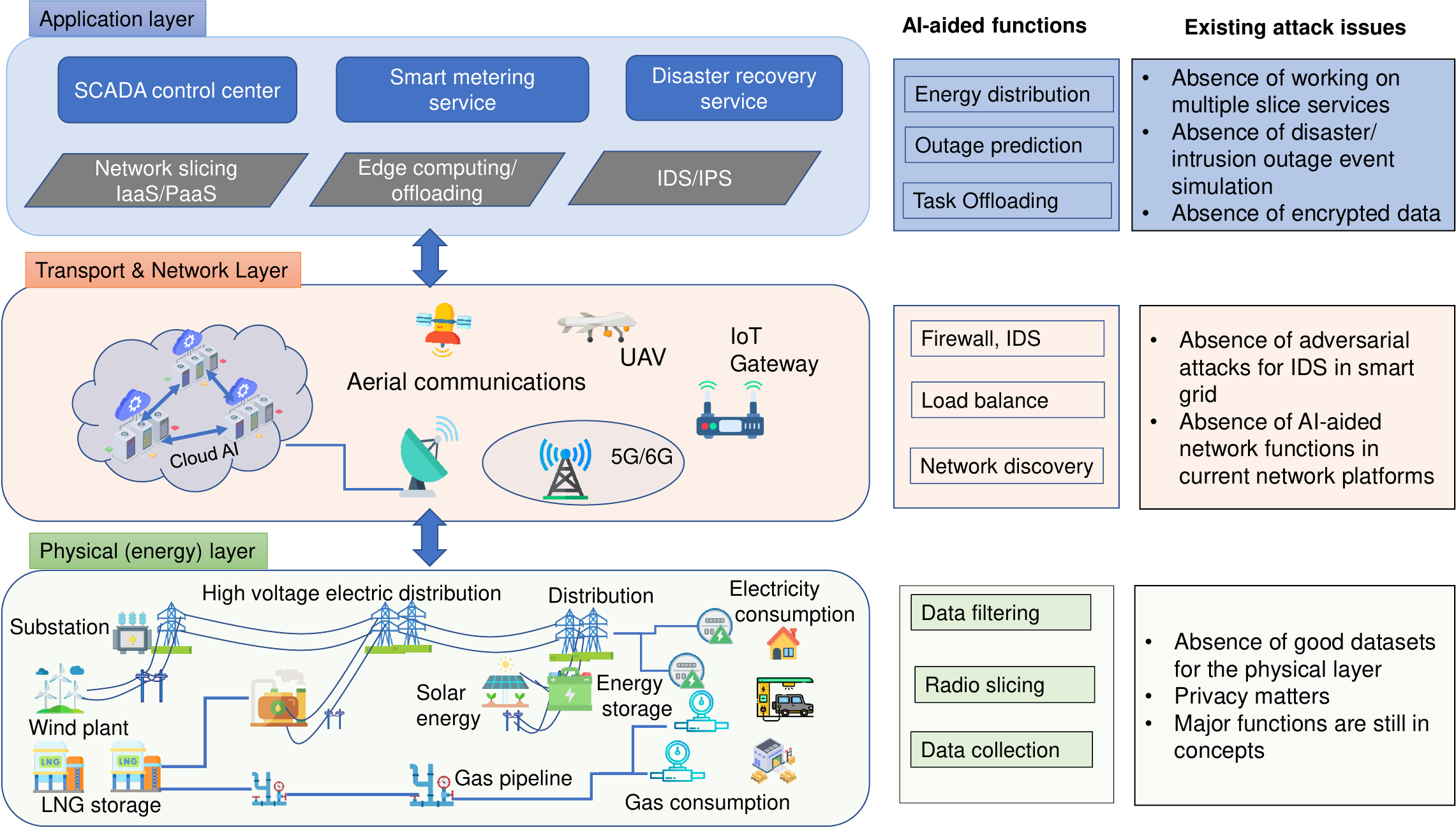}
	\end{center}
	\centering
	\caption{Summary of typical AI-aided functions in \ac{SG2} infrastructure with 5G/6G core networks, and remaining issues for the attacks/defenses. Accordingly, many functions of all layers in \ac{SG2}, such as radio slicing, outage prediction, and energy distribution, will be equipped with intelligent capability from advanced AI models.}
	\label{fig:ai-aided-function}
\end{figure*}

While AI in computer vision and natural language processing rapidly grows, AI for \ac{SG2} has just gotten the attention \cite{HAO2022123, Song21}. In essence, smart grid management functions can benefit much from AI learning capability, e.g., reliable electric distribution scheduling, autonomous energy management operations, voltage stability assessment, optimal dynamic pricing trading, and accurate outage forecast. Fig.~\ref{fig:ai-aided-function} shows the summary of typical AI-aided functions in \ac{SG2} infrastructure. 
In these applications, AI models can help to \textbf{increase the accuracy}: on/off switchgear decisions, electric price charging, electric generating rate limit (to avoid grid overload), and network congestion avoidance. 

However, data poisoning, model interference, and adversarial attacks are the biggest security concerns. Fig.~\ref{fig:adversarial-attacks} summarizes three attacks and their target in each phase of AI learning process.  Given AI's reliance on enormous amounts of data for training, data poisoning is a cyberattack aimed at tainting training datasets with outlier data points. Further, model interference attack involves an adversary manipulating the inputs but can also, outputs, or parameters of a machine learning model to degrade its performance or to cause it to produce erroneous results \cite{Nguyen21}. Both attacks assumes that the attacker must have full permission to access the dataset/learning model or control over the data pipeline, i.e., the write privilege to insert new data points/classes (the public repository or compromised equipment). By contrast, adversarial attacks generate adversarial samples on electricity reading to increase an unanticipated outcome such as misclassification. There are two adversarial attacks: black-box and white-box. White-box adversarial attacks have full access to the target model's architecture and parameters, allowing precise crafting of adversarial examples, while black-box adversarial attacks only interact with the model's inputs and outputs, lacking detailed knowledge of the internal workings. For example, the authors in \cite{Takiddin23} proposes two adversarial attacks, namely nearest neighbor perturbation (NNP) and nearest neighbor distance (NND), to fool electricity reading and steal energy by iteratively generating adversarial samples. The authors in \cite{Jiwei22} introduce a signal-agnostic adversarial attack method to generate adversarial examples and degrade the performance of CNN-based power quality classification in smart grids.

Recently, the researchers in \cite{Santos22} proposed a novel method to generate adversarial perturbations successfully against power allocation in communication networks with few random samples. In another study, the authors \cite{wang2021adversarial} create adversarial reinforcement learning against dynamic channel access and power control by manipulated state information in deep Q-network (DQN). The work \cite{bahramali2021robust} proposed to use the perturbation generator model (PGM) to attack grid-based wireless systems. The PGM is strong enough to generate input-agnostic perturbations, with constraints like power, undetectability, and robustness to ensure effective and stealthy perturbations. Also, there is a significant negative correlation between attack power and channel estimation model performance. As attack power improves, the model's vulnerability decreases. The authors in \cite{Song21} and \cite{Song24} found that universal adversarial perturbations can damage the power grid state estimation with the same magnitude as the widely studied false data injection attacks do. In another case, the authors in \cite{WangYu23} introduce a novel destabilizing attack to target inverters in microgrids by altering the power control gains and their impact on small-signal stability. The studies in \cite{Junfei21} and \cite{Jiwei22} present two different stealthy adversarial attacks that target deep learning-based smart metering to inject false data or bypass the energy consumption counters. To enhance the stealthy of the attacks, the attacker can propose to infiltrate the demand side management \cite{Youssef23} by a combination of false data injection, such as changing the thermostat set points of heating, ventilation, and air conditioning (HVAC) systems and electric water heaters (EWHs), and adversarial perturbations. Unlike adversarial attacks on image classification tasks, adversarial attacks on smart metering may mislead not just human intuition but also power system error-checking procedures \cite{ZHANG2023121405}.

The common point of all the adversarial attacks is that the attacker tries to influence the deep learning model's decision by manipulating environmental factors (noise) or spoofing signal requests (transmission power). By degrading AI-based power grid management and scheduling or radio resource allocation performance, the stability for the \ac{SG2} control is negatively affected, particularly in \ac{SG2} interdependent control and communications, particularly with digital twin features \cite{SIFAT2023}. The authors in \cite{Sadegh19} and \cite{Takiddin23} showed both white-box and black-box adversarial attacks against electricity theft detectors. They created universal adversarial perturbations based on an electricity reading and its nearby readings. Compared to the prior attacks, the attacker significantly needs less transmit power to create misclassification, which reveals a fundamental flaw in DL-based solutions to communication systems. The authors in \cite{Ren24} propose variational auto-encoder-based adversarial attacks with the masqueraded malicious dataset to mislead dynamic pricing systems.


\subsection{Trustworthy AI for power grid management functions and smart resilience in \ac{SG2}}

\begin{figure*}[t]
    		\begin{center}
    			\includegraphics[width=1\linewidth]{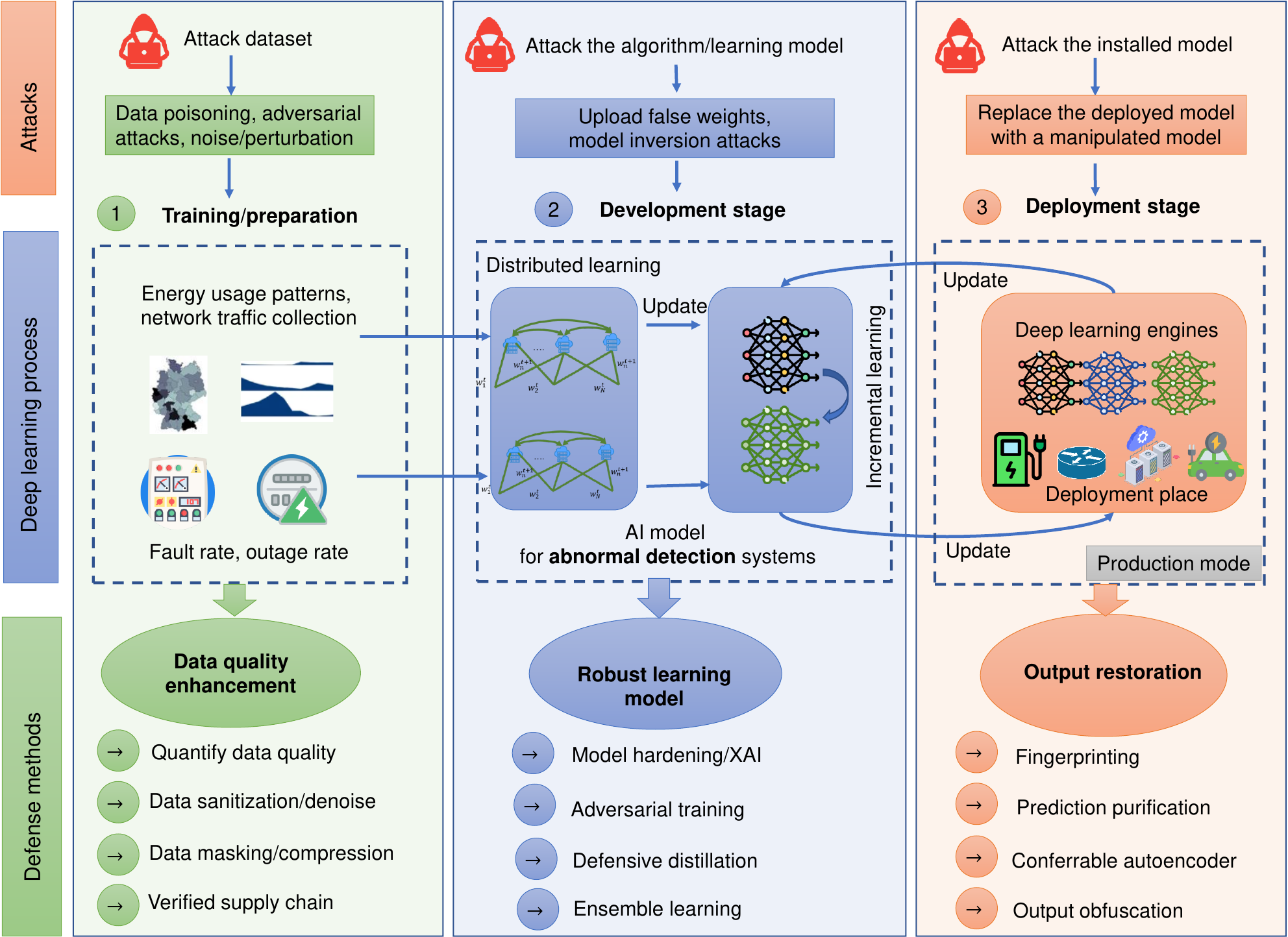}
    		\end{center}
    	\centering
    	 	\caption{Typical protection approaches against security attacks against AI-aided functions and energy management in \ac{SG2}. In essence, there should have a comprehensive protection strategy for all stages, from input data, training, to deployment stage (partially referred to materials in our prior work \cite{Nguyen21}).}
    	 	\label{fig:adversarial-attacks}
      \end{figure*}

The best protection model against the AI-targeted attacks is to provide a comprehensive approach, from protecting data for AI training to improve a neural network's or AI model's robustness. Fig.~\ref{fig:adversarial-attacks} summarizes three fundamental defense approaches to maintain trustworthy AI: protect input data against data poisoning (enhance data quality), protect AI design models against adversarial attacks (model protection), and protect running AI engines against model interference (output restoration). Each protection model is associated with a special stage: data preparation, AI development, and deployment. Since the attack methods are evolving, the defense methods are also expected to develop in parallel or beyond to keep up with the adversary. This subsection overviews defense strategies encompassing the optimization of model robustness, manipulation of model inputs, and assessment of the functional repercussions arising from neighboring weight interactions.

\subsubsection {Data quality enhancement}

Enhancing data quality is essential for developing trustworthy AI and preventing data poisoning attacks. High-quality data ensures that AI models are trained on accurate, relevant, and unbiased information, which improves their reliability and performance \cite{Nguyen21}. The basic idea behind data quality enhancement is to apply preprocessing strategies to the input examples to remove the influence of adversarial perturbation or outlier data points without changing the objects in the original picture of the examples. There are several methods for data quality enhancement: data sanitization (anomaly detection) and verified supply chain. For example, the authors in \cite {Addepalli2020Adversarial} introduced Structure-To-Signal Network (S2SNet), a protection mechanism based on gradient masking. The primary goal of this method is to convert category-related information into structural information, manage the information in the gradient of input data, and then encode the structural section utilized for classification and erase unrelated sections to remove adversarial perturbations. This method can be applied to the data layer with the feature extraction aspects or a part of the digital twins or AI-aided controllers in smart grids with interactive scheduling aspects. For verified supply chain, blockchain technology can be used to create a secure and transparent record of data provenance, ensuring that all data used for training models comes from verified and trustworthy sources \cite{Charithri21}.
	
Data masking/denoising and supply chain verification are also efficient approaches. For instance, in smart grids, personal data of consumers and payment bills can be masked to ensure no exposure of social security numbers and consumer home addresses while still allowing for the analysis of consumption patterns. Errors during the data preparation and collection should be avoid. The researchers \cite{Liao2018Defense} introduced a novel denoising approach, the so-called High-level Representation Guided Denoiser (HGD), to address the predicament of propagating residual errors. This approach uses the U-Net architecture to prevent mistakes from escalating in the input data. The HGD method outperforms an alternative approach when subjected to a limited subset of training data, thanks to the exclusive use of fine-tuning, a targeted mechanism to mitigate error amplification in the presence of adversarial examples. Another similar work is \cite{Elsisi23} where the authors propose to exploit continuous wavelet transform to build resilience models for mitigating adversarial attacks. Some other techniques are available in public outlier detection libraries \cite{PYOD}: Z-score, Isolation Forest, Variational Autoencoding Gaussian Mixture Model (VAEGMM), Prophet Time Series Outlier Detector.

\subsubsection{Model protection: optimize the model robustness and explainable AI}

There are many methods to optimize the learning model's robustness: model hardening, adversarial training/certifiable training. For example, the authors in \cite{Addepalli2020Towards} introduced a new Bit Plane Feature Consistency (BPFC)-based technique. Their primary goal is to generate a general impression from features in the higher-bit planes and then fine-tune their predictions using the lower-bit planes. By enforcing consistency within the lower bit planes to enhance predictive accuracy across diverse quantified images, the efficiency of deep neural networks (DNN) in adversarial contexts attains higher performance relative to conventionally trained models. Nonetheless, a constraint inherent to this approach pertains to the careful selection of specific lower-bit planes to tune their predictions to align with the required robustness of the DNN model. In another work, Kannan et al.\cite{Kannan2018Adversarial} offer a mixed-minibatch PGD (M-PGD) adversarial training approach that combines a logit pairing method with PGD adversarial training. M-PGD includes two pairing methods: (1) pairs a clean sample with an adversarial sample, and (2) pairs a clean sample with another clean sample. In \ac{SG2}, the adversarial training could mean training AI models for electric demand or metering data that includes possible faults or anomalies. 

For non-adversarial training approach, the authors in \cite{Huang23} introduce a causal structure learning model that recognizes the causal links in observational data and exploit the causal relationship between samples, outputs, and ground truth labels to limit the effect of adversarial attacks. Another prominent adversarial training technique is defensive distillation \cite{papernot2016distill}, which aims to reduce the effectiveness of adversarial samples on DNNs through a teacher-student learning process. This technique involves training a teacher model on benign data to produce soft probability distributions as targets. These soft targets are then used to train a student model, which helps the model generalize better and become more resistant to adversarial perturbations. The other studies in this category are \cite{Farajzadeh-Zanjani21,Takiddin232} where adversarial training anomaly detectors are robust against false data attacks and data poisoning. Another technique is certifiable training that refers to a set of techniques aimed at providing formal guarantees about the robustness of deep neural networks (DNNs) against adversarial attacks. Unlike traditional adversarial training methods that empirically improve robustness by training on adversarial examples, certifiable training focuses on creating models that can be mathematically proven to withstand certain types of perturbations within a specified range. For instance, the authors in \cite{wong2018provable} developed a method that uses convex relaxations to create an outer polytope that encloses all possible perturbations of the input. By optimizing this polytope during training, the model learns to maintain correct classifications within this bounded region, even processing with adversarial samples.

Besides adversarial training, ensemble ML models may increase resilience by pooling the findings of varied models to make better predictions. Because an attack that bypasses one model does not always bypass the others, ensembles can increase the learning's robustness against adversarial samples \cite{strauss2018ensemblemethodsdefenseadversarial}. Finally, Explainable AI (XAI) can indeed play a crucial role in preventing adversarial attacks and enhancing the trustworthiness of AI systems. For \ac{SG2}, explainability can assist operators in understanding why a certain decision was made, such as adjusting power distribution in response to demand fluctuations or charge/discharging decisions. This transparency can help detect and mitigate adversarial attacks by revealing unexpected or unusual model behaviors, thus enhancing the overall security and trust in \ac{SG2}. For example, the authors in \cite{GunningAha2019} summarized and emphasized that XAI methods contribute to building more robust and trustworthy AI systems by elucidating model behavior and improving human-AI interaction, particularly in US Defense Advanced Research Projects Agency (DARPA)'s programs. As a critical infrastructure to national security, in \ac{SG2}, XAI should be prioritized in AI-powered management tasks to balance the trade-off between the reliability/trust of decision-making and the high accuracy of the AI technique.

\subsubsection{Output restoration: weight optimization}

Output restoration is a key technique that aims to prevent adversarial attacks by ensuring the reliability of system outputs through fingerprinting, prediction purification, and output obfuscation. For example, fingerprinting involves embedding unique, traceable markers within the output data of grid systems, allowing operators to detect unauthorized changes or anomalies. The authors in \cite{FLARE23} proved a case to verify whether a suspected AI policy is an illegitimate copy of another (victim) policy. In \ac{SG2}, by embedding fingerprints in power usage reports, any tampering can be quickly identified and addressed. Prediction purification corrects outputs that are suspected of being influenced by adversarial inputs, particularly in model inversion and membership inference attacks \cite{yang2020defendingmodelinversionmembership}. For example, if an unexpected spike in energy demand is detected, purification methods can cross-verify with historical data and auxiliary models to filter out any potential malicious input. Output obfuscation adds controlled randomness to the outputs, such as slightly varying the reported energy usage data, making it difficult for attackers to predict and exploit the system's behavior \cite{athalye2018obfuscatedgradientsfalsesense}. These techniques collectively enhance the security and robustness of smart grid operations against adversarial threats.

\subsection{AI and Superintelligence for enhancing security protection models: intelligent IDS/IPS and abnormal detection}

As summarized in previous sections, many conventional solutions, such as firewalls and intrusion detection systems, have been the main forces to prevent security attacks. AI makes such systems more powerful and intelligent. While older security measures (such as signature-based intrusion detection) have been widely used, they have difficulty dealing with sophisticated threats in \ac{SG2} distributed energy resources and microgrids. Several studies \cite{ZHOU2022,LI20232,Nina21} highlighted that AI will be a core player in enhancing the efficiency and accuracy of intrusion detection in smart battery storage management of \ac{SG2}. AI can considerably assist in security defense by enhancing the detection engines' performance. For example, the authors in \cite{GALLEGO2023} advocated using deep reinforcement learning (DRL) to enhance secure peer-to-peer energy trading models. AI is a favored technique for detecting eavesdropping and DoS attacks in the network layer of the smart grid. With the large-scale learning, AI is particularly suitable for the following tasks:

\begin{itemize}
    \item  Abnormal behavior/intrusion detection in \ac{SG2}, such as convolutional neural networks (CNNs)/region search to detect electricity theft \cite{Donghuan19,Khan22,Liqiang23}, Spiking Neural Networks (SNN)/Temporal Failure Propagation Graph (TFPG) and federated learning to detect anomaly metering traffic and network intrusion activities \cite{Junho19,Chih-Che21,Chatzimiltis23} or in moving target defense (MTD)\cite{Yexiang22}.
     
    \item Isolating attack traffic and unauthorized access in \ac{SG2} enabling technologies, such as unsupervised machine learning to detect DDoS attacks \cite{Tianyang23} or ensemble learning to detect intrusion traffic \cite{Houda22} in software-defined \ac{SG2} networks. Another method is to exploit neural networks to enhance false data detection against state-of-charge estimation in energy storage systems \cite{Nina21,Zhuang21,Habib2023}.  
    
    \item Optimizing fault diagnosis and electrical power outage production in \ac{SG2}, such as distribution transformer fault prediction using deep neural networks \cite{Mhdawi20} and multi-data-source hybrid deep learning-based predictors to accurately predict energy usage \cite{Badr23}.
    
    \item Price charging prediction and load balance scheduling, such as deep reinforcement learning-based pricing strategy for profit maximization \cite{Chuang22} and smart isolated microgrids \cite{Jiaju23}. 
    
\end{itemize}

\begin{figure}[t]
	\begin{center}
		\includegraphics[width=1\linewidth]{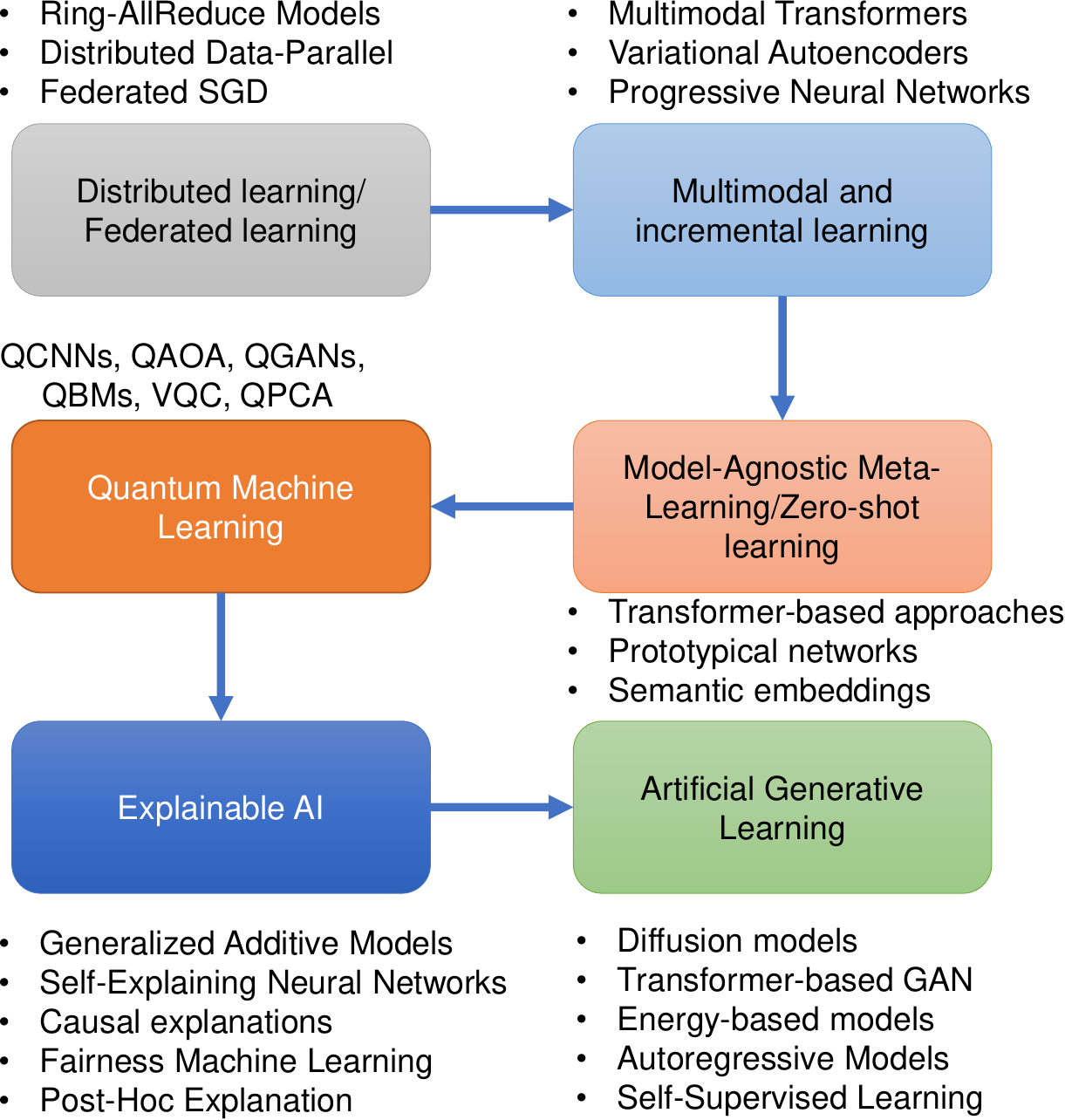}
	\end{center}
	\centering
	\caption{Summary of current AI models and future vision for intelligent cybersecurity in \ac{SG2}. As a result, the most dependable AI models will be favored for essential applications, and \ac{SG2} could be a primary objective. Several concepts in this vision have yet to be implemented fully as comprehensive solutions or security defense for power grid systems, e.g., quantum machine learning, AGI, explainable AI.}
	\label{fig:AI-promising-models}
\end{figure}

Inspired by recent progress in AI \cite{Salmon23}, AI for smart grid and renewable energy sources \cite{Bose17,Mohammadi22,MACHLEV22, ZHOU2022}, and related security \cite{Yayla22}, Fig.~\ref{fig:AI-promising-models}  summarizes our vision of AI promising models for cybersecurity, including for \ac{SG2}, in the coming years. Given the impending arrival of \ac{AGI} \cite{schuett2023best}, AI models in the aforementioned applications should be able to self-learn from their surroundings and adjust their capability to deal with unknown inputs. AGI for smart grids is likely a milestone in applying AI to critical infrastructure. Given strengths in learning from extremely large inputs and automatically optimizing choices based on continuous feedback from the environment, distributed learning, multimodal and incremental learning, quantum machine learning, meta-learning/zero-shot learning, and the combination of DRL and GAN \cite{strickland2023drlgan} will be top choices in this area. Similar to 6G, generative learning will be a crucial step toward the idea of ``AI-building AI'' or ``security design by AI'' vision \cite{Nguyen21}. However, the vision for AI and AGI popularity in \ac{SG2} raises new fears that the damage can be more catastrophic if AI-based functions are out of control due to adversarial attacks. For example, AGI can be used to design a sophisticated phishing attack by exploiting a public information and reports of the companies. AGI can expose the grid's operational protocols and identifies critical nodes through careless questions from the employees. 

Robust AI models with adversarial training and AGI safety and governance practices should be followed \cite{schuett2023best,Salmon23}. Furthermore, current AI models require large-scale data and expensive hardware accelerators with Graphics Processing Units (GPU) for training. To balance net zero and environment-friendly goals, training models may require a new approach. As a result, distributed learning and federated learning, which can coordinate the learning process across millions of distributed devices (local agents) to improve the quality of the centralized learning model (global agents), will likely be the top candidates. Finally, the blackbox of how a deep learning model works under specific conditions has been a concern for applying AI in critical fields like smart grids. As explained earlier, explainable AI (XAI)) models \cite{Yayla22} have been a major target for continuing work in cyber trust to reduce the risk of AI making unjustified judgments, particularly in controversial decisions, such as turning off crucial electric gates for particular communities to save for others or cutting off the electricity to prevent AI out of control. AGI for designing explainable AI can be a promising research direction. 


\section{Lessons Learned and Discussions on Future Research}
\label{sec:lessons-learned}

 \ac{SG2} will be a significant upgrade of the conventional smart grid framework, which passed through decades of research. The upgrade will highlight the automation, efficiency, and intelligence. Three primary trends to realize these goals are (1) empowering AI into many core components of smart grids, (2) upgrading communication technology infrastructure, and (3) developing new energy sources and trading, e.g., DERs, BESS. However, new advancements create new areas of security risk. This section summarizes key lessons learned through the survey and highlights promising research directions.

\subsection{Lessons learned from the survey}

From the survey, unified authentication, consistent security solutions and standards, and AI-aided security for \ac{SG2} emerging technologies such as new energy sources, electrified transportation, and P2P trading are the primary major upgrade targets and the central of ongoing research efforts. In summary, ten lessons learned from the review on security issues and defense methods for \ac{SG2} are as follows.

  \textcircled{\raisebox{-0.9pt}{1}} \textit{The transition from fossil energy to renewable energy era requires an evolution of digitizing management systems and enabling connectivity for all stages from energy generation, transmission, and distribution to consumption in \ac{SG2}}. Meanwhile, the networkization comes up with cybersecurity concerns, e.g., cascading failures of connectivity, DDoS. Once \ac{SG2} elements are connected, data collection from each stage can provide rich information for optimizing the operations of the entire power grid, e.g., by exploiting the computing capability of cloud-based platforms and software-defined control methods. However, the interdependence of control and communication technologies creates a new route for the attackers to remotely access and sabotage the power grids, particularly in cyberwarfare.
       
   \textcircled{\raisebox{-0.9pt}{2}} \textit{The diversity of smart metering devices can mitigate the problem of supply chains but potentially cause security fragmentation due to their incompatibility in using security protection standards}. For example, the meter devices can be built based on NB-IoT, LoRa, WiFi, or Zigbee \cite{Tala22}. However, these communication technologies are incompatible with all security designs, let alone their own vulnerabilities. Maintaining consistency and compatibility for communications between consumers and energy distributors is challenging but should be carefully considered in modernizing a smart grid platform. Maintaining a reliable supply chain for vital materials in smart grid infrastructure is also an important issue, given that grid security implies national safety.
       
   \textcircled{\raisebox{-0.9pt}{3}} \textit{The efficient authentication matter is one of the biggest challenges in \ac{SG2}}: Given the diversity of communication technologies for smart metering and industrial control standards, the authentication mechanisms are expected to vary. However, the heterogeneity in the protection level (e.g., weak authentication in LoRAWAN vs strong authentication in 5G core networks) may pose risks for the entire connected smart grids. The weak security network segment can become a vehicle to intrude on the core smart grid system. A unified authentication framework, e.g., via authentication, authorization, and accounting (AAA) services in a well-protected control center, can be a promising model. For example, unified authentication can significantly simplify the administration of user credentials in the era of multiple energy providers. The authentication server and protocols can be upgraded quickly for all devices.
       
   \textcircled{\raisebox{-0.9pt}{4}} \textit{Unlike IT systems, holistic protection in \ac{SG2} requires both digital and physical aspects of security}. For example, firewalls/IDS/Encryption (digital) are to secure communications, while wall/fencing/shield cages/supply chains (physical) are to protect power lines and components against physical attacks (remote drones, jamming vehicles). Most recorded attacks in US and other countries are software-based (malware, ransomware). However, with the popularity of drones and autonomous vehicles, non-traditional attacks like suicide UAVs, jamming vehicles, or supply chain breakage have been new severe threats that cannot be ignored while building the protection schemes for \ac{SG2}. In this way, building distributed microgrids and emergency stocks of critical grid components will be critical. Further, national regular drills/rigid testings of collaborative efforts among industry stakeholders in energy crisis scenarios are essential.   

   \textcircled{\raisebox{-0.9pt}{5}} \textit{Evolution for communication systems in smart grids often lags behind that for IT and mobile networks}: Given the importance of the energy systems in our society, technologies like networking and security protocols must usually be verified, particularly at stable and durable capacity, before they can be deployed in \ac{SG2} commercially. For example, while 5G technology is rapidly being deployed in mobile networks, its deployment in smart grids is delayed. Smart grids continue to rely on older 4G or even wired technologies \cite{abrahamsen2021communicationtechnologiessmartgrid} due to the extensive verification required to ensure 5G meets the high stability and reliability standards critical for energy systems, potentially slowing the adoption of advanced communication solutions in the energy sector. This slow-but-sure strategy has many advantages for maintaining stability and reliability in a critical field like \ac{SG2}. However, the lengthy verification and certification processes can delay the deployment of new technologies. This can hinder innovation and the ability to implement cutting-edge solutions quickly, potentially slowing down progress in rolling out new technologies. Finally, networking and attack techniques evolve quickly, and when a defense and communication technology is thoroughly verified and certified, it may already be somewhat outdated. Striking the right balance between thorough testing and timely deployment is crucial. Recently, the security-as-a-service and move-near-cloud models seem to be getting much attention, given their advantage of enabling fast innovation. 

    \textcircled{\raisebox{-0.9pt}{6}}\textit{AI-aided solutions for applications and core functions in \ac{SG2} are still in development with much potential for exploration}. Unlike computer vision and natural language processing, machine learning and deep learning research for improving performance in the smart grid's core functions, e.g., energy distribution, outage prediction, energy usage forecasting, and related security matters, are just at the early research stage. As we summarized in Fig.~\ref{fig:ai-aided-function}, there are many functions at all three layers where AI can help. However, currently, there are few research on this problem. The primary challenge is the lack of rigorous security datasets for key activities of \ac{SG2} and potential attacks, particularly in \ac{SG2} emerging technologies, e.g., electric vehicle charging. This is because of the concern for system safety or data leakage. Adversarial attacks and defense approaches against AI-powered energy management are still at the early research stage. Finally, the potential for using AI to enhance \ac{SG2} security is great, but it is not a magic wand that can solve all security challenges. Non-AI innovations, such as quantum-safe security protocols, continue to get significant industry support and are a part behind \ac{SG2} 's success.

    \textcircled{\raisebox{-0.9pt}{7}} \textit{Information interoperability has emerged as a vital factor in securing the heterogeneous networks of \ac{SG2}}: Energy distributors, energy generators, regulatory bodies, and technology providers often act in their interests, given the concern of secret trade and national security. However, cross-sector collaboration can help to identify emerging threats and vulnerabilities and enable the development of standardized security frameworks that can adapt to evolving challenges. The main challenge is that it is unclear how these sectors can share the information safely, given the privacy concerns and responsibility litigation. To address this, a secure API service can be created, where each sector has controlled access to specific data based on their role and privilege level. For instance, energy distributors can access threat intelligence data provided by cybersecurity firms, while regulatory bodies/policymakers can oversight compliance and incident reports without accessing sensitive operational details. This system is akin to Lawful Interception Services in 5G networks, where law enforcement agencies can access communication data through secure, predefined channels without compromising the privacy and security of the network as a whole. 

    \textcircled{\raisebox{-0.9pt}{8}} \textit{Blockchain can be the key player in securing communications and optimizing data exchange for peer-to-peer energy trading and distributed energy sources in \ac{SG2}}. The goal is to maintain the trust among distributed components of smart grids along with privacy preservation. Indeed, there have been many blockchain-based studies \cite{Park23,Hassan22,Kang17, AlSkaif22,Gai19,Kaur21} for this purpose. However, such a vision is unlikely to be realized until blockchain's expensive computing and security flaws (e.g., 51\% attacks, DAO attacks \cite{Shabani22}) are addressed \cite{Archana19}. Another major challenge is the scalability of the blockchain. Imagine that the blockchain is deployed in managing energy transactions in a large urban area such as New York City. In such a setting, millions of households and businesses would participate in a decentralized energy market, trading electricity produced from renewable sources. A blockchain network, e.g., Ethereum, has a limit of transaction processing, can process about 7 to 15 transactions per second \cite{LI2023103686} only. This is small compared to the thousands of transactions per second required in a large-scale busy smart grid. Additionally, the heavy energy consumption of blockchain networks, especially those using proof-of-work-based consensus mechanisms, would require significant upgrades on high-performance computing infrastructure, which is costly to invest in. Further, the latency issues of transaction confirmation can be up to 10 minutes \cite{LI2023103686} which is not suitable for the real-time processing needs of a smart grid. Furthermore, the storage requirements for storing a full ledger on every node would be massive, making it impossible for nodes to keep up with the fast development of transaction data. Currently there are several prominent solutions, such as sharding and layer-2 scaling protocols \cite{LI2023103686}, their architecture is complicated and has yet to be tested for large-scale deployment. As a result, blockchain in its present form faces severe scaling issues for \ac{SG2}, further studies on this matter will be an interesting topic.

    \textcircled{\raisebox{-0.9pt}{9}} \textit{Firewalls, IDS, DMZ, will not lose their roles in \ac{SG2}, but intelligence capabilities are mostly the most-looking-forward features}. For years, these platforms have established their repute for safeguarding networks from many threats and network intrusions. They are still significant players in \ac{SG2} and beyond. However, in order to retain detection efficiency in a complex environment with various connection technologies, these legacy systems need substantial enhancements in both predictive and adversarial attack resistance capabilities. AI-aided IDS/IPS are then the central efforts for the coming years.

    \textcircled{\raisebox{-0.9pt}{10}}  \textit{Privacy concerns, mainly when energy systems collect and process consumer data, remain a pressing challenge}. For example, the AI system requires rich data collected from consumers (energy consumption/charging habits) and infrastructure (electricity load in each region) to optimize their prediction models. However, these data can be potentially leaked, probably by sloppy management or insider attacks. Striking the right balance between data utilization for optimization and preserving consumer privacy demands robust data anonymization, encryption, and consent mechanisms. Addressing these challenges requires close collaboration with regulatory bodies to develop and enforce policies that protect consumer data without hindering energy innovation.

   In conclusion, reaching \ac{SG2} with full operation can be a long road ahead. Lessons learned from failures of the current security architecture along with security issues in new energy trading, new bidirectional relationship among distributed microgrids of \ac{SG2} highlight the need for comprehensive security frameworks tailored to the unique requirements of SG2, such as security interoperability for multiple energy sources, local power switch optimization for grid balancing, strong authentication mechanisms for decentralized microgrids, trustworthy AI for all AI-powered energy management systems, and enhancing well-proven technologies like IDS with AI capability.

\subsection{Open challenges and promising research directions}

Given the ongoing transition towards \ac{SG2} at the early stage, there are several open challenges and promising research directions for future studies. We summarize several typical ones as follows.

 \textcircled{\raisebox{-0.9pt}{1}} \textit{AI-aided solutions with resilience from blackouts or sudden outages}: As we mentioned earlier, there is a lack of AI research on core functions of smart grids (e.g., slicing, outage prediction, intelligent energy distribution, network balancing, firewall/IDS), given the challenges of collecting datasets and simulating the real environment once the system is operating. The AI-targeted attacks and defense research in this field are even less. Rigorous datasets in this field, particularly those that reflect the comprehensive activities or network traffic in metering, energy coordination, and disaster response, will be extremely welcomed by the research community. Without well-established datasets, enhancing the performance of AI-based solutions will be difficult. Besides, based on this baseline model, adversarial attacks can be further conducted to verify the resilience performance of AI models and then propose corresponding defense updates. Advanced AI models (as summarized in Fig.~\ref{fig:AI-promising-models}), such as few-shot learning/federated learning/multimodal incremental learning, can be another interesting topic to enhance AI capability on small data cases or privacy preservation.
      
 \textcircled{\raisebox{-0.9pt}{2}} \textit{Unified authentication for multiple microgrids:} The diversity of communication technologies for HAN/FAN (e.g., between the consumers and several energy providers) or distributed microgrids/energy storage technologies creates several different authentication techniques that may result in an inconsistency in accessing the system. A unified authentication can prevent this. For example, consumers may handle power service payments, car charging, battery swap, and P2P electricity trading with one identity and access point. Consumers can access different systems with one set of credentials via single sign-on (SSO), boosting convenience and minimizing password fatigue. However, the solution may need to consider many constraints, e.g., interoperability issues arise due to the varying network protocols and systems used by different grid operators, making seamless integration challenging. Secondly, the risk of a single point of failure increases, as a breach in the unified system could compromise the security of the entire grid. Additionally, the complexity of managing and synchronizing authentication data across multiple providers can lead to inefficiencies and vulnerabilities. The willingness of grid suppliers to share customer authentication information with competitors is a concern in business. Integrating EAP-TLS authentication to enable API calls and creating a uniform authentication system via these API calls might be a nice start. 

  \textcircled{\raisebox{-0.9pt}{3}} \textit{Developing custom 5G/6G private networks for \ac{SG2}}: Custom 5G/6G private networks can revolutionize network infrastructure's security in \ac{SG2} significantly, given the reputation of solid security protection and fast innovation in cellular network technologies. While 5G/6G promises enhanced connectivity and security, the complexity and scale of these networks demand for distributed energy sources in large areas, making it challenging for communication operators to justify the private investment for different grid operators. Those high costs can be the initial investment on private base stations, network components/routers, and 5G-based smart meters for millions of consumers. Additionally, the ongoing operational expenses, such as network management and periodic upgrades, further strain the financial resources of power grid operators. However, in SG2, the practicality of 5G/6G private networks is enhanced due to the maturity of end-to-end network slicing technologies, which allows for the efficient building of private networks (under the same hardware infrastructure). 
  For now, operators may still rely more realistic solutions by implementing radio slicing and VPN along with endpoint security on public communication networks that effectively balance cost, complexity, and security. 

  \textcircled{\raisebox{-0.9pt}{4}}\textit{Security defense for autonomous systems and green energy transactions}: The recent transition from gasoline cars to electric vehicles, automated cars, and new charging infrastructure creates a unique chance for software and network-based solutions to optimize the efficiency of data transmission as well as energy distribution. This will remind us of the importance of in-vehicle security protection techniques (abnormal behavior detection, thief charging detection) and cooperative verification (identity verification, Sybil attack detection) in well-connected smart vehicles and power grid infrastructure. However, this topic has not yet been well-explored. 

    \textcircled{\raisebox{-0.9pt}{5}} \textit{New hardware security techniques for preventing physical attacks}: Physical attacks are common in \ac{SG2}, and the consequence can be a catastrophe if important transformers are victims. Besides physical shields, hardware security techniques, such as trusted execution environment (TEE), physical unclonable function (PUF), post-quantum cryptography, chip-to-chip authentication, and tamper-evident packaging, can be critical solutions that can be enhanced in the future. Further studies and prototypes on these topics are welcome.

    \textcircled{\raisebox{-0.9pt}{6}} \textit{Enhancing privacy technologies with affordable costs}: Smart grids collect a vast amount of data about energy consumption and generation, and this data can be used to identify individuals and their activities. It is important to protect this data from unauthorized access and use, but doing so can be expensive. Promising solutions can be using secure multi-party computation (allows multiple parties to compute a function on their data without revealing their data to each other), differential privacy (adding noise to data without sacrificing its accuracy), or holomorphic encryption (allows computations to be performed on encrypted data) \cite{Nguyen21}. 

   \textcircled{\raisebox{-0.9pt}{7}} \textit{Digital Twin for speeding up security research innovation for \ac{SG2}}: Digital Twin (DT), a digital representation of a physical object, person, or process, contextualized in a digital version of its environment, can speed up security research innovation in \ac{SG2} by providing a safe and controlled environment for researchers to experiment and test new security solutions. DT can be used to simulate and analyze the behavior of the control systems, too. In the context of \ac{SG2}, DTs can be used to create virtual representations of the grid's infrastructure, including its power plants, transmission lines, and distribution systems. These DTs can then be used to simulate various security threats and test the effectiveness of different security solutions. However, this research direction is still in its infancy \cite{SIFAT2023}.

    \textcircled{\raisebox{-0.9pt}{8}} \textit{Supply chain security for microgrids}: As we mentioned earlier, the weakness in any part of the supply chain, e.g., shortage of electric transformers, could encourage the attackers to attack specific components (physical attacks or malware for the specific targets) to disable the grid operators. Given the growing fears of many trade wars, trade protectionism, and global conflicts, developing an efficient mechanism to monitor supply chain changes and suspicious activity to avoid relying on a single supplier for any critical component or service can help mitigate the risks and enhance the resilience. Besides technical solutions, building a list of trusted suppliers through vendor risk management protocols and trade initiatives can be a very helpful approach, particularly with microgrid operators.

\section{Conclusion}
\label{sec:conclusion}
	
  \ac{SG2} is still on the way to shaping its unique characteristics vs the current smart grid model (i.e., SG1). Through lessons learned from the limitations of SG1 security implementations and the remaining technical challenges in SG1 infrastructure, this study highlighted that security protection and many techniques in \ac{SG2} lag behind the fast innovation of overall networking and security technologies. Further, the market of core security functions (e.g., authentication/firewalls) is rich but fragmented in standards. Given the high dependence on communication networks to connect distributed microgrids in \ac{SG2}, potential cascading failures of connectivity poses many severe security concerns. Besides suffering conventional attacks such as ransomware/DoS attacks, there are a growing number of new threats in full operation \ac{SG2}, such as physical attacks against substations and energy storage or adversarial attacks against intelligent AI-empowered energy management. This work provided an overview of key security threats and prospective solution comparisons in the vision of three stakeholders in \ac{SG2}: power grid operator, communication network provider, and consumer. Security threats and protection models for emerging technologies that are expected to be deployed widely in SG2 are also presented. We found that many advanced security models (e.g., network slicing, scalable blockchain, post-quantum encryption, trustworthy AI for \ac{SG2}) are still in the early stage of the research-prototyping loop. Without these new features, SG2 likely relies on the existing platforms, which are mostly outdated and consist of security vulnerabilities. Finally, we believe holistic protection in \ac{SG2} requires collaborative efforts to implement the aforementioned protection and power switch optimization models from many stakeholders, policymakers, cybersecurity experts, and standardization bodies. Further, information interoperability has emerged as a vital factor in securing the heterogeneous networks of SG2.
  
\section{Acknowledgements}
The authors would like to thank anonymous reviewers for their helpful comments. The authors also thank the Energy Administration of the Ministry of Economic Affairs in Taiwan for its support. This work is also sponsored by the National Science Technology Council under Grant No 111-2222-E-194-007-MY2, 112-2221-E-194-017-MY3, and in part by the Advanced Institute of Manufacturing with High-Tech Innovations (AIM-HI) through the Featured Areas Research Center Program of the Ministry of Education.

    \bibliographystyle{ieeetr}
    \bibliography{References}

\end{document}